\documentclass[%
superscriptaddress,
nofootinbib,
twocolumn,
amsmath,amssymb,
aps,
prd,
a4paper
]{revtex4-2}

\usepackage[colorlinks=true,linkcolor=blue,citecolor=blue,linktocpage=false]{hyperref}

\usepackage{newtxtext,newtxmath}
\usepackage{graphicx}% Include figure files
\usepackage{dcolumn}% Align table columns on decimal point
\usepackage{bm}% bold math
\usepackage{color}
\usepackage{enumitem}
\usepackage{placeins}
\usepackage{multirow}

\begin{document}

\preprint{}

\title{
IR-Safe and IR-Resummed Bispectra Before and After Reconstruction in Unified Lagrangian Perturbation Theory
}

\author{Naonori Sugiyama}
\email{nao.s.sugiyama@gmail.com}
\affiliation{Independent Researcher, Tokyo, Japan}
\affiliation{National Astronomical Observatory of Japan, Mitaka, Tokyo 181-8588, Japan}%Lines break automatically or can be forced with \\
\thanks{Special Visiting Researcher (non-salaried)}

\date{\today}% It is always \today, today,
             %  but any date may be explicitly specified

\begin{abstract}
We develop a unified analytic framework for modeling the real-space dark matter bispectrum, including both auto and cross bispectra constructed from pre- and post-reconstruction density fields, based on Unified Lagrangian Perturbation Theory (ULPT). ULPT reorganizes the standard Lagrangian approach by separating the Jacobian deviation, which generates the nonlinear source bispectrum, from the displacement-mapping effect that determines how long-wavelength bulk flows distort observed configurations. Within this structure, we derive general one-loop ULPT expressions for the bispectrum and analyze their infrared (IR) behavior, demonstrating exact, nonperturbative IR cancellation and the natural emergence of an IR-resummed description of baryon acoustic oscillation (BAO) damping. In particular, ULPT enables a detailed and fully analytic treatment of the wiggle--wiggle contribution to the IR-resummed bispectrum, whose structure has remained comparatively unexplored in previous approaches. We further construct IR-resummed models for the cross bispectra of all pre- and post-reconstruction combinations. For configurations in which the displacement fields differ, ULPT captures not only the overall exponential damping but also the more intricate BAO-scale modulation characteristic of mixed pre/post bispectra. Our results clarify the physical origin of nonlinear BAO suppression and provide a compact theoretical framework in which the full ULPT bispectrum, once implemented numerically through displacement-mapping convolution integrals, will automatically encode all nonlinear BAO and IR effects. The framework developed here therefore offers a unified and IR-safe foundation for interpreting next-generation bispectrum measurements.
\end{abstract}

%\keywords{Suggested keywords}%Use showkeys class option if keyword
                              %display desired
\maketitle

\section{Introduction}
\label{sec:intro}

The large-scale structure of the Universe encodes a wealth of cosmological information. Most of this information is captured by the two-point correlation function or its Fourier counterpart, the power spectrum, which serves as a primary observable in large-scale structure analyses. In contrast, the bispectrum arises from nonlinear effects such as gravitational evolution~\cite{Bernardeau:2001qr}, redshift-space distortions (RSDs)~\cite{Kaiser:1987qv}, and galaxy bias~\cite{Desjacques:2016bnm}, thereby providing additional constraints beyond those accessible from the power spectrum alone. Recent studies have also shown that the reconstruction technique~\cite{Eisenstein:2006nk}, originally developed to sharpen the baryon acoustic oscillation (BAO) feature~\cite{Sunyaev:1970eu,Peebles:1970ag}, improves parameter constraints more broadly by enhancing the information content of both the power spectrum~\cite{Hikage:2020fte,Wang:2022nlx} and the bispectrum~\cite{Shirasaki:2020vkk}. To fully exploit the potential of these statistics, it is essential to incorporate all relevant nonlinear effects, including gravitational dynamics, RSD, galaxy bias, and reconstruction, within a consistent and unified theoretical framework.

A key conceptual ingredient in the perturbative modeling of cosmological statistics, such as the power spectrum and bispectrum, is the treatment of infrared (IR) effects induced by long-wavelength bulk flows. In the limit of extremely large-scale displacements, these flows do not alter the shape of small-scale correlation functions, a property known as IR safety or IR cancellation~\cite{Jain:1995kx,Scoccimarro:1995if,Kehagias:2013yd,Peloso:2013zw,Sugiyama:2013pwa,Sugiyama:2013gza,Blas:2013bpa,Blas:2015qsi,Lewandowski:2017kes}.

In contrast, long-wavelength modes can have nonperturbative effects that give rise to distinct IR phenomena. These include the smearing of the BAO feature~\cite{Baldauf:2015xfa,Blas:2016sfa} and the exponential damping that arises when correlating density fields with different large-scale displacements. Representative examples of this cross-correlation damping include the cross spectrum between pre- and post-reconstruction fields~\cite{Wang:2022nlx,Sugiyama:2024eye}, the cross correlation between nonlinear and linear density fields, often referred to as the propagator~\cite{Crocce:2007dt}, and the cross spectrum of density fields evaluated at different redshifts~\cite{Chisari:2019tig}. To obtain robust predictions for the power spectrum and bispectrum, these IR effects must be resummed in a nonperturbative manner.

To construct a theoretical framework that achieves nonperturbative IR safety, it is advantageous to adopt Lagrangian perturbation theory (LPT), in which the displacement field serves as the fundamental variable. A recent study~\cite{Chen:2024pyp} investigated the IR behavior of the bispectrum within the Zel'dovich approximation (ZA), an LPT formulation based on the linear displacement field. Their results demonstrated that the full Lagrangian calculation provides an IR-safe representation of the bispectrum and clarified its connection to standard IR-resummation schemes, including the Gaussian damping of BAO wiggles~\cite{Blas:2016sfa}.

As an alternative IR-safe formulation, we recently proposed the Unified Lagrangian Perturbation Theory (ULPT)~\cite{Sugiyama:2025ntz}. This framework reorganizes standard LPT so that Lagrangian bias, RSD, and density-field reconstruction can all be treated consistently within a single formalism. A central feature of ULPT is the explicit decomposition of the density contrast into two physically distinct components: the Jacobian deviation, which captures intrinsic nonlinear growth including bias effects, and the displacement-mapping effect, which accounts for large-scale convective coordinate shifts. This structural separation guarantees IR safety for observable statistics such as the power spectrum and bispectrum. As a consequence, ULPT exhibits exact nonperturbative IR cancellation, naturally reproduces IR-resummed descriptions of BAO damping, and predicts exponential suppression in cross spectra between fields with different large-scale displacements, including the pre/post-reconstruction cross spectrum.

Beyond its structural advantages for understanding IR effects, the usefulness of ULPT has also been demonstrated in accurately predicting the power-spectrum shape for both dark matter and biased tracers. For dark matter~\cite{Sugiyama:2025myq}, the one-loop ULPT predictions match high-precision emulators such as Dark Emulator~\cite{Nishimichi:2018etk} and Euclid Emulator~2~\cite{Euclid:2020rfv} at the 2--3\% level up to $k \simeq 0.4\,h\,\mathrm{Mpc}^{-1}$ for $z \geq 0.5$, without introducing any nuisance parameters beyond the cosmological parameters. For biased tracers such as halos and galaxies~\cite{Sugiyama:2025juk}, ULPT naturally provides a renormalization-free bias sector built solely from Galileon-type operators~\cite{Chan:2012jj,Assassi:2014fva}. In each halo mass and redshift bin, a single set of renormalization-free bias parameters simultaneously fits the halo--halo and halo--matter power spectra, achieving subpercent-level accuracy up to $k \simeq 0.3\,h\,\mathrm{Mpc}^{-1}$ for $b_1 \in [0.8, 2.0]$ and up to $k \simeq 0.2\,h\,\mathrm{Mpc}^{-1}$ for $b_1 \sim 3$.

A fast numerical implementation of ULPT is available in the \texttt{\small ULPTKIT} Python package\footnote{\url{https://github.com/naonori/ulptkit}}, which uses \texttt{FAST-PT}~\cite{McEwen:2016fjn,Fang:2016wcf} together with \texttt{FFTL{\small OG}}~\cite{Hamilton:1999uv,mcfit} to carry out the computations required by the ULPT formulation. With these tools, \texttt{\small ULPTKIT} computes the one-loop ULPT power spectrum within 1--2 seconds.

In this work, we take the first step toward extending ULPT to the bispectrum. We derive a general expression for the ULPT bispectrum of dark matter in real space, valid for arbitrary triangular configurations and applicable to both pre- and post-reconstruction cases, and we present its one-loop form. We further demonstrate that the bispectrum inherits the IR-safety properties of ULPT, including exact IR cancellation, the natural emergence of an IR-resummed bispectrum model, and exponential damping in cross bispectra between pre- and post-reconstruction fields.

Because ULPT consistently describes both pre- and post-reconstruction fields within a single theoretical framework, it naturally leads to an IR-resummed bispectrum model applicable at either stage of reconstruction. Previous studies~\cite{Wang:2022nlx,Hikage:2020fte} have shown that, in the context of the power spectrum, both the post-reconstruction spectrum and the cross spectrum between pre- and post-reconstruction fields carry additional cosmological information beyond that contained in the pre-reconstruction spectrum alone. Extending this idea to the bispectrum is therefore expected to enhance the constraining power of future reconstruction-based analyses, as suggested in Ref.~\cite{Shirasaki:2020vkk}. Developing a theoretical model for the post-reconstruction bispectrum within ULPT thus provides a consistent foundation for detailed analyses of reconstructed data in upcoming surveys.

Existing IR-resummed bispectrum models commonly apply the wiggle/no-wiggle decomposition, which induces exponential damping in several components, including the wiggle--wiggle and wiggle--no-wiggle terms. Among these, the wiggle--wiggle contribution, representing the product of the wiggle parts, shows slight discrepancies between previous formulations such as Refs.~\cite{Blas:2016sfa,Sugiyama:2020uil}. Our ULPT-based formulation provides new insight into the origin and structure of this term from the perspective of ULPT.

A numerical evaluation of ULPT bispectrum shapes is left for future work, since the goal of the present study is to establish the analytic foundation.

Throughout this work, we adopt a fiducial cosmology consistent with the Planck 2015 best-fit $\Lambda$CDM model~\cite{Planck:2015fie}. The cosmological parameters are specified as follows: physical baryon density $\omega_b = 0.02225$, physical cold dark matter density $\omega_c = 0.1198$, dark energy density $\Omega_{\mathrm{de}} = 0.6844$, scalar spectral index $n_s = 0.9645$, amplitude of primordial curvature perturbations $\ln(10^{10} A_s) = 3.094$, dark energy equation-of-state parameter $w_0 = -1$, and total neutrino mass $\sum m_\nu = 0.06\,\mathrm{eV}$. The Hubble parameter is then determined to be $h = 0.6727$ from the flatness condition.

This paper is organized as follows. In Sec.~\ref{sec:recon_density}, we briefly review the unified Lagrangian formulation of density fluctuations before and after reconstruction within the ULPT framework. In Sec.~\ref{sec:bispectrum}, we derive the general expression for the real-space dark matter bispectrum in ULPT. The one-loop ULPT bispectrum and its perturbative structure are presented in Sec.~\ref{sec:ULPT_bispec_1loop}. In Sec.~\ref{sec:IRcancel}, we analyze the IR properties of the bispectrum and demonstrate exact nonperturbative IR cancellation for configurations in which all density fields share the same displacement. In Secs.~\ref{sec:IR_resum} and \ref{sec:cross_bispectrum}, we construct IR-resummed bispectrum models and investigate in detail the BAO damping and phase modulation for both auto and mixed pre/post cross bispectra. Future prospects and possible extensions are discussed in Sec.~\ref{sec:future}. Finally, we summarize our main conclusions in Sec.~\ref{sec:conclusion}. An intuitive physical interpretation of the universal exponential damping factor appearing in mixed pre/post bispectra is provided in Appendix~\ref{sec:appendix_universal}.

\section{Unified Lagrangian Formulation of Density Fluctuations Before and After Reconstruction}
\label{sec:recon_density}

In this section, we summarize the expressions for the dark matter density fluctuations before and after reconstruction within the ULPT framework. A complete derivation of these results is presented in Ref.~\cite{Sugiyama:2025ntz}.

\subsection{Dark matter density contrast}
\label{subsec:delta_dm}

We begin by reviewing the ULPT expression for the dark matter density contrast. The transformation from Lagrangian coordinates $\boldsymbol{q}$ to Eulerian coordinates $\boldsymbol{x}$ is defined by
\begin{equation}
    \boldsymbol{x} = \boldsymbol{q} + \boldsymbol{\Psi}(\boldsymbol{q}),
    \label{eq:xqPsi}
\end{equation}
where $\boldsymbol{\Psi}(\boldsymbol{q})$ denotes the Lagrangian displacement field. Under this mapping, the volume element transforms according to the Jacobian determinant
\begin{equation}
    d^3x = J(\boldsymbol{q})\, d^3q, \qquad
    J(\boldsymbol{q}) \equiv \det\!\left( \frac{\partial \boldsymbol{x}}{\partial \boldsymbol{q}} \right).
    \label{eq:jacobian}
\end{equation}

Mass conservation requires that the matter density $\rho(\boldsymbol{x})$ and its homogeneous background value $\bar{\rho}$ satisfy
\begin{equation}
    \rho(\boldsymbol{x})\, d^3x = \bar{\rho}\, d^3q .
\end{equation}
Introducing the density contrast through
\begin{equation}
    \rho(\boldsymbol{x}) = \bar{\rho}\,[1+\delta(\boldsymbol{x})],
\end{equation}
we obtain
\begin{equation}
    \delta(\boldsymbol{q} + \boldsymbol{\Psi}(\boldsymbol{q})) = \frac{1}{J(\boldsymbol{q})} - 1 .
    \label{eq:delta_L}
\end{equation}

A convenient starting point for expressing the Eulerian density field in terms of Lagrangian coordinates is the identity
\begin{equation}
    \delta(\boldsymbol{x}) = \int d^3x'\, \delta(\boldsymbol{x}')\, \delta_{\rm D}(\boldsymbol{x} - \boldsymbol{x}'),
\end{equation}
where $\delta_{\rm D}$ is the Dirac delta function. Changing variables via $\boldsymbol{x}' = \boldsymbol{q} + \boldsymbol{\Psi}(\boldsymbol{q})$ and applying Eqs.~\eqref{eq:jacobian} and \eqref{eq:delta_L} yields
\begin{equation}
    \delta(\boldsymbol{x}) =
    \int d^3q\, \delta_{\rm J}(\boldsymbol{q})\,
    \delta_{\rm D}(\boldsymbol{x} - \boldsymbol{q} - \boldsymbol{\Psi}(\boldsymbol{q})),
    \label{eq:delta_ULPT}
\end{equation}
where we have introduced the Jacobian deviation
\begin{equation}
    \delta_{\rm J}(\boldsymbol{q}) \equiv 1 - J(\boldsymbol{q}) .
\end{equation}

Alternatively, Eq.~\eqref{eq:delta_ULPT} may be rewritten entirely in Eulerian coordinates as
\begin{equation}
    \delta(\boldsymbol{x})
    = \sum_{n=0}^\infty \frac{(-1)^n}{n!}
    \partial_{i_1} \cdots \partial_{i_n}
    \!\left[
        \Psi_{i_1}(\boldsymbol{x}) \cdots \Psi_{i_n}(\boldsymbol{x})\,
        \delta_{\rm J}(\boldsymbol{x})
    \right],
    \label{eq:d_dJ_Psi}
\end{equation}
where $\partial_i \equiv \partial/\partial x_i$. For $n=0$, the expression reduces to the Jacobian deviation alone, reflecting the intrinsic growth contribution.

Taking the Fourier transform of Eq.~\eqref{eq:delta_ULPT}, we obtain
\begin{equation}
    \widetilde{\delta}(\boldsymbol{k})
    = \int d^3q\,
    e^{-i\boldsymbol{k}\cdot\boldsymbol{q}}\,
    e^{-i\boldsymbol{k}\cdot\boldsymbol{\Psi}(\boldsymbol{q})}\,
    \delta_{\rm J}(\boldsymbol{q}),
    \label{eq:delta_ULPT_F}
\end{equation}
where tildes denote Fourier-space quantities.

Equation~\eqref{eq:delta_ULPT_F} highlights the essential structure of ULPT: the density contrast naturally decomposes into two physically distinct components. The \textit{Jacobian deviation} $\delta_{\rm J}$ encapsulates the intrinsic linear and nonlinear growth, while the \textit{displacement-mapping effect}, encoded in the exponential term involving $\boldsymbol{\Psi}$, describes the nonlinear convective remapping of the density field.

\subsection{Reconstructed density contrast}
\label{subsec:delta_rec}

The density-field reconstruction procedure~\cite{Eisenstein:2006nk} modifies the observed density field by applying a displacement vector computed from a smoothed version of the density contrast. The reconstruction shift vector $\boldsymbol{t}(\boldsymbol{x})$ is defined as
\begin{equation}
    \boldsymbol{t}(\boldsymbol{x})
    = i \int \frac{d^3k}{(2\pi)^3}\,
    e^{i\boldsymbol{k}\cdot\boldsymbol{x}}\,
    \boldsymbol{R}(\boldsymbol{k})\, \widetilde{\delta}(\boldsymbol{k}),
    \label{eq:t}
\end{equation}
where $\boldsymbol{R}(\boldsymbol{k})$ is the reconstruction vector kernel,
\begin{equation}
    \boldsymbol{R}(\boldsymbol{k}) = -\frac{\boldsymbol{k}}{k^2}\, W_{\rm G}(kR),
    \label{eq:R}
\end{equation}
and $W_{\rm G}(kR) = \exp(-k^2 R^2 / 2)$ is the Gaussian smoothing kernel with characteristic scale $R$.

The reconstructed density field $\delta_{\rm rec}(\boldsymbol{x}_{\rm rec})$ is obtained by convolving the original density contrast with the shifted coordinate frame~\cite{Shirasaki:2020vkk,Sugiyama:2020uil}:
\begin{equation}
    \delta_{\rm rec}(\boldsymbol{x}_{\rm rec})
    = \int d^3x\,
    \delta(\boldsymbol{x})\,
    \delta_{\rm D}(\boldsymbol{x}_{\rm rec} - \boldsymbol{x} - \boldsymbol{t}(\boldsymbol{x})).
\end{equation}
Substituting the ULPT expression for $\delta(\boldsymbol{x})$ from Eq.~\eqref{eq:delta_ULPT}, we obtain
\begin{equation}
    \delta_{\rm rec}(\boldsymbol{x}_{\rm rec})
    = \int d^3q\,
    \delta_{\rm J}(\boldsymbol{q})\,
    \delta_{\rm D}(\boldsymbol{x}_{\rm rec}
    - \boldsymbol{q} - \boldsymbol{\Psi}_{\rm rec}(\boldsymbol{q})),
    \label{eq:delta_rec}
\end{equation}
where the reconstructed displacement field is defined as
\begin{equation}
    \boldsymbol{\Psi}_{\rm rec}(\boldsymbol{q})
    = \boldsymbol{\Psi}(\boldsymbol{q})
    + \boldsymbol{t}(\boldsymbol{q} + \boldsymbol{\Psi}(\boldsymbol{q})).
    \label{eq:Psi_rec}
\end{equation}

In Fourier space, the reconstructed density contrast becomes
\begin{equation}
    \widetilde{\delta}_{\rm rec}(\boldsymbol{k})
    = \int d^3q\,
    e^{-i\boldsymbol{k}\cdot\boldsymbol{\Psi}_{\rm rec}(\boldsymbol{q})}\,
    \delta_{\rm J}(\boldsymbol{q}),
    \label{Eq:delta_rec_F}
\end{equation}
which mirrors the structure of the pre-reconstruction expression.

Importantly, the reconstruction procedure alters only the displacement field; the Jacobian deviation $\delta_{\rm J}$ remains unchanged. As a result, the decomposition into the intrinsic growth component $\delta_{\rm J}$ and the displacement-mapping contribution persists after reconstruction, preserving the ULPT formulation across both pre- and post-reconstructed density fields.

\subsection{Bias and RSD within the ULPT framework}

As shown in Ref.~\cite{Sugiyama:2025ntz}, the ULPT framework consistently incorporates not only reconstruction but also galaxy bias and RSD. In the presence of bias, the Lagrangian bias fluctuation $\delta_{\rm b}(\boldsymbol{q})$ is added to the Jacobian deviation, forming the combined source term $\delta_{\rm J} + \delta_{\rm b}$. Thus, bias modifies only the intrinsic growth component encoded in the Jacobian deviation while leaving the displacement-mapping effect unchanged. Furthermore, ULPT naturally yields a renormalization-free formulation of galaxy bias, eliminating the need for counterterms at the operator level. Its practical effectiveness has been demonstrated in the power spectrum analysis of Ref.~\cite{Sugiyama:2025juk}, where accurate predictions were obtained using only a minimal set of bias parameters. This success indicates that ULPT-based bias modeling can be robustly extended to higher-order statistics such as the bispectrum.

In contrast, RSDs modify the displacement field through the contribution of the line-of-sight peculiar velocity. Since the Jacobian deviation is defined directly in terms of the displacement field, both the displacement-mapping effect and the Jacobian deviation are affected. The ULPT framework is designed to accommodate these complexities. By expressing all contributions to the density contrast, including bias, RSD, and reconstruction, in terms of two fundamental components, namely the Jacobian deviation and the displacement field, ULPT provides a unified and internally consistent description. This structural coherence is the basis for referring to the framework as ``Unified.''

Although the present work focuses on the dark matter field in real space, both before and after reconstruction, and does not explicitly include bias or RSD, the mathematical structure of the bispectrum and the resulting BAO behavior derived in this paper remain valid when these effects are taken into account due to the unified nature of the formulation.

\section{Bispectrum in ULPT}
\label{sec:bispectrum}

In this section, we derive the general expression for the bispectrum of real-space dark matter density fluctuations within the ULPT framework.

\subsection{General expression for the bispectrum}

The bispectrum is defined as the ensemble average of the product of three Fourier-transformed density fluctuations:
\begin{equation}
    \langle \tilde{\delta}(\boldsymbol{k}_1)
    \tilde{\delta}(\boldsymbol{k}_2)
    \tilde{\delta}(\boldsymbol{k}_3)  \rangle
    = (2\pi)^3\delta_{\rm D}(\boldsymbol{k}_{123})\, B(\boldsymbol{k}_1,\boldsymbol{k}_2,\boldsymbol{k}_3),
\end{equation}
where $\boldsymbol{k}_{123} \equiv \boldsymbol{k}_1+\boldsymbol{k}_2+\boldsymbol{k}_3$.

The bispectrum is symmetric under permutations of the wavevectors and satisfies the triangle condition $\boldsymbol{k}_{123} = \boldsymbol{0}$. As a result, it depends on only two of the three wavevectors and is fully specified by any two independent wavevectors forming a triangle. For convenience, it is often expressed as a sum over cyclic permutations:
\begin{equation}
    B(\boldsymbol{k}_1,\boldsymbol{k}_2,\boldsymbol{k}_3)
    = B_{12}(\boldsymbol{k}_1,\boldsymbol{k}_2) + B_{23}(\boldsymbol{k}_2,\boldsymbol{k}_3) + B_{31}(\boldsymbol{k}_3,\boldsymbol{k}_1),
    \label{eq:B12_23_31}
\end{equation}
where $B_{\alpha\beta}(\boldsymbol{k}_{\alpha}, \boldsymbol{k}_{\beta})$ is defined for $(\alpha, \beta) = (1,2), (2,3), (3,1)$ and is symmetric under exchange of its arguments.

To compute the bispectrum in ULPT, we begin with Eq.~\eqref{eq:delta_ULPT_F}, which expresses the density contrast in terms of the displacement field and Jacobian deviation. The corresponding ensemble average of three Fourier modes is given by
\begin{align}
    & \langle \tilde{\delta}(\boldsymbol{k}_1)
    \tilde{\delta}(\boldsymbol{k}_2)
    \tilde{\delta}(\boldsymbol{k}_3)  \rangle \nonumber \\
    &= \int d^3q_1\, e^{-i \boldsymbol{k}_1 \cdot \boldsymbol{q}_1}
        \int d^3q_2\, e^{-i \boldsymbol{k}_2 \cdot \boldsymbol{q}_2}
        \int d^3q_3\, e^{-i \boldsymbol{k}_3 \cdot \boldsymbol{q}_3} \nonumber \\
    &\quad \times
    \Big\langle
    e^{-i\boldsymbol{k}_1 \cdot \boldsymbol{\Psi}(\boldsymbol{q}_1)
      -i\boldsymbol{k}_2 \cdot \boldsymbol{\Psi}(\boldsymbol{q}_2)
      -i\boldsymbol{k}_3 \cdot \boldsymbol{\Psi}(\boldsymbol{q}_3)}
    \delta_{\rm J}(\boldsymbol{q}_1)\delta_{\rm J}(\boldsymbol{q}_2)\delta_{\rm J}(\boldsymbol{q}_3)
    \Big\rangle.
    \label{eq:B}
\end{align}

Defining
\begin{align}
    X &= -i \boldsymbol{k}_1 \cdot \boldsymbol{\Psi}(\boldsymbol{q}_1)
         -i \boldsymbol{k}_2 \cdot \boldsymbol{\Psi}(\boldsymbol{q}_2)
         -i \boldsymbol{k}_3 \cdot \boldsymbol{\Psi}(\boldsymbol{q}_3), \nonumber \\
    Y_i &= \delta_{\rm J}(\boldsymbol{q}_i), \quad \text{for } i = 1,2,3,
\end{align}
we can write the ensemble average as
\begin{align}
    \langle e^X Y_1 Y_2 Y_3 \rangle
    &= \langle e^X \rangle
    \Big[
        \langle e^X Y_1 Y_2 Y_3 \rangle_{\rm c}
        + \langle e^X Y_1 Y_2 \rangle_{\rm c} \langle e^X Y_3 \rangle_{\rm c} \nonumber \\
    &\quad
        + \langle e^X Y_1 Y_3 \rangle_{\rm c} \langle e^X Y_2 \rangle_{\rm c}
        + \langle e^X Y_2 Y_3 \rangle_{\rm c} \langle e^X Y_1 \rangle_{\rm c} \nonumber \\
    &\quad
        + \langle e^X Y_1 \rangle_{\rm c} \langle e^X Y_2 \rangle_{\rm c} \langle e^X Y_3 \rangle_{\rm c}
    \Big],
\end{align}
where the subscript ``c'' denotes the connected part of the ensemble average.

We define the \emph{three-point source correlation function}, which encodes the correlated structure of the Jacobian deviations, as
\begin{align}
    \zeta_{\rm J}(\boldsymbol{q}_1,\boldsymbol{q}_2,\boldsymbol{q}_3)
    &\equiv \langle e^X Y_1 Y_2 Y_3 \rangle_{\rm c}
     + \langle e^X Y_1 Y_2 \rangle_{\rm c} \langle e^X Y_3 \rangle_{\rm c} \nonumber \\
    &\quad + \langle e^X Y_1 Y_3 \rangle_{\rm c} \langle e^X Y_2 \rangle_{\rm c}
           + \langle e^X Y_2 Y_3 \rangle_{\rm c} \langle e^X Y_1 \rangle_{\rm c} \nonumber \\
    &\quad + \langle e^X Y_1 \rangle_{\rm c} \langle e^X Y_2 \rangle_{\rm c} \langle e^X Y_3 \rangle_{\rm c}.
    \label{eq:source_3PCF}
\end{align}

Since cosmological perturbation theory is naturally formulated in Fourier space, it is convenient to work with the corresponding \emph{source bispectrum} $B_{\rm J}$ defined by
\begin{align}
    & \zeta_{\rm J}(\boldsymbol{q}_1,\boldsymbol{q}_2,\boldsymbol{q}_3) \nonumber \\
    &=
    \int \frac{d^3k_1}{(2\pi)^3} e^{i\boldsymbol{k}_1\cdot\boldsymbol{q}_1}
    \int \frac{d^3k_2}{(2\pi)^3} e^{i\boldsymbol{k}_2\cdot\boldsymbol{q}_2}
    \int \frac{d^3k_3}{(2\pi)^3} e^{i\boldsymbol{k}_3\cdot\boldsymbol{q}_3} \nonumber \\
    &\quad \times
    (2\pi)^3\delta_{\rm D}(\boldsymbol{k}_{123})\, B_{\rm J}(\boldsymbol{k}_1,\boldsymbol{k}_2,\boldsymbol{k}_3).
\end{align}

As with the full bispectrum in Eq.~\eqref{eq:B12_23_31}, the source bispectrum can be decomposed into a sum over pairwise contributions:
\begin{equation}
    B_{\rm J}(\boldsymbol{k}_1,\boldsymbol{k}_2,\boldsymbol{k}_3)
    = \sum_{(\alpha,\beta)=(1,2),(2,3),(3,1)}\, B_{{\rm J}_{\alpha\beta}}(\boldsymbol{k}_{\alpha},\boldsymbol{k}_{\beta}).
\end{equation}

Similarly, the three-point source correlation function can be written as a sum over terms depending on two relative separation vectors:
\begin{equation}
    \zeta_{\rm J}(\boldsymbol{q}_1,\boldsymbol{q}_2,\boldsymbol{q}_3)
    = \sum_{(\alpha,\beta)=(1,2),(2,3),(3,1)}\, \zeta_{{\rm J}_{\alpha\beta}}(\boldsymbol{r}_{\alpha}, \boldsymbol{r}_{\beta}),
\end{equation}
where the relative position vectors $\boldsymbol{r}_{\alpha}$ and $\boldsymbol{r}_{\beta}$ are defined as follows for each index pair:
\begin{equation}
(\boldsymbol{r}_{\alpha}, \boldsymbol{r}_{\beta}) =
\begin{cases}
(\boldsymbol{q}_1 - \boldsymbol{q}_3,\ \boldsymbol{q}_2 - \boldsymbol{q}_3), & \text{for } (\alpha,\beta) = (1,2), \\
(\boldsymbol{q}_2 - \boldsymbol{q}_1,\ \boldsymbol{q}_3 - \boldsymbol{q}_1), & \text{for } (\alpha,\beta) = (2,3), \\
(\boldsymbol{q}_3 - \boldsymbol{q}_2,\ \boldsymbol{q}_1 - \boldsymbol{q}_2), & \text{for } (\alpha,\beta) = (3,1).
\end{cases}
\label{eq:rarb}
\end{equation}

Each of the pairwise source correlation functions $\zeta_{{\rm J}_{\alpha\beta}}$ can be evaluated via a double Fourier transform:
\begin{equation}
    \zeta_{{\rm J}_{\alpha\beta}}(\boldsymbol{r}_{\alpha},\boldsymbol{r}_{\beta})
    =
    \int \frac{d^3k_{\alpha}}{(2\pi)^3} e^{i\boldsymbol{k}_{\alpha}\cdot\boldsymbol{r}_{\alpha}}
    \int \frac{d^3k_{\beta}}{(2\pi)^3} e^{i\boldsymbol{k}_{\beta}\cdot\boldsymbol{r}_{\beta}}
    B_{{\rm J}_{\alpha\beta}}(\boldsymbol{k}_{\alpha},\boldsymbol{k}_{\beta}).
\end{equation}

The factor $\langle e^X \rangle$ describes the impact of large-scale displacement fields that are uncorrelated with the Jacobian deviation and is referred to as the \emph{displacement-mapping factor}. Its logarithm admits the cumulant expansion
\begin{equation}
    \ln \langle e^X \rangle =
    \sum_{m=2}^\infty \frac{1}{m!} \langle X^m \rangle_{\rm c},
\end{equation}
where the $m=1$ term vanishes by construction due to $\langle \boldsymbol{\Psi} \rangle = 0$.

To characterize the scale dependence associated with each component $\zeta_{{\rm J}_{\alpha\beta}}$ of the source correlation function, we decompose
\begin{equation}
    \ln \langle e^X \rangle_{\alpha\beta}
    = -\overline{\Gamma}_{\alpha\beta}(\boldsymbol{k}_{\alpha},\boldsymbol{k}_{\beta})
    + \Gamma_{\alpha\beta}(\boldsymbol{k}_{\alpha},\boldsymbol{k}_{\beta},
    \boldsymbol{r}_{\alpha},\boldsymbol{r}_{\beta}),
    \label{eq:eX}
\end{equation}
where $\overline{\Gamma}_{\alpha\beta}$ is evaluated at coincident points, $\boldsymbol{q}_1 = \boldsymbol{q}_2 = \boldsymbol{q}_3$. In configurations where all density fields are governed by the same displacement field, such as in the pre-reconstruction case, this expression simplifies to
\begin{equation}
    \overline{\Gamma}_{\alpha\beta}(\boldsymbol{k}_{\alpha},\boldsymbol{k}_\beta)
    = \Gamma_{\alpha\beta}(\boldsymbol{k}_{\alpha},\boldsymbol{k}_{\beta},
    \boldsymbol{r}_{\alpha}=\boldsymbol{0},\boldsymbol{r}_{\beta}=\boldsymbol{0}).
    \label{eq:Gamma_relation}
\end{equation}

As shown in Eq.~\eqref{eq:Gamma_relation}, when all evaluation points coincide (i.e., $\boldsymbol{q}_1 = \boldsymbol{q}_2 = \boldsymbol{q}_3$), the displacement-mapping factor reduces to unity and therefore has no effect on the bispectrum. This configuration represents a spatially uniform large-scale motion, whose contribution cancels due to translational invariance. This is the standard \emph{IR cancellation} mechanism, which will be discussed in detail in Section~\ref{sec:IRcancel}.

However, IR cancellation does not hold for cross bispectra involving distinct displacement fields. In these cases, the residual displacement-mapping factor introduces a nontrivial scale dependence and yields an overall exponential damping of the bispectrum. This phenomenon will be further examined in Section~\ref{sec:cross_bispectrum}.

The bispectrum component $B_{\alpha\beta}$ in the ULPT framework is given by
\begin{align}
    & B_{\rm ULPT,\alpha\beta}(\boldsymbol{k}_{\alpha},\boldsymbol{k}_{\beta}) \nonumber \\
    &= e^{-\overline{\Gamma}_{\alpha\beta}(\boldsymbol{k}_{\alpha},\boldsymbol{k}_{\beta})}
    \int d^3r_{\alpha}\, e^{-i \boldsymbol{k}_{\alpha} \cdot \boldsymbol{r}_{\alpha}}
    \int d^3r_{\beta}\, e^{-i \boldsymbol{k}_{\beta} \cdot \boldsymbol{r}_{\beta}} \nonumber \\
    &\quad \times
    e^{\Gamma_{\alpha\beta}(\boldsymbol{k}_{\alpha},\boldsymbol{k}_{\beta},
    \boldsymbol{r}_{\alpha},\boldsymbol{r}_{\beta})}
    \zeta_{{\rm J}_{\alpha\beta}}(\boldsymbol{r}_{\alpha},\boldsymbol{r}_{\beta}),
    \label{eq:B_ULPT_general}
\end{align}
where the relative separation vectors $\boldsymbol{r}_{\alpha}$ and $\boldsymbol{r}_{\beta}$ are defined in Eq.~\eqref{eq:rarb}. The full bispectrum is then obtained by summing over all three pairwise components, as in Eq.~\eqref{eq:B12_23_31}, corresponding to $(\alpha,\beta) = (1,2), (2,3), (3,1)$.

A key advantage of ULPT is its structural universality: the decomposition in Eq.~\eqref{eq:B_ULPT_general} remains valid even in the presence of galaxy bias, RSD, or density-field reconstruction. These effects are consistently incorporated at the level of the density fluctuation field~\cite{Sugiyama:2025ntz} and enter the formulation through appropriate modifications of the displacement-mapping factor $\Gamma$ and the three-point source correlation function $\zeta_{\rm J}$.

\subsection{Multipole expansion and double Hankel transforms}

When statistical isotropy holds, the three-point source correlation function can be expanded in Legendre polynomials of the angle between $\boldsymbol{r}_{\alpha}$ and $\boldsymbol{r}_{\beta}$:
\begin{equation}
    \zeta_{{\rm J}_{\alpha\beta}}(\boldsymbol{r}_{\alpha},\boldsymbol{r}_{\beta})
    = \sum_{\ell=0}^{\infty}\zeta_{{{\rm J}_{\alpha\beta}},\ell}(r_{\alpha},r_{\beta})
    \,{\cal L}_{\ell}(\hat{r}_{\alpha}\!\cdot\!\hat{r}_{\beta}),
\end{equation}
where ${\cal L}_{\ell}$ denotes the Legendre polynomial of order $\ell$ and hats indicate unit vectors. An analogous expansion holds for the source bispectrum,
\begin{equation}
    B_{{\rm J}_{\alpha\beta}}(\boldsymbol{k}_{\alpha},\boldsymbol{k}_{\beta})
    = \sum_{\ell=0}^{\infty}B_{{{\rm J}_{\alpha\beta}},\ell}(k_{\alpha},k_{\beta})
    \,{\cal L}_{\ell}(\hat{k}_{\alpha}\!\cdot\!\hat{k}_{\beta}).
\end{equation}

The multipole coefficients in configuration space and Fourier space are related through a pair of double Hankel transforms involving spherical Bessel functions $j_\ell$. For any such multipole pair $(f_\ell,\tilde f_\ell)$, we have
\begin{align}
    f_{\ell}(r_{\alpha},r_{\beta})
    &= (-1)^{\ell}\!\int\!\frac{dk_{\alpha}\,k_{\alpha}^2}{2\pi^2}
       \int\!\frac{dk_{\beta}\,k_{\beta}^2}{2\pi^2} \nonumber \\
     &  \quad \times j_{\ell}(k_{\alpha} r_{\alpha})\, j_{\ell}(k_{\beta} r_{\beta})
       \,\tilde{f}_{\ell}(k_{\alpha},k_{\beta}),
    \label{eq:double_hankel_f}
\\
    \tilde{f}_{\ell}(k_{\alpha},k_{\beta})
    &= (-1)^{\ell}(4\pi)^2 \!\int\!dr_{\alpha}\,r_{\alpha}^2
       \int\!dr_{\beta}\,r_{\beta}^2 \nonumber \\
    &\quad \times   j_{\ell}(k_{\alpha} r_{\alpha})\, j_{\ell}(k_{\beta} r_{\beta})
       \, f_{\ell}(r_{\alpha},r_{\beta}),
    \label{eq:double_hankel_ft}
\end{align}
such that applying Eqs.~\eqref{eq:double_hankel_f}--\eqref{eq:double_hankel_ft} to $(f_\ell,\tilde f_\ell)= (\zeta_{{\rm J}_{\alpha\beta},\ell},B_{{\rm J}_{\alpha\beta},\ell})$ yields the desired relations between the source-correlation and source-bispectrum multipoles. In practice, it is often convenient to compute $B_{{\rm J}_{\alpha\beta},\ell}$ first and then obtain $\zeta_{{\rm J}_{\alpha\beta},\ell}$ through the inverse transform.

In redshift space, peculiar velocities induce an explicit dependence on the line-of-sight direction $\hat{n}$, making the bispectrum anisotropic. In this case, the Legendre expansion used in real space is naturally replaced by a more general decomposition based on the tripolar spherical harmonic (TripoSH) basis~\cite{Varshalovich:1988ifq}. This formalism provides a systematic treatment of angular dependencies and yields a direct connection to observational estimators. For further details on the TripoSH approach in the context of three-point statistics in redshift space, we refer the reader to Refs.~\cite{Sugiyama:2018yzo,Sugiyama:2019ike,Sugiyama:2023tes,Sugiyama:2023zvd}.

\subsection{Notation and conventions}
\label{subsec:notation_use}

For auto bispectra of either the pre-reconstruction or post-reconstruction fields, the three components $B_{12}$, $B_{23}$, and $B_{31}$ are related by cyclic permutations and thus share the same structure, differing only by relabeling of arguments. Therefore, throughout Sections~\ref{sec:ULPT_bispec_1loop}, \ref{sec:IRcancel}, and \ref{sec:IR_resum}, we focus exclusively on the $B_{12}$ component without loss of generality. For notational simplicity, we omit the subscript ``12'' from quantities such as $B_{{\rm ULPT},12}$, $\zeta_{{\rm J}_{12}}$, and $\Gamma_{12}$, and simply write them as $B_{\rm ULPT}$, $\zeta_{\rm J}$, and $\Gamma$, respectively. The full bispectrum is then obtained by summing over all three cyclic permutations, as defined in Eq.~\eqref{eq:B12_23_31}.

In Section~\ref{sec:cross_bispectrum}, where we analyze cross bispectra involving both pre- and post-reconstruction fields, we restore the full subscript notation (e.g., $B_{12}$, $B_{23}$, and $B_{31}$) to explicitly distinguish the different pairwise contributions.

\section{One-loop ULPT bispectrum}
\label{sec:ULPT_bispec_1loop}

In this section, we derive explicit analytic expressions for the one-loop bispectrum within the ULPT framework.

\subsection{General structure}

In cosmological perturbation theory, statistical quantities are expanded perturbatively with respect to the linear matter power spectrum \(P_{\rm lin}\). For the bispectrum, the leading-order (tree-level) contribution scales as \({\cal O}(P_{\rm lin}^2)\), while the next-to-leading (one-loop) contribution scales as \({\cal O}(P_{\rm lin}^3)\). Throughout this work, we adopt the following notation for perturbative order: the superscript ``(lin)'' denotes linear-order quantities of \({\cal O}(P_{\rm lin})\); ``(tree)'' denotes leading-order terms of \({\cal O}(P_{\rm lin}^2)\); and ``(1-loop)'' denotes next-to-leading order terms of \({\cal O}(P_{\rm lin}^3)\).

Within ULPT, the one-loop bispectrum contains the full one-loop contribution of standard perturbation theory (SPT)~\cite{Bernardeau:2001qr}, while also incorporating additional higher-order, nonperturbative corrections associated with preserving the exponential structure of the displacement-mapping factor. Correspondingly, the functions \(\Gamma\), \(\zeta_{\rm J}\), and \(B_{\rm J}\) entering the ULPT bispectrum are expanded perturbatively as
\begin{align}
    \Gamma & = \Gamma^{(\rm lin)}, \nonumber \\
    \zeta_{\rm J} & = \zeta_{\rm J}^{(\text{tree+1-loop})}
    = \zeta_{\rm J}^{(\rm tree)} + \zeta_{\rm J}^{(\text{1-loop})}, \nonumber \\
    B_{\rm J} & = B_{\rm J}^{(\text{tree+1-loop})}
    = B_{\rm J}^{(\rm tree)} + B_{\rm J}^{(\text{1-loop})}.
\end{align}

The one-loop ULPT bispectrum is given by
\begin{align}
    B_{\rm ULPT}(\boldsymbol{k}_1,\boldsymbol{k}_2)
    &= e^{-\overline{\Gamma}^{\rm (lin)}(\boldsymbol{k}_1,\boldsymbol{k}_2)}
    \int d^3r_1\, e^{-i \boldsymbol{k}_1 \cdot \boldsymbol{r}_1}
    \int d^3r_2\, e^{-i \boldsymbol{k}_2 \cdot \boldsymbol{r}_2} \nonumber \\
    &\quad \times
    e^{\Gamma^{\rm (lin)}(\boldsymbol{k}_1,\boldsymbol{k}_2,\boldsymbol{r}_1,\boldsymbol{r}_2)}
    \,\zeta_{\rm J}^{\rm (tree+1\mbox{-}loop)}(\boldsymbol{r}_1,\boldsymbol{r}_2).
    \label{eq:ULPT_1loop}
\end{align}

Alternatively, it can be decomposed as
\begin{align}
    B_{\rm ULPT}(\boldsymbol{k}_1,\boldsymbol{k}_2)
    &= B_{\rm J}^{\rm (tree+1\mbox{-}loop)}(\boldsymbol{k}_1,\boldsymbol{k}_2)
    + \sum_{n=1}^{\infty} B_{\rm DM}^{(\text{$n$-loop})}(\boldsymbol{k}_1,\boldsymbol{k}_2),
\end{align}
where the displacement-mapping (DM) contribution is defined by
\begin{align}
    \sum_{n=1}^{\infty} B_{\rm DM}^{(\text{$n$-loop})}(\boldsymbol{k}_1,\boldsymbol{k}_2)
    &= \int d^3r_1\, e^{-i \boldsymbol{k}_1 \cdot \boldsymbol{r}_1}
       \int d^3r_2\, e^{-i \boldsymbol{k}_2 \cdot \boldsymbol{r}_2} \nonumber \\
    &\quad \times
    \left[
        e^{-\overline{\Gamma}^{\rm (lin)}(\boldsymbol{k}_1,\boldsymbol{k}_2)
        + \Gamma^{\rm (lin)}(\boldsymbol{k}_1,\boldsymbol{k}_2,\boldsymbol{r}_1,\boldsymbol{r}_2)}
        - 1
    \right] \nonumber \\
    &\quad \times
    \zeta_{\rm J}^{\rm (tree+1\mbox{-}loop)}(\boldsymbol{r}_1,\boldsymbol{r}_2).
\end{align}
The DM contribution begins at one-loop order.

The one-loop DM term is
\begin{align}
    B_{\rm DM}^{\rm (1\mbox{-}loop)}(\boldsymbol{k}_1,\boldsymbol{k}_2)
    &= -\overline{\Gamma}^{\rm (lin)}(\boldsymbol{k}_1,\boldsymbol{k}_2)\,
       B_{\rm J}^{\rm (tree)}(\boldsymbol{k}_1,\boldsymbol{k}_2) \nonumber \\
    &\quad +
    \int d^3r_1\, e^{-i \boldsymbol{k}_1 \cdot \boldsymbol{r}_1}
    \int d^3r_2\, e^{-i \boldsymbol{k}_2 \cdot \boldsymbol{r}_2} \nonumber \\
    &\quad \times
    \Gamma^{\rm (lin)}(\boldsymbol{k}_1,\boldsymbol{k}_2,\boldsymbol{r}_1,\boldsymbol{r}_2)\,
    \zeta_{\rm J}^{\rm (tree)}(\boldsymbol{r}_1,\boldsymbol{r}_2).
    \label{eq:DM_1loop}
\end{align}

Higher-order DM corrections (\(n\ge2\)) are given by
\begin{align}
    & B_{\rm DM}^{(\text{$n \ge 2$-loop})}(\boldsymbol{k}_1,\boldsymbol{k}_2) \nonumber \\
    &= \int d^3r_1\, e^{-i \boldsymbol{k}_1 \cdot \boldsymbol{r}_1}
       \int d^3r_2\, e^{-i \boldsymbol{k}_2 \cdot \boldsymbol{r}_2} \nonumber \\
    &\quad \times \Bigg[
        \frac{1}{n!}\left( -\overline{\Gamma}^{\rm (lin)}
        + \Gamma^{\rm (lin)} \right)^{n}
        \zeta_{\rm J}^{\rm (tree)}(\boldsymbol{r}_1,\boldsymbol{r}_2)
        \nonumber \\
    &\qquad +
        \frac{1}{(n-1)!}\left( -\overline{\Gamma}^{\rm (lin)}
        + \Gamma^{\rm (lin)} \right)^{n-1}
        \zeta_{\rm J}^{\rm (1\mbox{-}loop)}(\boldsymbol{r}_1,\boldsymbol{r}_2)
    \Bigg],
    \label{eq:B_DM_nloop}
\end{align}
where we have suppressed the arguments of \(\Gamma^{\rm (lin)}\) for brevity.

For comparison, the SPT bispectrum truncated at one-loop order is
\begin{equation}
    B_{\rm SPT}(\boldsymbol{k}_1,\boldsymbol{k}_2)
    = B_{\rm J}^{(\text{tree+1-loop})}(\boldsymbol{k}_1,\boldsymbol{k}_2)
    + B_{\rm DM}^{\rm (1\mbox{-}loop)}(\boldsymbol{k}_1,\boldsymbol{k}_2).
    \label{eq:B_SPT_J_DM}
\end{equation}

These expressions highlight the essential difference between the one-loop ULPT and SPT bispectra: although both contain the same one-loop source bispectrum, SPT truncates the DM contribution at one-loop order, whereas ULPT retains the full exponential structure of the displacement-mapping factor and therefore evaluates the DM term nonperturbatively.

The SPT bispectrum expressed in terms of \(B_{\rm J}\) and \(B_{\rm DM}\) immediately yields several useful relations. At tree level,
\begin{equation}
    B_{\rm J}^{(\rm tree)}(\boldsymbol{k}_1,\boldsymbol{k}_2)
    = 
    B_{\rm SPT}^{(\rm tree)}(\boldsymbol{k}_1,\boldsymbol{k}_2),
    \label{eq:B_SPT_J_tree}
\end{equation}
and at one-loop order,
\begin{align}
    B_{\rm J}^{(\text{1-loop})}(\boldsymbol{k}_1,\boldsymbol{k}_2)
    = B_{\rm SPT}^{(\text{1-loop})}(\boldsymbol{k}_1,\boldsymbol{k}_2)
    - B_{\rm DM}^{(\text{1-loop})}(\boldsymbol{k}_1,\boldsymbol{k}_2).
    \label{eq:BJ_1loop}
\end{align}
This result shows that one does not need to evaluate the full expression in Eq.~\eqref{eq:source_3PCF} to obtain the source bispectrum. Instead, existing one-loop SPT bispectrum results may be used together with the known DM term in Eq.~\eqref{eq:DM_1loop}. Consequently, for computing the one-loop source bispectrum, the only new quantity requiring explicit evaluation is the one-loop DM contribution.

\subsection{Displacement-mapping factor}
\label{subsec:DM_factor}

To evaluate the one-loop ULPT bispectrum, we first compute the linear-order contribution to the exponent of the displacement-mapping factor. The first-order (linear) displacement vector in perturbation theory is given by
\begin{equation}
    \boldsymbol{\Psi}^{(1)}(\boldsymbol{q})
    = i \int \frac{d^3k}{(2\pi)^3} \, e^{i\boldsymbol{k}\cdot\boldsymbol{q}} \,
      \frac{\boldsymbol{k}}{k^2} \, \tilde{\delta}^{(1)}(\boldsymbol{k}),
\end{equation}
where $\tilde{\delta}^{(1)}(\boldsymbol{k})$ denotes the Fourier transform of the linear dark matter density fluctuation.

For later convenience, we define the linear-order displacement correlation tensor as
\begin{align}
    I_{ij}(\boldsymbol{r})
    = \big\langle \Psi^{(1)}_i(\boldsymbol{q}) \, \Psi^{(1)}_j(\boldsymbol{q}') \big\rangle_{\rm c},
    \label{eq:Iij_def_start}
\end{align}
where $\boldsymbol{r} = \boldsymbol{q} - \boldsymbol{q}'$, and the indices $i$ and $j$ denote Cartesian components of the displacement field, with $i,j = 1,2,3$.

The tensor $I_{ij}$ can be written explicitly as
\begin{equation}
    I_{ij}(\boldsymbol{r})
    = \int \frac{d^3p}{(2\pi)^3}\,
       e^{i\boldsymbol{p}\cdot\boldsymbol{r}}\,
       \hat{p}_i \hat{p}_j\,\frac{P_{\rm lin}(p)}{p^2},
    \label{eq:Iij_def_1}
\end{equation}
where the linear matter power spectrum is defined through
\begin{equation}
    \big\langle \tilde{\delta}^{(1)}(\boldsymbol{k})\,\tilde{\delta}^{(1)}(\boldsymbol{k}') \big\rangle
    = (2\pi)^3 \delta_{\rm D}(\boldsymbol{k}+\boldsymbol{k}')\, P_{\rm lin}(k).
\end{equation}
Evaluating the angular integration analytically, Eq.~\eqref{eq:Iij_def_1} reduces to
\begin{equation}
    I_{ij}(\boldsymbol{r})
    = \delta_{ij}\,\sigma_0^2(r)
    + \left(\frac{3\hat{r}_i \hat{r}_j - \delta_{ij}}{2}  \right)\,
        2\sigma_2^2(r),
    \label{eq:Iij_def_2}
\end{equation}
where $\delta_{ij}$ is the Kronecker delta. The functions $\sigma_\ell^2(r)$ are defined by
\begin{align}
    \sigma_{\ell}^2(r)
    = \frac{1}{3}\, i^{\ell}
      \int \frac{dp}{2\pi^2}\, j_{\ell}(pr)\, P_{\rm lin}(p),
    \label{eq:sigma_def}
\end{align}
with $j_{\ell}$ denoting the spherical Bessel function of order $\ell$.

In the limit $r \to 0$, we have $j_{0}(pr)\to 1$ and $j_{2}(pr)\to 0$.
Therefore, the displacement correlation tensor becomes
\begin{align}
    I_{ij}(\boldsymbol{r}\to \boldsymbol{0})
    = \delta_{ij}\,\bar{\sigma}^2,
    \label{eq:Iij_r0}
\end{align}
where the variance $\bar{\sigma}^2$ is given by
\begin{align}
    \bar{\sigma}^2
    = \sigma_0^2(r=0)
    = \frac{1}{3}\int \frac{dp}{2\pi^2}\,P_{\rm lin}(p).
    \label{eq:sigma_bar}
\end{align}

Using Eq.~\eqref{eq:Iij_def_start}, the linear-order exponent of the displacement-mapping factor can be written as
\begin{align}
    \overline{\Gamma}^{\rm (lin)}(\boldsymbol{k}_1,\boldsymbol{k}_2)
    & = \frac{1}{2}\left( k_1^2 + k_2^2 + k_3^2 \right)\bar{\sigma}^2 , \nonumber \\
    \Gamma^{\rm (lin)}(\boldsymbol{k}_1,\boldsymbol{k}_2,\boldsymbol{r}_1,\boldsymbol{r}_2)
    & = - k_1^i k_3^j I_{ij}(\boldsymbol{r}_1)
        - k_2^i k_3^j I_{ij}(\boldsymbol{r}_2) \nonumber \\
    &\quad - k_1^i k_2^j I_{ij}(\boldsymbol{r}_{1}-\boldsymbol{r}_2).
    \label{eq:Gamma_lin}
\end{align}

Employing the triangle condition $\boldsymbol{k}_1+\boldsymbol{k}_2+\boldsymbol{k}_3=\boldsymbol{0}$, the exponent can be recast as
\begin{equation}
    \overline{\Gamma}^{\rm (lin)}(\boldsymbol{k}_1,\boldsymbol{k}_2)
    = -\left( \boldsymbol{k}_1\cdot\boldsymbol{k}_3
    + \boldsymbol{k}_2\cdot\boldsymbol{k}_3
    + \boldsymbol{k}_1\cdot\boldsymbol{k}_2 \right)\bar{\sigma}^2 .
    \label{eq:Gamma_triangle}
\end{equation}
This expression makes it straightforward to verify explicitly, using Eq.~\eqref{eq:Iij_r0}, that Eq.~\eqref{eq:Gamma_relation} holds at linear order:
\begin{align}
    \overline{\Gamma}^{\rm (lin)}(\boldsymbol{k}_1,\boldsymbol{k}_2)
    = \Gamma^{\rm (lin)}(\boldsymbol{k}_1,\boldsymbol{k}_2,\boldsymbol{r}_1\to\boldsymbol{0},\boldsymbol{r}_2\to\boldsymbol{0}).
    \label{eq:Gamma_relation_lin}
\end{align}

\subsection{Source bispectrum}

In this subsection, we present explicit expressions for the one-loop corrections to the source bispectrum. As a starting point, we recall the general expression for the $n$th-order solution of the dark matter density fluctuation in Fourier space:
\begin{align}
    \tilde{\delta}^{(n)}(\boldsymbol{k})
    &= \int \frac{d^3p_1}{(2\pi)^3} \cdots \int \frac{d^3p_n}{(2\pi)^3}
    (2\pi)^3 \delta_{\rm D}(\boldsymbol{k} - \boldsymbol{p}_{1n}) \nonumber \\
    &\quad \times
    F_n(\boldsymbol{p}_1,\ldots,\boldsymbol{p}_n)\,
    \tilde{\delta}^{(1)}(\boldsymbol{p}_1)\cdots\tilde{\delta}^{(1)}(\boldsymbol{p}_n),
    \label{eq:delta_n}
\end{align}
where $\boldsymbol{p}_{1n}\equiv \boldsymbol{p}_1+\cdots+\boldsymbol{p}_n$. The nonlinear kernel functions $F_n$ can be systematically obtained for arbitrary $n$ via the well-known recursion relations~\cite{Bernardeau:2001qr}.

At tree level in SPT, the bispectrum is constructed from the second-order kernel $F_2$ together with the linear power spectrum, and is given by
\begin{equation}
    B_{\rm SPT}^{(\rm tree)}(\boldsymbol{k}_1,\boldsymbol{k}_2)
    = 2\,F_2(\boldsymbol{k}_1,\boldsymbol{k}_2)\,P_{\rm lin}(k_1)P_{\rm lin}(k_2).
    \label{eq:Btree}
\end{equation}
As noted in Sec.~\ref{subsec:notation_use}, we focus on the $B_{12}$ component of the full bispectrum defined in Eq.~\eqref{eq:B12_23_31}, which depends only on $(\boldsymbol{k}_1,\boldsymbol{k}_2)$ and we omit the two cyclic permutations for brevity.

For convenience, we classify the one-loop bispectrum corrections according to the appearance of mode-coupling (MC) integrals. As a preliminary example, let us recall the case of the power spectrum, which can be generically decomposed as~\cite{Crocce:2005xy,Crocce:2007dt}
\begin{equation}
    P(k) = G^2(k)\,P_{\rm lin}(k) + P_{\rm MC}(k).
    \label{eq:G_MC}
\end{equation}
Here the first term, proportional to the linear power spectrum, does not involve any MC integrals, and its prefactor $G(k)$ is referred to as the propagator. The second term $P_{\rm MC}(k)$ represents contributions that do include MC integrals. At one-loop order in SPT, the power spectrum correction consists of the $P^{(13)}$ term, arising from the product of first- and third-order density fluctuations, and the $P^{(22)}$ term, arising from the product of two second-order fluctuations. These are naturally identified as $P^{(13)} \in G^2 P_{\rm lin}$ and $P^{(22)} \in P_{\rm MC}$.

For the bispectrum, the situation is more involved due to its dependence on two independent wavevectors, \(\boldsymbol{k}_1\) and \(\boldsymbol{k}_2\). We therefore introduce the following classification. If both $\boldsymbol{k}_1$ and $\boldsymbol{k}_2$ enter without any MC integrals, we denote the contribution as the ``GG'' term, where ``G'' is inherited from the propagator structure in the power spectrum. If one of the two wavevectors involves MC integrals while the other does not, we denote the terms as ``GM'' and ``MG,'' depending on which argument carries the MC dependence. If both $\boldsymbol{k}_1$ and $\boldsymbol{k}_2$ involve MC integrals, the contribution is denoted as the ``MM'' term. However, the MM term first appears at two-loop order and is absent at one-loop order in the bispectrum. Finally, when the scale dependences of $\boldsymbol{k}_1$ and $\boldsymbol{k}_2$ are both connected through a common MC integral, we refer to the resulting contribution as the ``MMM'' term. Taken together, these classifications lead to the following general decomposition of the nonlinear bispectrum~\cite{Sugiyama:2020uil,Sugiyama:2024qsw}:
\begin{align}
    B(\boldsymbol{k}_1,\boldsymbol{k}_2)
    & = B_{\rm GG}(\boldsymbol{k}_1,\boldsymbol{k}_2)\,
        P_{\rm lin}(k_1)\,P_{\rm lin}(k_2)\nonumber \\
    & \quad + B_{\rm GM}(\boldsymbol{k}_1,\boldsymbol{k}_2)\,P_{\rm lin}(k_1)
          + B_{\rm MG}(\boldsymbol{k}_1,\boldsymbol{k}_2)\,P_{\rm lin}(k_2)\nonumber \\
    & \quad + B_{\rm MM}(\boldsymbol{k}_1,\boldsymbol{k}_2)
        + B_{\rm MMM}(\boldsymbol{k}_1,\boldsymbol{k}_2).
\end{align}

Within this scheme, the tree-level bispectrum in SPT, Eq.~\eqref{eq:Btree}, corresponds to
\begin{equation}
    B_{\rm SPT,\,GG}^{(\rm tree)}(\boldsymbol{k}_1,\boldsymbol{k}_2)
    = 2\,F_2(\boldsymbol{k}_1,\boldsymbol{k}_2),
\end{equation}
so that only the GG contribution appears at tree level.

At one-loop order in SPT, the bispectrum corrections can be written as
\begin{widetext}
\begin{align}
    B_{\rm SPT,\,GG}^{(1\text{-loop})}(\boldsymbol{k}_1,\boldsymbol{k}_2)
    & = 12 \int \frac{d^3p}{(2\pi)^3}
        F_4(\boldsymbol{k}_1,\boldsymbol{k}_2,\boldsymbol{p},-\boldsymbol{p})\,
        P_{\rm lin}(p) \nonumber \\
    & \quad + 6\,F_2(\boldsymbol{k}_1,\boldsymbol{k}_2)
        \int \frac{d^3p}{(2\pi)^3}
        \left[ F_3(\boldsymbol{k}_1,\boldsymbol{p},-\boldsymbol{p})
             + F_3(\boldsymbol{k}_2,\boldsymbol{p},-\boldsymbol{p}) \right] P_{\rm lin}(p),  \nonumber \\
    B_{\rm SPT,\,GM}^{(1\text{-loop})}(\boldsymbol{k}_1,\boldsymbol{k}_2)
    & = 6 \int \frac{d^3p}{(2\pi)^3}
        F_2(\boldsymbol{p},\boldsymbol{k}_2-\boldsymbol{p})\,
        F_3(\boldsymbol{k}_1,\boldsymbol{p},\boldsymbol{k}_2-\boldsymbol{p})\,
        P_{\rm lin}(p)\,P_{\rm lin}(|\boldsymbol{k}_2-\boldsymbol{p}|),  \nonumber \\
    B_{\rm SPT,\,MG}^{(1\text{-loop})}(\boldsymbol{k}_1,\boldsymbol{k}_2)
    & = 6 \int \frac{d^3p}{(2\pi)^3}
        F_2(\boldsymbol{p},\boldsymbol{k}_1-\boldsymbol{p})\,
        F_3(\boldsymbol{k}_2,\boldsymbol{p},\boldsymbol{k}_1-\boldsymbol{p})\,
        P_{\rm lin}(p)\,P_{\rm lin}(|\boldsymbol{k}_1-\boldsymbol{p}|),  \nonumber \\
    B_{\rm SPT,\,MMM}^{(1\text{-loop})}(\boldsymbol{k}_1,\boldsymbol{k}_2)
    & = \frac{8}{3} \int \frac{d^3p}{(2\pi)^3}
        F_2(\boldsymbol{p},\boldsymbol{k}_1-\boldsymbol{p})\,
        F_2(-\boldsymbol{p},\boldsymbol{k}_2+\boldsymbol{p})\,
        F_2(\boldsymbol{k}_2+\boldsymbol{p},\boldsymbol{k}_1-\boldsymbol{p}) \nonumber \\
    & \qquad \times
        P_{\rm lin}(p)\,
        P_{\rm lin}(|\boldsymbol{k}_2+\boldsymbol{p}|)\,
        P_{\rm lin}(|\boldsymbol{k}_1-\boldsymbol{p}|).
    \label{eq:B_SPT}
\end{align}
\end{widetext}

Applying the same classification to the displacement-mapping (DM) contribution at one-loop order [Eq.~\eqref{eq:DM_1loop}], and substituting Eqs.~\eqref{eq:Iij_def_1} and \eqref{eq:Gamma_lin}, we obtain
\begin{widetext}
\begin{align}
    B_{\rm DM,\,GG}^{(1\text{-loop})}(\boldsymbol{k}_1,\boldsymbol{k}_2)
    & = - \frac{1}{2}\left(k_1^2 + k_2^2 + k_3^2\right)\,
    \bar{\sigma}^2\,\left[ 2\,F_2(\boldsymbol{k}_1,\boldsymbol{k}_2) \right], \nonumber \\
    B_{\rm DM,\,GM}^{(1\text{-loop})}(\boldsymbol{k}_1,\boldsymbol{k}_2)
    & = -2 \int \frac{d^3p}{(2\pi)^3}
        \left( \frac{\boldsymbol{k}_2 \cdot \boldsymbol{p}}{p^2} \right)
        \left( \frac{\boldsymbol{k}_3 \cdot \boldsymbol{p}}{p^2} \right)
        F_2(\boldsymbol{k}_2-\boldsymbol{p},\boldsymbol{k}_1)\,
        P_{\rm lin}(p)\,P_{\rm lin}(|\boldsymbol{k}_2-\boldsymbol{p}|),  \nonumber \\
    B_{\rm DM,\,MG}^{(1\text{-loop})}(\boldsymbol{k}_1,\boldsymbol{k}_2)
    & = -2 \int \frac{d^3p}{(2\pi)^3}
        \left( \frac{\boldsymbol{k}_1 \cdot \boldsymbol{p}}{p^2} \right)
        \left( \frac{\boldsymbol{k}_3 \cdot \boldsymbol{p}}{p^2} \right)
        F_2(\boldsymbol{k}_1-\boldsymbol{p},\boldsymbol{k}_2)\,
        P_{\rm lin}(p)\,P_{\rm lin}(|\boldsymbol{k}_1-\boldsymbol{p}|),  \nonumber \\
    B_{\rm DM,\,MMM}^{(1\text{-loop})}(\boldsymbol{k}_1,\boldsymbol{k}_2)
    & = -2 \int \frac{d^3p}{(2\pi)^3}
        \left( \frac{\boldsymbol{k}_1 \cdot \boldsymbol{p}}{p^2} \right)
        \left( \frac{\boldsymbol{k}_2 \cdot \boldsymbol{p}}{p^2} \right)
        F_2(\boldsymbol{k}_1-\boldsymbol{p},\boldsymbol{k}_2+\boldsymbol{p})\,
        P_{\rm lin}(p)\,
        P_{\rm lin}(|\boldsymbol{k}_1-\boldsymbol{p}|)\,
        P_{\rm lin}(|\boldsymbol{k}_2+\boldsymbol{p}|).
        \label{eq:DM_1loop_comp}
\end{align}
\end{widetext}

Finally, using Eq.~\eqref{eq:BJ_1loop}, the one-loop source bispectrum can be obtained from the SPT result by subtracting the corresponding DM contributions:
\begin{align}
    B_{\rm J}(\boldsymbol{k}_1,\boldsymbol{k}_2)
    & = B^{(\text{tree}+1\text{-loop})}_{\rm J,\,GG}(\boldsymbol{k}_1,\boldsymbol{k}_2)\,
        P_{\rm lin}(k_1)\,P_{\rm lin}(k_2)\nonumber \\
    & \quad + B^{(1\text{-loop})}_{\rm J,\,GM}(\boldsymbol{k}_1,\boldsymbol{k}_2)\,
        P_{\rm lin}(k_1) \nonumber \\
    & \quad + B^{(1\text{-loop})}_{\rm J,\,MG}(\boldsymbol{k}_1,\boldsymbol{k}_2)\,
        P_{\rm lin}(k_2)\nonumber \\
    & \quad + B^{(1\text{-loop})}_{\rm J,\,MMM}(\boldsymbol{k}_1,\boldsymbol{k}_2),
    \label{eq:B_J_1loop}
\end{align}
where
\begin{align}
    B_{\rm J,\,GG}^{(\text{tree})} & = B_{\rm GG}^{(\text{tree})}, \nonumber \\
    B^{(1\text{-loop})}_{\rm J,\,X}(\boldsymbol{k}_1,\boldsymbol{k}_2)
    & = B^{(1\text{-loop})}_{\rm SPT,\,X}(\boldsymbol{k}_1,\boldsymbol{k}_2)
      - B^{(1\text{-loop})}_{\rm DM,\,X}(\boldsymbol{k}_1,\boldsymbol{k}_2),
      \label{eq:B_J_X_1loop}
\end{align}
with ${\rm X} \in \{\text{GG, GM, MG, MMM}\}$.

\section{IR cancellation}
\label{sec:IRcancel}

In this section, we investigate how IR cancellation in the bispectrum is realized within the ULPT framework. Section~\ref{sec:intuitive} reviews the general mechanism of IR cancellation in three-point statistics. In Sec.~\ref{sec:source_bispec_IRcancel}, we examine the IR limits of the one-loop SPT bispectrum, the one-loop DM bispectrum, and the one-loop source bispectrum, illustrating how cancellations arise among these contributions. Finally, in Sec.~\ref{sec:ULPTbispec_IRcancel}, we demonstrate explicitly how the full ULPT bispectrum incorporates the mechanism of nonperturbative IR cancellation, showing that, in the IR limit, it consistently reduces to the corresponding SPT solution at the perturbative order of interest.

\subsection{Conditions for IR cancellation in the bispectrum}
\label{sec:intuitive}

It has been rigorously proven, using the recursive relations for the perturbative kernels $F_n$ and mathematical induction, that when the amplitudes of the last $n-m$ wavenumbers among $\{\boldsymbol{p}_1,\ldots,\boldsymbol{p}_n\}$ become sufficiently small, the following relation holds~\cite{Sugiyama:2013pwa}\footnote{Equation~\eqref{eq:F_n_lim} is valid only in $\Lambda$CDM-like universes where the recursion relations of $F_n$ hold. In certain modified gravity theories, such as degenerate higher-order scalar-tensor (DHOST) models~\cite{Hirano:2018uar}, this relation already fails at second order. See Appendix A of Ref.~\cite{Sugiyama:2024eye} for details.}:
\begin{align}
   &\lim_{p_{m+1},\,\ldots,\,p_n \to 0}
   F_n(\boldsymbol{p}_1,\ldots,\boldsymbol{p}_n) \nonumber \\
   &\quad = \frac{m!}{n!}
   \left( \frac{\boldsymbol{p}_{1m}\cdot\boldsymbol{p}_{m+1}}{p_{m+1}^2} \right)
   \cdots
   \left( \frac{\boldsymbol{p}_{1m}\cdot\boldsymbol{p}_{n}}{p_{n}^2} \right)
   F_{m}(\boldsymbol{p}_1,\ldots,\boldsymbol{p}_m),
   \label{eq:F_n_lim}
\end{align}
where $\boldsymbol{p}_{1m}=\boldsymbol{p}_1+\cdots+\boldsymbol{p}_m$.

This relation states that nonlinear contributions of order higher than $m$ originate solely from couplings to long-wavelength (infrared) modes $p_{m+1},\ldots,p_n \to 0$. We refer to this limiting procedure as the IR limit.

For the Dirac delta function, the corresponding IR limit is
\begin{equation}
    \lim_{p_{m+1},\,\ldots,\,p_n \to 0}
    \delta_{\rm D}(\boldsymbol{k}-\boldsymbol{p}_{1n})
    =
    \delta_{\rm D}(\boldsymbol{k}-\boldsymbol{p}_{1m}).
    \label{eq:deltaD_lim}
\end{equation}

Substituting Eqs.~\eqref{eq:F_n_lim} and \eqref{eq:deltaD_lim} into the expression for the $n$th-order density fluctuation [Eq.~\eqref{eq:delta_n}], and accounting for the combinatorial factor ${n \choose m} = n!/[m!\,(n-m)!]$, one finds that, in the IR limit, the dark matter density perturbation of any order satisfying $n \geq m$ takes the form
\begin{equation}
    \tilde{\delta}^{(n)}(\boldsymbol{k})
    = \frac{1}{(n-m)!}
    \left[ -i\boldsymbol{k}\cdot\overline{\boldsymbol{\Psi}}^{(1)} \right]^{n-m}
      \tilde{\delta}^{(m)}(\boldsymbol{k}),
\end{equation}
where $\overline{\boldsymbol{\Psi}}^{(1)}$ denotes the linear displacement vector evaluated at the origin,
\begin{equation}
    \overline{\boldsymbol{\Psi}}^{(1)} = \boldsymbol{\Psi}^{(1)}(\boldsymbol{q}=\boldsymbol{0}) = i \int \frac{d^3p}{(2\pi)^3} \frac{\boldsymbol{p}}{p^2} \,\tilde{\delta}^{(1)}(\boldsymbol{p}).
\end{equation}

Summing over all perturbative orders, the nonperturbative density contrast in the IR limit becomes
\begin{align}
    \tilde{\delta}(\boldsymbol{k}) & = \sum_{n=1}^{\infty} \tilde{\delta}^{(n)}(\boldsymbol{k}) \nonumber \\
    & \xrightarrow[\text{IR}]{}
    \sum_{n=1}^{m-1}\tilde{\delta}^{(n)}(\boldsymbol{k})
    + e^{-i\boldsymbol{k}\cdot\overline{\boldsymbol{\Psi}}^{(1)}}\, \tilde{\delta}^{(m)}(\boldsymbol{k}),
\end{align}
where the symbol $\xrightarrow[\text{IR}]{}$ denotes taking the IR limit. In configuration space this corresponds to
\begin{align}
    \delta(\boldsymbol{x}) \xrightarrow[\text{IR}]{}
    \sum_{n=1}^{m-1}\delta^{(n)}(\boldsymbol{x})
    + \delta^{(m)}(\boldsymbol{x}-\overline{\boldsymbol{\Psi}}^{(1)}).
\end{align}

For example, in the case $m=1$ one obtains
\begin{equation}
    \delta(\boldsymbol{x})\xrightarrow[\text{IR}]{}
    \delta^{(1)}(\boldsymbol{x}-\overline{\boldsymbol{\Psi}}^{(1)}),
    \label{eq:delta1_IR}
\end{equation}
implying that all nonlinear effects can be absorbed into a coordinate transformation induced by the large-scale IR flow.
Since the tree-level bispectrum is composed of two linear and one second-order fluctuation, we also consider the case $m=2$:
\begin{equation}
    \delta(\boldsymbol{x})\xrightarrow[\text{IR}]{}
    \delta^{(1)}(\boldsymbol{x}) + \delta^{(2)}(\boldsymbol{x}-\overline{\boldsymbol{\Psi}}^{(1)}).
    \label{eq:delta2_IR}
\end{equation}

Consequently, to compute the three-point correlation function, we apply Eq.~\eqref{eq:delta1_IR} to the first two points, $\boldsymbol{x}_1$ and $\boldsymbol{x}_2$, and Eq.~\eqref{eq:delta2_IR} to the third point $\boldsymbol{x}_3$. This yields
\begin{align}
    & \langle \delta(\boldsymbol{x}_1)\,\delta(\boldsymbol{x}_2)\,\delta(\boldsymbol{x}_3) \rangle
    \nonumber \\
    & \xrightarrow[\text{IR}]{}
    \Big\langle
    \delta^{(1)}(\boldsymbol{x}_1-\overline{\boldsymbol{\Psi}}^{(1)})
    \delta^{(1)}(\boldsymbol{x}_2-\overline{\boldsymbol{\Psi}}^{(1)})
    \delta^{(2)}(\boldsymbol{x}_3-\overline{\boldsymbol{\Psi}}^{(1)})
    \Big\rangle \nonumber \\
    & \quad \quad + \text{2 perms.}
    \nonumber \\
    & 
    \overset{\text{assump.}}{\longrightarrow}
    \Big\langle
    \delta^{(1)}(\boldsymbol{x}_1)\,\delta^{(1)}(\boldsymbol{x}_2)\,\delta^{(2)}(\boldsymbol{x}_3)
    \Big\rangle
    + \text{2 perms.},
    \label{eq:three_IR}
\end{align}
where ``2 perms.'' denotes the additional permutations $\boldsymbol{x}_1 \leftrightarrow \boldsymbol{x}_3$ and $\boldsymbol{x}_2 \leftrightarrow \boldsymbol{x}_3$. The second arrow follows from the assumption that the long-wavelength displacement field is statistically uncorrelated with the small-scale density fluctuations. Under this assumption, statistical homogeneity removes any dependence on $\overline{\boldsymbol{\Psi}}^{(1)}$, causing the contribution of the long-wavelength displacement to vanish.

When evaluating the IR-limit effects of cosmological statistics such as the three-point correlation function, we therefore retain only those terms in which the long-wavelength displacement is uncorrelated with the small-scale modes. Although this should be viewed as a physically intuitive approximation at this stage, we will explicitly show in Sec.~\ref{sec:source_bispec_IRcancel} that, in the IR limit, such uncorrelated contributions dominate the one-loop corrections. Consequently, the spurious IR dependence is removed entirely by statistical homogeneity, and the three-point function reduces to its tree-level expression. This elimination of the nonperturbative long-wavelength contribution is referred to as \emph{IR cancellation}.

In Fourier space, the bispectrum in the IR limit reduces to
\begin{align}
    & \langle \tilde{\delta}(\boldsymbol{k}_1) \tilde{\delta}(\boldsymbol{k}_2) \tilde{\delta}(\boldsymbol{k}_3) \rangle \nonumber \\
    & \xrightarrow[\text{IR}]{}
    \Big\langle e^{-i(\boldsymbol{k}_1+\boldsymbol{k}_2+\boldsymbol{k}_3)\cdot\overline{\boldsymbol{\Psi}}^{(1)}} \Big\rangle
    \langle \tilde{\delta}^{(1)}(\boldsymbol{k}_1) \tilde{\delta}^{(1)}(\boldsymbol{k}_2) \tilde{\delta}^{(2)}(\boldsymbol{k}_3) \rangle
    + \text{2 perms.} \nonumber \\
    & = \langle \tilde{\delta}^{(1)}(\boldsymbol{k}_1) \tilde{\delta}^{(1)}(\boldsymbol{k}_2) \tilde{\delta}^{(2)}(\boldsymbol{k}_3) \rangle
    + \text{2 perms.},
    \label{eq:fourier_IR}
\end{align}
where the cancellation arises from the triangle condition $\boldsymbol{k}_1+\boldsymbol{k}_2+\boldsymbol{k}_3=\boldsymbol{0}$, which enforces translational invariance. Thus, only the tree-level bispectrum survives.

For later convenience, we explicitly express the IR contribution in Eq.~\eqref{eq:fourier_IR} in terms of the displacement-mapping exponent $\overline{\Gamma}$:
\begin{equation}
    \Big\langle e^{-i(\boldsymbol{k}_1+\boldsymbol{k}_2+\boldsymbol{k}_3)\cdot\overline{\boldsymbol{\Psi}}^{(1)}} \Big\rangle
    = e^{-\overline{\Gamma}^{(\rm lin)}(\boldsymbol{k}_1,\boldsymbol{k}_2)}
      e^{\overline{\Gamma}^{(\rm lin)}(\boldsymbol{k}_1,\boldsymbol{k}_2)}
    = 1,
\end{equation}
where $\overline{\Gamma}^{(\rm lin)}$ is defined in Eq.~\eqref{eq:Gamma_lin}. Therefore, the bispectrum in the IR limit can be written as
\begin{align}
    B(\boldsymbol{k}_1,\boldsymbol{k}_2)
    & \xrightarrow[\text{IR}]{}
    e^{-\overline{\Gamma}^{(\rm lin)}(\boldsymbol{k}_1,\boldsymbol{k}_2)}
    e^{\overline{\Gamma}^{(\rm lin)}(\boldsymbol{k}_1,\boldsymbol{k}_2)}
    B_{\rm SPT}^{(\rm tree)}(\boldsymbol{k}_1,\boldsymbol{k}_2) \nonumber \\
    & = B_{\rm SPT}^{(\rm tree)}(\boldsymbol{k}_1,\boldsymbol{k}_2).
    \label{eq:bispec_IR}
\end{align}
This explicitly demonstrates the cancellation of IR effects in the bispectrum.

In summary, the conditions under which IR cancellation occurs can be stated as follows:
\begin{enumerate}
    \item \label{cond:IR1} For perturbative contributions up to a given order, one retains the exact expressions.
    For higher-order terms, the IR limit is taken so that only the effect of the large-scale IR flow is extracted, reducing the long-wavelength contributions to those of a spatially uniform displacement vector.
    \item \label{cond:IR2} When computing statistical quantities such as two- or three-point correlation functions, only the effect in which the long-wavelength displacement field is uncorrelated with the short-wavelength density fluctuations is retained.
    Under this assumption, statistical homogeneity ensures that spurious IR contributions from orders higher than the one of interest are completely canceled, while the exact contributions up to the chosen order (e.g., tree level or one loop) remain unaffected.
\end{enumerate}

\subsection{One-loop bispectrum in the IR Limit}
\label{sec:source_bispec_IRcancel}

In this subsection, we investigate the infrared (IR) behavior of the various one-loop contributions to the bispectrum: the standard SPT bispectrum, the displacement-mapping (DM) bispectrum, and the source bispectrum. Our objective is to evaluate their IR limits explicitly and to demonstrate how the resulting contributions cancel.

\subsubsection*{IR limit of the SPT bispectrum}

We begin by examining the SPT bispectrum. Taking the IR limit $p \rightarrow 0$ for the integration variable $\boldsymbol{p}$ in Eq.~\eqref{eq:B_SPT}, and applying the kernel relation in Eq.~\eqref{eq:F_n_lim}, yields the following behavior. The contribution $B^{\text{1-loop}}_{\mathrm{SPT,GM}}$ must be evaluated carefully, because the transformation $\boldsymbol{p}' = \boldsymbol{k}_2 - \boldsymbol{p}$ implies that both poles at $\boldsymbol{p} \rightarrow 0$ and $\boldsymbol{k}_2 - \boldsymbol{p} \rightarrow 0$ are relevant. This leads to a factor of two in the IR limit. An analogous doubling occurs for $B^{\text{1-loop}}_{\mathrm{SPT,MG}}$, whose integrand contains poles at $\boldsymbol{p} \rightarrow 0$ and $|\boldsymbol{k}_1 - \boldsymbol{p}|\rightarrow 0$. Finally, the term $B^{\text{1-loop}}_{\mathrm{SPT,MMM}}$ contains three distinct poles: $\boldsymbol{p} \rightarrow 0$, $\boldsymbol{k}_1 - \boldsymbol{p} \rightarrow 0$, and $\boldsymbol{k}_2 + \boldsymbol{p} \rightarrow 0$, resulting in a multiplicative factor of three.

With these considerations, the IR limits of the one-loop SPT contributions become
\begin{align}
    & B_{\rm SPT,\,GG}^{(1\text{-loop})}(\boldsymbol{k}_1,\boldsymbol{k}_2)\,
    P_{\rm lin}(k_1)\,P_{\rm lin}(k_2) \nonumber \\
    & \xrightarrow[\text{IR}]{} -\frac{1}{2}
      \left(k_1^2 + k_2^2 + k_3^2\right)\,\bar{\sigma}^2
      \left[2\,F_2(\boldsymbol{k}_1,\boldsymbol{k}_2)\right]
      P_{\rm lin}(k_1)\,P_{\rm lin}(k_2)
      \nonumber \\
    & B_{\rm SPT,\,GM}^{(1\text{-loop})}(\boldsymbol{k}_1,\boldsymbol{k}_2)\,
    P_{\rm lin}(k_1) \nonumber \\
    & \xrightarrow[\text{IR}]{} -(\boldsymbol{k}_2 \cdot \boldsymbol{k}_3)
      \bar{\sigma}^2
      \left[2\,F_2(\boldsymbol{k}_1,\boldsymbol{k}_2)\right]
      P_{\rm lin}(k_1)\,P_{\rm lin}(k_2)
      \nonumber \\
    & B_{\rm SPT,\,MG}^{(1\text{-loop})}(\boldsymbol{k}_1,\boldsymbol{k}_2)\,
    P_{\rm lin}(k_2) \nonumber \\
    & \xrightarrow[\text{IR}]{} -(\boldsymbol{k}_1 \cdot \boldsymbol{k}_3)
      \bar{\sigma}^2
      \left[2\,F_2(\boldsymbol{k}_1,\boldsymbol{k}_2)\right]
      P_{\rm lin}(k_1)\,P_{\rm lin}(k_2)
      \nonumber \\
    & B_{\rm SPT,\,MMM}^{(1\text{-loop})}(\boldsymbol{k}_1,\boldsymbol{k}_2) \nonumber \\
    & \xrightarrow[\text{IR}]{} -(\boldsymbol{k}_1 \cdot \boldsymbol{k}_2)
      \bar{\sigma}^2
      \left[2\,F_2(\boldsymbol{k}_1,\boldsymbol{k}_2)\right]
      P_{\rm lin}(k_1)\,P_{\rm lin}(k_2).
    \label{eq:SPT_B_IR}
\end{align}

Using the triangle relation
$\boldsymbol{k}_1 + \boldsymbol{k}_2 + \boldsymbol{k}_3 = 0$, which
implies
\(
k_1^2 + k_2^2 + k_3^2 = -2(
  \boldsymbol{k}_1 \cdot \boldsymbol{k}_3 +
  \boldsymbol{k}_2 \cdot \boldsymbol{k}_3 +
  \boldsymbol{k}_1 \cdot \boldsymbol{k}_2 ),
\)
one finds that the contributions in Eq.~\eqref{eq:SPT_B_IR} cancel exactly when summed:
\begin{align}
    B_{\rm SPT}^{(\text{1-loop})}(\boldsymbol{k}_1,\boldsymbol{k}_2)
    & \xrightarrow[\text{IR}]{}
      \left[-\overline{\Gamma}(\boldsymbol{k}_1,\boldsymbol{k}_2)
      + \overline{\Gamma}(\boldsymbol{k}_1,\boldsymbol{k}_2)\right]
      B_{\rm SPT}^{(\text{tree})}(\boldsymbol{k}_1,\boldsymbol{k}_2)
      \nonumber \\
    & = 0.
\end{align}
This represents the one-loop truncation of the nonperturbative IR cancellation formula in Eq.~\eqref{eq:bispec_IR}.

\subsubsection*{IR limit of the DM bispectrum}

A completely analogous analysis applies to the DM term. Taking the IR limit $\boldsymbol{p} \rightarrow 0$ in Eq.~\eqref{eq:DM_1loop_comp} yields
\begin{align}
    & B_{\rm DM,\,GG}^{(1\text{-loop})}(\boldsymbol{k}_1,\boldsymbol{k}_2)\,
    P_{\rm lin}(k_1)\,P_{\rm lin}(k_2) \nonumber \\
    &= -\frac{1}{2}(k_1^2 + k_2^2 + k_3^2)\,
       \bar{\sigma}^2\,[2\,F(\boldsymbol{k}_1,\boldsymbol{k}_2)]
       P_{\rm lin}(k_1)\,P_{\rm lin}(k_2)
       \nonumber \\
    & B_{\rm DM,\,GM}^{(1\text{-loop})}(\boldsymbol{k}_1,\boldsymbol{k}_2)\,
    P_{\rm lin}(k_1) \nonumber \\
    & \xrightarrow[\boldsymbol{p}\to 0]{}
       -(\boldsymbol{k}_2 \cdot \boldsymbol{k}_3)\,
       \bar{\sigma}^2\,[2\,F(\boldsymbol{k}_1,\boldsymbol{k}_2)]
       P_{\rm lin}(k_1)\,P_{\rm lin}(k_2)
       \nonumber \\
    & B_{\rm DM,\,MG}^{(1\text{-loop})}(\boldsymbol{k}_1,\boldsymbol{k}_2)\,
    P_{\rm lin}(k_2) \nonumber \\
    & \xrightarrow[\boldsymbol{p}\to 0]{}
       -(\boldsymbol{k}_1 \cdot \boldsymbol{k}_3)\,
       \bar{\sigma}^2\,[2\,F(\boldsymbol{k}_1,\boldsymbol{k}_2)]
       P_{\rm lin}(k_1)\,P_{\rm lin}(k_2)
       \nonumber \\
    & B_{\rm DM,\,MMM}^{(1\text{-loop})}(\boldsymbol{k}_1,\boldsymbol{k}_2) \nonumber \\
    & \xrightarrow[\boldsymbol{p}\to 0]{}
       -(\boldsymbol{k}_1 \cdot \boldsymbol{k}_2)\,
       \bar{\sigma}^2\,[2\,F(\boldsymbol{k}_1,\boldsymbol{k}_2)]
       P_{\rm lin}(k_1)\,P_{\rm lin}(k_2).
    \label{eq:DM_B_IR}
\end{align}

The GG contribution requires no IR treatment, as it contains no mode-coupling integrals. Summing all contributions again yields complete cancellation:
\begin{align}
    B_{\rm DM}^{(\text{1-loop})}(\boldsymbol{k}_1,\boldsymbol{k}_2)
    & \xrightarrow[\text{IR}]{}
      \left[-\overline{\Gamma}(\boldsymbol{k}_1,\boldsymbol{k}_2)
      + \overline{\Gamma}(\boldsymbol{k}_1,\boldsymbol{k}_2)\right]
      B_{\rm SPT}^{(\text{tree})}(\boldsymbol{k}_1,\boldsymbol{k}_2)
      \nonumber \\
    &= 0.
\end{align}

Comparing Eqs.~\eqref{eq:SPT_B_IR} and \eqref{eq:DM_B_IR}, we find that the IR limits of the SPT and DM bispectra are identical. Since Eq.~\eqref{eq:B_SPT_J_DM} shows that the one-loop SPT bispectrum splits into the sum of the source bispectrum and the DM term, it follows that all IR-sensitive contributions originate exclusively from the DM term. The source bispectrum is therefore intrinsically short-wavelength and contains no IR effect describable as a uniform coordinate transformation:
\begin{equation}
    B_{{\rm J},X}^{(\text{1-loop})}(\boldsymbol{k}_1,\boldsymbol{k}_2)
    \xrightarrow[\text{IR}]{} 0,
    \qquad
    X \in \{\mathrm{GG},\mathrm{GM},\mathrm{MG},\mathrm{MMM}\}.
\end{equation}

\subsubsection*{Implications for IR cancellation}

The DM term describes the action of the displacement-mapping operator on the source bispectrum. The displacement-mapping factor itself contains only those contributions from the displacement field that are statistically uncorrelated with the Jacobian deviation. Thus, the fact that IR-dominant contributions arise solely from the DM term provides a one-loop realization of condition~(\ref{cond:IR2}) for IR cancellation: in the IR limit, only contributions in which the long-wavelength displacement is uncorrelated with short-scale density fluctuations are retained.

Taking the IR limit of the DM term corresponds to evaluating the displacement-mapping factor at coincident points, $\boldsymbol{r}_1 \rightarrow 0$ and $\boldsymbol{r}_2 \rightarrow 0$. Using Eq.~\eqref{eq:DM_1loop}, we obtain
\begin{align}
    & B_{\rm DM}^{(\text{1-loop})}(\boldsymbol{k}_1,\boldsymbol{k}_2) \nonumber \\
    & \xrightarrow[\text{IR}]{}
      -\overline{\Gamma}^{(\mathrm{lin})}(\boldsymbol{k}_1,\boldsymbol{k}_2)
      B_{\rm J}^{(\mathrm{tree})}(\boldsymbol{k}_1,\boldsymbol{k}_2)
      \nonumber \\
    &\quad +
      \int d^3 r_1\,e^{-i \boldsymbol{k}_1\cdot\boldsymbol{r}_1}
      \int d^3 r_2\,e^{-i \boldsymbol{k}_2\cdot\boldsymbol{r}_2} \nonumber \\
      & \quad \times
      \Gamma^{(\mathrm{lin})}(\boldsymbol{k}_1,\boldsymbol{k}_2,
      \boldsymbol{r}_1 \rightarrow 0,\boldsymbol{r}_2 \rightarrow 0)
      \zeta_{\rm J}^{(\mathrm{tree})}(\boldsymbol{r}_1,\boldsymbol{r}_2)
      \nonumber \\
    &= \left[-\overline{\Gamma}^{(\mathrm{lin})}(\boldsymbol{k}_1,\boldsymbol{k}_2)
      + \overline{\Gamma}^{(\mathrm{lin})}(\boldsymbol{k}_1,\boldsymbol{k}_2)
      \right]
      B_{\rm SPT}^{(\mathrm{tree})}(\boldsymbol{k}_1,\boldsymbol{k}_2)
      \nonumber \\
    &= 0.
\end{align}

This result illustrates condition~(\ref{cond:IR1}): taking the IR limit isolates the contribution of a spatially uniform long-wavelength displacement vector. The evaluation involves only coincident points, $\boldsymbol{r}_1 \rightarrow 0$ and $\boldsymbol{r}_2 \rightarrow 0$, so that the separations between distinct points effectively vanish.

\subsection{IR cancellation in ULPT}
\label{sec:ULPTbispec_IRcancel}

In this subsection, we demonstrate explicitly how the nonperturbative IR cancellation arises in the one-loop ULPT bispectrum. By construction, the ULPT bispectrum reproduces the SPT prediction exactly up to one-loop order, while retaining only the displacement-mapping contributions from higher perturbative orders ($n \geq 2$). Accordingly, the one-loop ULPT bispectrum may be written in terms of SPT quantities as
\begin{equation}
    B_{\rm ULPT}(\boldsymbol{k}_1,\boldsymbol{k}_2)
    = B_{\rm SPT}^{(\mathrm{tree + 1\text{-}loop})}(\boldsymbol{k}_1,\boldsymbol{k}_2)
    + \sum_{n=2}^{\infty}
      B_{\rm DM}^{(\mathrm{n\text{-}loop})}(\boldsymbol{k}_1,\boldsymbol{k}_2),
\end{equation}
where $B_{\rm DM}^{(n\geq2\text{-loop})}$ is given by Eq.~\eqref{eq:B_DM_nloop}.

Taking the IR limit of the higher-order DM contributions corresponds to evaluating the displacement-mapping factor at coincident points, $\boldsymbol{r}_1 \to 0$ and $\boldsymbol{r}_2 \to 0$, in Eq.~\eqref{eq:B_DM_nloop}. Under this limit, all such contributions vanish:
\begin{equation}
    B_{\rm DM}^{(\mathrm{n\geq2\text{-}loop})}(\boldsymbol{k}_1,\boldsymbol{k}_2)
    \xrightarrow[\mathrm{IR}]{} 0.
\end{equation}
Thus, only the SPT terms up to the perturbative order of interest (here one-loop) survive, yielding
\begin{equation}
    B_{\rm ULPT}(\boldsymbol{k}_1,\boldsymbol{k}_2)
    \xrightarrow[\mathrm{IR}]{}
    B_{\rm SPT}^{(\mathrm{tree + 1\text{-}loop})}(\boldsymbol{k}_1,\boldsymbol{k}_2).
\end{equation}

In this manner, the ULPT bispectrum realizes the IR cancellation mechanism in a fully nonperturbative fashion. When the IR limit is taken, the one-loop ULPT bispectrum consistently reduces to the corresponding one-loop SPT result, thereby satisfying the expected infrared behavior.

\section{IR-Resummed Modeling}
\label{sec:IR_resum}

A key strength of ULPT is its ability to incorporate IR effects in a fully nonperturbative manner. Consequently, ULPT naturally provides an IR-resummed description that captures the nonlinear damping of BAO. In conventional approaches, IR resummation is implemented by decomposing the linear matter power spectrum into a ``wiggle'' component, $P_{\rm w}$, which contains the BAO feature, and a smooth ``no-wiggle'' component, $P_{\rm nw}$~\cite{Eisenstein:1997ik}, such that $P_{\rm lin} = P_{\rm w} + P_{\rm nw}$. The nonperturbative IR effects are resummed only for the wiggle part, ensuring the correct modeling of BAO damping, while the broadband shape is described using the standard SPT prediction based on the no-wiggle component.

In this section, we begin in Sec.~\ref{sec:history} by reviewing the historical development of IR-resummed models for the power spectrum, followed by a ULPT-based derivation in Sec.~\ref{sec:IR_resummed_power}. Section~\ref{sec:TSPT} summarizes the IR-resummed bispectrum model formulated within time-sliced perturbation theory (TSPT)~\cite{Blas:2016sfa}, and Sec.~\ref{sec:NoMC} presents an alternative bispectrum model in which the wiggle contribution in the mode-coupling term is neglected~\cite{Sugiyama:2020uil}. Incorporating these insights, we introduce in Sec.~\ref{sec:ULPT_bispec_IR} a one-loop IR-resummed bispectrum model derived from ULPT. Finally, Sec.~\ref{sec:ULPT_bispec_IR_recon} extends this formulation to the post-reconstruction bispectrum.

\subsection{Historical development of the IR-resummed power spectrum}
\label{sec:history}

The earliest form of the IR-resummed power spectrum model was proposed in Ref.~\cite{Eisenstein:2006nj}, where the nonlinear damping of the BAO feature was shown to be well described by the factor $e^{-k^2\bar{\sigma}^2} P_{\rm lin}(k)$. To recover the small-scale linear spectrum without oscillatory features, one simply adds the no-wiggle power spectrum $P_{\rm nw}$ multiplied by the complement of this smearing factor, $1 - e^{-k^2\bar{\sigma}^2}$. This template model has proven extremely useful and has been widely adopted in BAO data analyses (see, e.g., the recent results of Ref.~\cite{Chen:2024tfp} and references therein).

In Ref.~\cite{Sugiyama:2013gza}, the power spectrum was decomposed into the propagator and mode-coupling contributions according to Eq.~\eqref{eq:G_MC}, and their IR limits were derived explicitly as
\begin{align}
    G^2(k)\,P_{\rm lin}(k) & \xrightarrow[\text{IR}]{}
    e^{-k^2\bar{\sigma}^2}\,P_{\rm lin}(k), \nonumber \\
    P_{\rm MC}(k) & \xrightarrow[\text{IR}]{}
    \left(1 - e^{-k^2\bar{\sigma}^2}\right)\,P_{\rm lin}(k).
    \label{eq:G_P_MC_IR}
\end{align}
Based on the empirical observation that the wiggle component in the mode-coupling term is negligibly small, the linear spectrum appearing in $P_{\rm MC}$ was replaced by the no-wiggle spectrum, thereby defining the approximate no-wiggle mode-coupling contribution
\begin{equation}
    P_{\rm MC, nw}(k) \approx
    \left(1 - e^{-k^2\bar{\sigma}^2}\right)\,P_{\rm nw}(k).
\end{equation}
Adding this no-wiggle mode-coupling term to the propagator contribution reproduces the Eisenstein template model~\cite{Eisenstein:2006nj,Sugiyama:2013gza}:
\begin{equation}
    P_{\rm IR\text{-}resum}(k)
    = e^{-k^2\bar{\sigma}^2}\, P_{\rm w}(k) + P_{\rm nw}(k),
    \label{eq:IR_resummed_power_1}
\end{equation}
which was subsequently extended to two-loop order in the same work.

While it is intuitively expected that the superposition of many modes in the mode-coupling integrals tends to smooth out oscillatory structures such as BAO, this behavior is not guaranteed \emph{a priori} for arbitrary functions. Subsequent studies~\cite{Baldauf:2015xfa,Blas:2016sfa} provided a detailed theoretical analysis of the wiggle contribution to the mode-coupling term and derived
\begin{align}
    P_{\rm MC}(k)
    &= \left(1 - e^{-k^2\bar{\sigma}^2}\right)\,P_{\rm nw}(k) \nonumber \\
    &\quad + \left(e^{-k^2\sigma_{\rm BAO}^2}
    - e^{-k^2\bar{\sigma}^2}\right)\,P_{\rm w}(k),
    \label{eq:P_MC_BAO}
\end{align}
where
\begin{equation}
    \sigma_{\rm BAO}^2
    = \frac{1}{3} \int \frac{dp}{2\pi^2}
    \left[1 - j_0(pr_{\rm BAO}) + 2 j_2(pr_{\rm BAO})\right]
    P_{\rm lin}(p),
    \label{eq:sigma_BAO}
\end{equation}
with $r_{\rm BAO} \simeq 105\,h^{-1}\,{\rm Mpc}$ denoting the characteristic BAO scale. Combining Eq.~\eqref{eq:P_MC_BAO} with the propagator term in Eq.~\eqref{eq:G_P_MC_IR}, one obtains
\begin{equation}
    P_{\rm IR\text{-}resum}(k)
    = e^{-k^2\sigma_{\rm BAO}^2}\, P_{\rm w}(k) + P_{\rm nw}(k).
    \label{eq:IR_resummed_power_2}
\end{equation}

Comparing Eqs.~\eqref{eq:IR_resummed_power_1} and \eqref{eq:IR_resummed_power_2}, the only difference lies in whether the BAO damping is characterized by $\bar{\sigma}^2$ or $\sigma_{\rm BAO}^2$. As emphasized in Ref.~\cite{Blas:2016sfa}, these two parameters take nearly identical values in $\Lambda$CDM models, rendering the wiggle contribution in the mode-coupling term effectively negligible. For the fiducial Planck 2015 cosmology at $z=0$, we obtain $\bar{\sigma}^2 = 35.77\,(h^{-1}\,{\rm Mpc})^2$ and $\sigma_{\rm BAO}^2 = 36.3\,(h^{-1}\,{\rm Mpc})^2$, corresponding to a difference of only $1.46\%$. In this case, the prefactor $(e^{-k^2\sigma_{\rm BAO}^2} - e^{-k^2\bar{\sigma}^2})$ in Eq.~\eqref{eq:P_MC_BAO} remains below $0.005$ over the range $k = 0.01$--$0.3\,h\,{\rm Mpc}^{-1}$, quantitatively confirming that the wiggle contribution in the mode-coupling term is strongly suppressed. Therefore, in realistic $\Lambda$CDM-like universes, neglecting this contribution constitutes an excellent approximation.

An additional advantage of using $\sigma_{\rm BAO}^2$ is that the integrand in its definition contains the factor
$[1 - j_0(pr_{\rm BAO}) + 2j_2(pr_{\rm BAO})]$, which behaves as $p^2$ in the limit $p \to 0$, ensuring rapid decay and improved numerical convergence. However, since in $\Lambda$CDM cosmologies the linear power spectrum itself scales as $P_{\rm lin}(k)\propto k^{n_s}$ with $n_s\simeq 0.96$ as $k\to 0$, the integral defining $\bar{\sigma}^2$ is already convergent. Consequently, the practical numerical advantage of using $\sigma_{\rm BAO}^2$ is modest.

More recently, IR-resummed models have also been developed for the reconstructed power spectrum. In early implementations of density-field reconstruction, the nonlinear BAO damping was often modeled as a superposition of two Gaussian exponentials~\cite{Padmanabhan:2008dd}. However, recent studies~\cite{Sugiyama:2024eye,Chen:2024tfp} have shown that, when IR effects are treated consistently, the BAO damping after reconstruction can also be described by a single Gaussian exponential, in close analogy with the pre-reconstruction case. The linear IR-resummed model for the reconstructed power spectrum is then given by
\begin{equation}
    P_{\rm IR\text{-}resum,rec}(k)
    = e^{-k^2\sigma_{\rm BAO,rec}^2}\, P_{\rm w}(k) + P_{\rm nw}(k),
    \label{eq:IR_resummed_power_recon}
\end{equation}
with the damping scale
\begin{align}
    \sigma_{\rm BAO,rec}^2
    &= \frac{1}{3} \int \frac{dp}{2\pi^2}
    \left[1 - j_0(pr_{\rm BAO}) + 2 j_2(pr_{\rm BAO})\right] \nonumber \\
    & \quad \times
    \left[1 - W_{\rm G}(pR)\right]^2
    P_{\rm lin}(p).
    \label{eq:sigma_BAO_recon}
\end{align}
As pointed out in Ref.~\cite{Sugiyama:2024eye}, the Gaussian smoothing kernel $W_{\rm G}$ introduced during reconstruction suppresses the integrand at large scales ($p \to 0$), thereby reducing the contributions of the spherical Bessel functions $j_0$ and $j_2$. This suppression further diminishes the wiggle part in the mode-coupling term, making its contribution even more negligible after reconstruction. Under this improved approximation, the IR-resummed power spectrum may equivalently be written as
\begin{equation}
    P_{\rm IR\text{-}resum,rec}(k)
    = e^{-k^2\bar{\sigma}_{\rm rec}^2}\, P_{\rm w}(k) + P_{\rm nw}(k),
\end{equation}
with
\begin{equation}
    \bar{\sigma}_{\rm rec}^2
    = \frac{1}{3} \int \frac{dp}{2\pi^2}
    \left[1 - W_{\rm G}(pR)\right]^2 P_{\rm lin}(p).
    \label{eq:sigma_bar_rec}
\end{equation}
For the fiducial $\Lambda$CDM cosmology at $z=0$ with smoothing scale $R = 15\,h^{-1}\,{\rm Mpc}$, we obtain $\bar{\sigma}_{\rm rec}^2 = 12.471\,(h^{-1}\,{\rm Mpc})^2$ and $\sigma_{\rm BAO,rec}^2 = 12.448\,(h^{-1}\,{\rm Mpc})^2$. The relative difference of only $0.18\%$ demonstrates that neglecting the wiggle contribution in the mode-coupling term is an even better approximation in the post-reconstruction case than in the pre-reconstruction one.

As a further extension, Ref.~\cite{Sugiyama:2024ggt} developed a one-loop IR-resummed model for the reconstructed halo power spectrum, incorporating reconstruction-specific effects such as shot noise. This formulation is likewise based on the approximation that the wiggle contribution in the mode-coupling term can safely be neglected.

\subsection{IR-resummed power spectrum within ULPT}
\label{sec:IR_resummed_power}

In this subsection, we demonstrate how ULPT naturally incorporates an IR-resummed description of the power spectrum. A more detailed derivation is presented in Refs.~\cite{Sugiyama:2025ntz,Sugiyama:2025myq}.

Within ULPT, the dark matter power spectrum is given by
\begin{equation}
    P_{\rm ULPT}(k)
    = e^{-\overline{\Sigma}(k)}
    \int d^3r\, e^{-i\boldsymbol{k}\cdot\boldsymbol{r}}\,
    e^{\Sigma(\boldsymbol{k},\boldsymbol{r})}\,\xi_{\rm J}(r),
    \label{eq:ULPT_power}
\end{equation}
where
\begin{equation}
    -\overline{\Sigma}(k) + \Sigma(\boldsymbol{k},\boldsymbol{r})
    =
    \sum_{m=2}^\infty \frac{1}{m!}
    \left\langle
    \left[ -i\boldsymbol{k}\cdot\left( \boldsymbol{\Psi}(\boldsymbol{q}_1)
    - \boldsymbol{\Psi}(\boldsymbol{q}_2)\right)\right]^m
    \right\rangle_{\rm c},
\end{equation}
with $\boldsymbol{r}=\boldsymbol{q}_1-\boldsymbol{q}_2$, and where $\xi_{\rm J}(r)$ denotes the source two-point correlation function.

The source correlation function can be decomposed into wiggle and no-wiggle components,
\begin{equation}
    \xi_{\rm J}(r)
    =
    \xi_{\rm J, w}(r) +
    \xi_{\rm J, nw}(r).
\end{equation}
Substituting this decomposition into Eq.~\eqref{eq:ULPT_power}, we find that the no-wiggle component primarily determines the broadband shape of the power spectrum. In line with the philosophy of IR resummation, this broadband shape is described by the SPT prediction. Accordingly, we define
\begin{align}
    P_{\rm ULPT, nw}(k)
    & =  e^{-\overline{\Sigma}(k)}
    \int d^3r\, e^{-i\boldsymbol{k}\cdot\boldsymbol{r}}\,
    e^{\Sigma(\boldsymbol{k},\boldsymbol{r})}\,\xi_{\rm J, nw}(r) \nonumber \\
    & \approx P_{\rm SPT, nw}(k),
    \label{eq:SPT_power_nw}
\end{align}
where $P_{\rm SPT, nw}$ denotes the truncated SPT solution at a given $n$-loop order, constructed using the no-wiggle linear matter power spectrum.

For the wiggle contribution, we simplify the convolution integrals associated with the displacement-mapping factor by adopting the following approximation. The wiggle part of the source correlation function is sharply localized around the BAO scale and negligible elsewhere, effectively behaving like a delta function in configuration space. This allows us to fix the scale dependence of the displacement-mapping factor at $r = r_{\rm BAO} = 105\,h^{-1}\,{\rm Mpc}$. For simplicity, we furthermore set the angular dependence to $\mu = \hat{k} \cdot \hat{r} = 1$ as a representative choice. Under these approximations, the displacement-mapping factor can be factored out of the integration, yielding
\begin{align}
    P_{\rm ULPT, w}(k)
    & \sim e^{-\overline{\Sigma}(k)}\,e^{\Sigma(k, r= r_{\rm BAO}, \mu=1)}\,
    \int d^3r\, e^{-i\boldsymbol{k}\cdot\boldsymbol{r}}\,
    \xi_{\rm J,w}(r) \nonumber \\
    & = e^{-\overline{\Sigma}(k)}\,e^{\Sigma(k, r=r_{\rm BAO}, \mu=1)}\,
    P_{\rm J,w}(k),
    \label{eq:ULPT_power_w}
\end{align}
where $P_{\rm J,w}(k)$ is the wiggle part of the source power spectrum.

Combining the two pieces, the general ULPT expression for the IR-resummed power spectrum becomes
\begin{equation}
    P_{\rm IR\text{-}resum}(k)
    = e^{-\overline{\Sigma}(k)+\Sigma(k, r= r_{\rm BAO}, \mu=1)}\, P_{\rm J,w}(k)
    + P_{\rm SPT, nw}(k).
\end{equation}

Evaluating $\Sigma$ at linear order yields
\begin{equation}
    -\overline{\Sigma}^{(\rm lin)}(k)+\Sigma^{(\rm lin)}(k, r= r_{\rm BAO}, \mu=1)
    = -k^2 \sigma_{\rm BAO}^2,
\end{equation}
where
\begin{equation}
    \sigma_{\rm BAO}^2 =  \bar{\sigma}^2 - \sigma_0^2(r_{\rm BAO}) - 2 \sigma_2^2(r_{\rm BAO}),
    \label{eq:sigma_BAO_ver2}
\end{equation}
as defined in Eq.~\eqref{eq:sigma_BAO}.

In the simplest case, if $P_{\rm J,w}$ and $P_{\rm SPT, nw}$ are both evaluated at linear order, they reduce to the wiggle and no-wiggle linear power spectra, $P_{\rm w}$ and $P_{\rm nw}$, respectively. This immediately reproduces the linear IR-resummed model of Eq.~\eqref{eq:IR_resummed_power_2}.

A major advantage of ULPT is its ability to provide a unified description of density fluctuations both before and after reconstruction. As shown in Eq.~\eqref{eq:delta_rec}, the effect of reconstruction modifies only the displacement vector: the pre-reconstruction displacement $\boldsymbol{\Psi}$ in Eq.~\eqref{eq:delta_ULPT} is simply replaced by the reconstructed displacement $\boldsymbol{\Psi}_{\rm rec}$ [Eq.~\eqref{eq:Psi_rec}]. Since the mathematical structure of the density field remains unchanged, the derivation of the IR-resummed model for the pre-reconstruction case can be directly extended to the post-reconstruction case without further modification.

Defining the exponent of the reconstructed displacement-mapping factor as
\begin{align}
   & -\overline{\Sigma}_{\rm rec}(k) + \Sigma_{\rm rec}(\boldsymbol{k},\boldsymbol{r}) \nonumber \\
   & =
    \sum_{m=2}^\infty \frac{1}{m!}
    \left\langle
    \left[ -i\boldsymbol{k}\cdot\left( \boldsymbol{\Psi}_{\rm rec}(\boldsymbol{q}_1)
    - \boldsymbol{\Psi}_{\rm rec}(\boldsymbol{q}_2)\right)\right]^m
    \right\rangle_{\rm c},
\end{align}
one can immediately evaluate, at linear order, the reconstructed BAO damping scale $\sigma_{\rm BAO, rec}^2$ as defined in Eq.~\eqref{eq:sigma_BAO_recon}:
\begin{equation}
    -\overline{\Sigma}_{\rm rec}^{(\rm lin)}(k)
    +\Sigma_{\rm rec}^{(\rm lin)}(k, r= r_{\rm BAO}, \mu=1)
    = -k^2 \sigma_{\rm BAO, rec}^2.
\end{equation}
Thus, the reconstructed IR-resummed model of Eq.~\eqref{eq:IR_resummed_power_recon} is obtained in a straightforward manner.

\subsection{IR-resummed bispectrum in time-sliced perturbation theory}
\label{sec:TSPT}

The first formulation of an IR-resummed bispectrum was developed within the framework of time-sliced perturbation theory (TSPT)~\cite{Blas:2016sfa}. In this approach, the analysis is restricted to the leading-order contributions in the BAO feature, namely terms of order ${\cal O}(P_{\rm w})$, while higher-order terms such as ${\cal O}(P_{\rm w}^2)$ are systematically neglected. Under this approximation, the tree-level IR-resummed bispectrum takes the form
\begin{align}
    B_{\rm IR-resum}(\boldsymbol{k}_1,\boldsymbol{k}_2)
    &= 2\,F_2(\boldsymbol{k}_1,\boldsymbol{k}_2)\Big[
    P_{\rm nw}(k_1)P_{\rm nw}(k_2) \nonumber \\
    & \quad +
    e^{-k_1^2\sigma_{\rm BAO}^2}P_{\rm w}(k_1)P_{\rm nw}(k_2) \nonumber \\
    & \quad +
    + e^{-k_2^2\sigma_{\rm BAO}^2}P_{\rm nw}(k_1)P_{\rm w}(k_2)
    \Big].
    \label{eq:bispec_IR_resum_1}
\end{align}

Furthermore, Ref.~\cite{Blas:2016sfa} proposed a general prescription applicable to arbitrary $n$-point correlation functions $C_n$, namely,
\begin{equation}
    C_n(\boldsymbol{k}_1,\cdots,\boldsymbol{k}_n)
    = C_n^{(\rm tree)}\Big[
    P_{\rm lin}\to P_{\rm nw} + e^{-k^2\sigma_{\rm BAO}^2} P_{\rm w}
    \Big],
\end{equation}
where the replacement rule in brackets indicates that every occurrence of the linear power spectrum in the tree-level $n$-point function is replaced by the sum of a no-wiggle component and an exponentially damped wiggle component.

Applying this prescription to the bispectrum yields
\begin{align}
    B_{\rm IR-resum}(\boldsymbol{k}_1,\boldsymbol{k}_2)
    &= 2\,F_2(\boldsymbol{k}_1,\boldsymbol{k}_2)
    \Big[P_{\rm nw}(k_1)+e^{-k_1^2\sigma_{\rm BAO}^2}P_{\rm w}(k_1)\Big] \nonumber \\
    &\quad \times
    \Big[P_{\rm nw}(k_2)+e^{-k_2^2\sigma_{\rm BAO}^2}P_{\rm w}(k_2)\Big].
    \label{eq:bispec_IR_resum_2}
\end{align}
Expanding Eq.~\eqref{eq:bispec_IR_resum_2} to leading order in $P_{\rm w}$ reproduces Eq.~\eqref{eq:bispec_IR_resum_1}.

It should be noted, however, that contributions of order ${\cal O}(P_{\rm w}^2)$, such as the term proportional to
$e^{-(k_1^2+k_2^2)\sigma_{\rm BAO}^2} P_{\rm w}(k_1)P_{\rm w}(k_2)$, are not formally justified within the TSPT framework, since the expansion is explicitly truncated at ${\cal O}(P_{\rm w})$. Nevertheless, such IR-resummed bispectrum models have been successfully applied to galaxy bispectrum analyses (e.g., Ref.~\cite{DAmico:2022osl}), yielding consistent fits and thereby demonstrating their practical utility in cosmological data modeling.

\subsection{IR-resummed bispectrum with neglected wiggle contributions in the mode-coupling terms}
\label{sec:NoMC}

An alternative approach to IR resummation was proposed in Ref.~\cite{Sugiyama:2020uil}, motivated by earlier studies of the power spectrum. In this framework, the wiggle components appearing in the mode-coupling integrals of the bispectrum are assumed to be negligible. Incorporating this approximation and retaining contributions up to ${\cal O}(P_{\rm w}^2)$, the IR-resummed bispectrum model is given by
\begin{align}
    B_{\rm IR-resum}(\boldsymbol{k}_1,\boldsymbol{k}_2)
    &= 2\,F_2(\boldsymbol{k}_1,\boldsymbol{k}_2)\Big[
       P_{\rm nw}(k_1)P_{\rm nw}(k_2) \nonumber \\
    & \quad + e^{-k_1^2\bar{\sigma}^2}P_{\rm w}(k_1)P_{\rm nw}(k_2) \nonumber \\
    & \quad + e^{-k_2^2\bar{\sigma}^2}P_{\rm nw}(k_1)P_{\rm w}(k_2) \nonumber \\
    & \quad + e^{-\tfrac{1}{2}(k_1^2+k_2^2+k_3^2)\bar{\sigma}^2}
    P_{\rm w}(k_1)P_{\rm w}(k_2)\Big].
    \label{eq:bispec_IR_resum_nao}
\end{align}

Comparing Eq.~\eqref{eq:bispec_IR_resum_2} with Eq.~\eqref{eq:bispec_IR_resum_nao}, we find that the ${\cal O}(P_{\rm w})$ terms have the same structure, with the damping scale $\sigma_{\rm BAO}^2$ replaced by $\bar{\sigma}^2$. This is directly analogous to the situation for the power spectrum, where neglecting wiggle contributions in the mode-coupling integrals leads to $\bar{\sigma}^2 \simeq \sigma_{\rm BAO}^2$ in $\Lambda$CDM cosmologies. Therefore, the approximation is well justified at ${\cal O}(P_{\rm w})$.

At ${\cal O}(P_{\rm w}^2)$, however, a new scale dependence emerges. Using the identity
$k_1^2 + k_2^2 + k_3^2 = 2(k_1^2 + k_2^2 + \boldsymbol{k}_1 \cdot \boldsymbol{k}_2)$,
the exponential factor in Eq.~\eqref{eq:bispec_IR_resum_nao} introduces a cross term
$e^{-\boldsymbol{k}_1 \cdot \boldsymbol{k}_2\,\bar{\sigma}^2}$,
which is absent from Eq.~\eqref{eq:bispec_IR_resum_2}. This reflects a subtle but genuine difference between the two resummation schemes beyond leading order.

This model has been widely used in a variety of bispectrum analyses, including anisotropic BAO measurements with RSD~\cite{Sugiyama:2020uil}, tests of modified gravity~\cite{Sugiyama:2023tes}, and studies of large-scale structure consistency relations~\cite{Sugiyama:2023zvd}.

More recently, Ref.~\cite{Sugiyama:2024qsw} extended this approximation, namely the neglect of the BAO signal in the mode-coupling integrals, to reconstructed fields. It was shown that the same structural form of the IR-resummed bispectrum remains valid after reconstruction. In this extension, the BAO damping scale and the kernel functions are consistently replaced by their reconstructed counterparts, while the overall functional structure of the model is preserved:
\begin{align}
    B_{\rm IR-resum,\,rec}(\boldsymbol{k}_1,\boldsymbol{k}_2)
    &= 2\,F_{{\rm rec},2}(\boldsymbol{k}_1,\boldsymbol{k}_2)
       \Big[P_{\rm nw}(k_1)P_{\rm nw}(k_2) \nonumber \\
    &\quad + e^{-k_1^2\bar{\sigma}_{\rm rec}^2}P_{\rm w}(k_1)P_{\rm nw}(k_2) \nonumber \\
    &\quad + e^{-k_2^2\bar{\sigma}_{\rm rec}^2}P_{\rm nw}(k_1)P_{\rm w}(k_2) \nonumber \\
    &\quad + e^{-\tfrac{1}{2}(k_1^2+k_2^2+k_3^2)
        \bar{\sigma}_{\rm rec}^2}
        P_{\rm w}(k_1)P_{\rm w}(k_2)\Big].
    \label{eq:bispec_IR_resum_nao_rec}
\end{align}
Here, the second-order nonlinear kernel after reconstruction is given by~\cite{Schmittfull:2015mja}
\begin{equation}
    F_{{\rm rec},2}(\boldsymbol{k}_1,\boldsymbol{k}_2)
    = F_2(\boldsymbol{k}_1,\boldsymbol{k}_2)
    + \frac{1}{2}\left[ \boldsymbol{k}\cdot\boldsymbol{R}(\boldsymbol{k}_1) + \boldsymbol{k}\cdot\boldsymbol{R}(\boldsymbol{k}_2)  \right],
        \label{eq:F2_rec}
\end{equation}
where $\boldsymbol{R}(\boldsymbol{k})$ is the reconstruction vector kernel introduced in Eq.~\eqref{eq:R}.

\subsection{Derivation of the IR-resummed bispectrum within ULPT}
\label{sec:ULPT_bispec_IR}

Both IR-resummed bispectrum models presented in Eqs.~\eqref{eq:bispec_IR_resum_2} and \eqref{eq:bispec_IR_resum_nao} have been shown to provide mutually consistent descriptions of observational data. The origin of this agreement is twofold. First, in $\Lambda$CDM cosmologies one finds $\bar{\sigma}^2 \simeq \sigma_{\rm BAO}^2$, so that the small numerical difference between these parameters has only a negligible impact on the predicted damping of BAO features. Second, the ${\cal O}(P_{\rm w}^2)$ contributions are generally subdominant relative to those of ${\cal O}(P_{\rm w})$, implying that the distinct scale dependence appearing at ${\cal O}(P_{\rm w}^2)$ does not significantly affect the overall performance of the models.

Nevertheless, a distinctive property of the bispectrum is that ${\cal O}(P_{\rm w}^2)$ terms already appear at the lowest nontrivial order, namely at tree level in SPT, because the tree-level bispectrum is constructed from products of two linear power spectra, $P_{\rm lin}^2$. A precise theoretical understanding of the BAO damping associated with these terms is therefore essential for constructing a more accurate and systematically consistent bispectrum model. The absence of such a formulation in previous studies reflects the lack of a sufficiently detailed theoretical framework for BAO damping in the bispectrum.

In this subsection, we address this issue by deriving a more consistent IR-resummed bispectrum model directly within the ULPT framework. This provides a theoretically well-motivated and nonperturbative description of BAO damping in the bispectrum, including both ${\cal O}(P_{\rm w})$ and ${\cal O}(P_{\rm w}^2)$ contributions on equal footing.

\subsubsection{Decomposition of the bispectrum into wiggle and no-wiggle components}
\label{sec:decom}

The three-point source correlation function $\zeta_{\rm J}(\boldsymbol{r}_1,\boldsymbol{r}_2)$ depends on two spatial scales, $r_1 = |\boldsymbol{r}_1|$ and $r_2 = |\boldsymbol{r}_2|$. This allows one to define wiggle and no-wiggle components with respect to each of these two scales. Accordingly, we decompose the correlation function into four distinct components:
\begin{align}
    \zeta_{ \text{J}}(\boldsymbol{r}_1,\boldsymbol{r}_2) &=
    \zeta_{ \text{J, w--w}}(\boldsymbol{r}_1,\boldsymbol{r}_2)
    + \zeta_{ \text{J, w--nw}}(\boldsymbol{r}_1,\boldsymbol{r}_2) \nonumber \\
    &\quad + \zeta_{ \text{J, nw--w}}(\boldsymbol{r}_1,\boldsymbol{r}_2)
    + \zeta_{ \text{J, nw--nw}}(\boldsymbol{r}_1,\boldsymbol{r}_2),
\end{align}
where the subscript notation ``X--Y'' indicates the presence (w) or absence (nw) of BAO wiggles in the $r_1$ and $r_2$ dependence, respectively. That is, X denotes the wiggle content along $r_1$ and Y along $r_2$, with $\text{X--Y} \in \{\text{w--w},\, \text{w--nw},\, \text{nw--w},\, \text{nw--nw}\}$. For example, ``w--nw'' refers to a component that exhibits oscillations in the $r_1$ direction but remains smooth in $r_2$.

Substituting this decomposition into the one-loop ULPT bispectrum expression given in Eq.~\eqref{eq:ULPT_1loop}, the bispectrum naturally separates into four corresponding contributions:
\begin{align}
    B_{\rm ULPT}(\boldsymbol{k}_1,\boldsymbol{k}_2)
    &= B_{\text{w--w}}(\boldsymbol{k}_1,\boldsymbol{k}_2)
    + B_{\text{w--nw}}(\boldsymbol{k}_1,\boldsymbol{k}_2) \nonumber \\
    &\quad + B_{\text{nw--w}}(\boldsymbol{k}_1,\boldsymbol{k}_2)
    + B_{ \text{nw--nw}}(\boldsymbol{k}_1,\boldsymbol{k}_2),
\end{align}
with each component given by
\begin{align}
    B_{ \text{X--Y}}(\boldsymbol{k}_1,\boldsymbol{k}_2)
    &= e^{-\overline{\Gamma}^{\rm (lin)}(\boldsymbol{k}_1,\boldsymbol{k}_2)}
    \int d^3r_1\, e^{-i \boldsymbol{k}_1 \cdot \boldsymbol{r}_1}
    \int d^3r_2\, e^{-i \boldsymbol{k}_2 \cdot \boldsymbol{r}_2} \nonumber \\
    &\quad \times
    e^{\Gamma^{\rm (lin)}(\boldsymbol{k}_1,\boldsymbol{k}_2,\boldsymbol{r}_1,\boldsymbol{r}_2)}
    \zeta^{(\text{tree + 1-loop})}_{ \text{J, X--Y}}(\boldsymbol{r}_1,\boldsymbol{r}_2),
\end{align}
where $\text{X--Y} \in \{\text{w--w},\, \text{w--nw},\, \text{nw--w},\, \text{nw--nw}\}$.

This decomposition provides a systematic basis for constructing the IR-resummed bispectrum model. As in the case of the power spectrum, the central idea is to nonperturbatively resum the long-wavelength contributions that affect the BAO signal, while evaluating the smooth broadband terms in the IR limit such that the nonperturbative IR effects cancel. In this way, the nonlinear damping of BAO features is properly incorporated, while the smooth broadband shape of the bispectrum is preserved in agreement with the SPT prediction.

\subsubsection{No-wiggle--no-wiggle contribution}

We first consider the $\text{nw--nw}$ contribution, which is free from any BAO features. For this term, we take the IR limit of the displacement-mapping exponent, under which all higher-order contributions beyond one loop cancel as a consequence of the IR cancellation mechanism discussed in Sec.~\ref{sec:ULPTbispec_IRcancel}.

As a result, the $\text{nw--nw}$ component reduces to the standard SPT bispectrum evaluated with the no-wiggle linear power spectrum,
\begin{equation}
    B_{\text{nw--nw}}(\boldsymbol{k}_1,\boldsymbol{k}_2)
    \xrightarrow[\text{IR}]{}
    B^{(\text{tree+1-loop})}_{\rm SPT,\,nw}(\boldsymbol{k}_1,\boldsymbol{k}_2),
\end{equation}
where $B_{\rm SPT,\,nw}$ denotes the SPT bispectrum up to one-loop order computed using the no-wiggle linear power spectrum as its input.

\subsubsection{Wiggle--no-wiggle and no-wiggle--wiggle contributions}

We next consider the mixed contributions, $\text{w--nw}$ and $\text{nw--w}$, in which the source correlation function exhibits a BAO feature with respect to only one of the two separations, $r_1$ or $r_2$, while remaining smooth with respect to the other.

For the $\text{w--nw}$ contribution, the source correlation function $\zeta_{\text{J, w--nw}}(\boldsymbol{r}_1,\boldsymbol{r}_2)$ develops a localized peak around $r_1 \simeq r_{\rm BAO}$ and is negligible elsewhere. This sharply localized behavior allows us to approximate the $r_1$ dependence as delta-function--like, and accordingly we fix the displacement-mapping factor at $r_1 = r_{\rm BAO}$. Following the standard approximation commonly adopted in power-spectrum resummation, we also align the direction of the BAO separation with the corresponding wavevector, setting $\hat{r}_1 = \hat{k}_1$.

For the remaining $r_2$ dependence, which does not contain any BAO feature, we take the IR limit $\boldsymbol{r}_2 \to \boldsymbol{0}$, so that all long-wavelength contributions cancel, as discussed in Sec.~\ref{sec:ULPTbispec_IRcancel}. Under these assumptions, the displacement correlation tensor defined in Eq.~\eqref{eq:Iij_def_2} yields
\begin{align}
    & k_1^i k_3^j I_{ij}(\boldsymbol{r}_1;\, r_1 = r_{\rm BAO},\, \hat{r}_1 = \hat{k}_1)
    \nonumber \\
    &\quad = (\boldsymbol{k}_1 \cdot \boldsymbol{k}_3)\,
    \big[\sigma_0^2(r_{\rm BAO}) + 2\sigma_2^2(r_{\rm BAO})\big], \\
    & k_2^i k_3^j I_{ij}(\boldsymbol{r}_2 \to \boldsymbol{0})
    = (\boldsymbol{k}_2 \cdot \boldsymbol{k}_3)\,\bar{\sigma}^2, \\
    & k_1^i k_2^j I_{ij}(\boldsymbol{r}_1 - \boldsymbol{r}_2;\,
    r_1 = r_{\rm BAO},\, \hat{r}_1 = \hat{k}_1,\, \boldsymbol{r}_2 \to \boldsymbol{0})
    \nonumber \\
    &\quad = (\boldsymbol{k}_1 \cdot \boldsymbol{k}_2)\,
    \big[\sigma_0^2(r_{\rm BAO}) + 2\sigma_2^2(r_{\rm BAO})\big].
\end{align}

Substituting these results into the linear displacement-mapping exponent $\Gamma^{\rm (lin)}$ defined in Eq.~\eqref{eq:Gamma_lin}, we obtain
\begin{align}
    & -\overline{\Gamma}^{\rm (lin)}(\boldsymbol{k}_1,\boldsymbol{k}_2)
    + \Gamma^{\rm (lin)}(\boldsymbol{k}_1,\boldsymbol{k}_2;\,
    r_1 = r_{\rm BAO},\, \hat{r}_1 = \hat{k}_1,\, \boldsymbol{r}_2 \to \boldsymbol{0})
    \nonumber \\
    & = -k_1^2 \sigma_{\rm BAO}^2,
\end{align}
where $\sigma_{\rm BAO}^2$ is given by Eq.~\eqref{eq:sigma_BAO_ver2}.

The $\text{nw--w}$ contribution can be treated in a completely analogous manner by exchanging $r_1$ and $r_2$. Specifically, we fix $r_2 = r_{\rm BAO}$ and $\hat{r}_2 = \hat{k}_2$, while taking the IR limit $\boldsymbol{r}_1 \to \boldsymbol{0}$. This leads to
\begin{align}
    & -\overline{\Gamma}^{\rm (lin)}(\boldsymbol{k}_1,\boldsymbol{k}_2)
    + \Gamma^{\rm (lin)}(\boldsymbol{k}_1,\boldsymbol{k}_2;\,
    \boldsymbol{r}_1 \to \boldsymbol{0},\, r_2 = r_{\rm BAO},\, \hat{r}_2 = \hat{k}_2)
    \nonumber \\
    & = -k_2^2 \sigma_{\rm BAO}^2.
\end{align}

In both cases, the scale dependence of the displacement-mapping factor becomes fixed and can be pulled out of the convolution integral in the ULPT bispectrum expression. Consequently, each mixed contribution reduces to a simple exponential damping factor multiplied by the corresponding source bispectrum,
\begin{align}
    B_{\text{w--nw}}(\boldsymbol{k}_1,\boldsymbol{k}_2)
    &= e^{-k_1^2 \sigma_{\rm BAO}^2}\,
    B_{\text{J, w--nw}}^{\rm (tree+1\mbox{-}loop)}(\boldsymbol{k}_1,\boldsymbol{k}_2), \\
    B_{\text{nw--w}}(\boldsymbol{k}_1,\boldsymbol{k}_2)
    &= e^{-k_2^2 \sigma_{\rm BAO}^2}\,
    B_{\text{J, nw--w}}^{\rm (tree+1\mbox{-}loop)}(\boldsymbol{k}_1,\boldsymbol{k}_2).
\end{align}

\subsubsection{Wiggle--wiggle contribution}

We now turn to the $\text{w--w}$ contribution, for which the source correlation function exhibits BAO features with respect to both separations, $r_1$ and $r_2$. In this case, the scale dependence of the displacement-mapping factor is fixed at $r_1 = r_2 = r_{\rm BAO}$. We also fix the angular dependence by aligning the BAO separation vectors with the corresponding wavevectors, namely $\hat{r}_1 = \hat{k}_1$ and $\hat{r}_2 = \hat{k}_2$.

Under these assumptions, the displacement correlation tensor yields
\begin{align}
    & k_1^i k_3^j I_{ij}(\boldsymbol{r}_1;\, r_1 = r_{\rm BAO},\, \hat{r}_1 = \hat{k}_1)
    \nonumber \\
    &\quad = (\boldsymbol{k}_1 \cdot \boldsymbol{k}_3)\!
    \left[\sigma_0^2(r_{\rm BAO}) + 2\sigma_2^2(r_{\rm BAO})\right], \\
    & k_2^i k_3^j I_{ij}(\boldsymbol{r}_2;\, r_2 = r_{\rm BAO},\, \hat{r}_2 = \hat{k}_2)
    \nonumber \\
    &\quad = (\boldsymbol{k}_2 \cdot \boldsymbol{k}_3)\!
    \left[\sigma_0^2(r_{\rm BAO}) + 2\sigma_2^2(r_{\rm BAO})\right].
\end{align}

The mixed tensor $I_{ij}(\boldsymbol{r}_1 - \boldsymbol{r}_2)$ requires additional care. Assuming $r_1 = r_2 = r_{\rm BAO}$ together with $\hat{r}_1 \cdot \hat{k}_1 = \hat{r}_2 \cdot \hat{k}_2 = 1$, we obtain
\begin{align}
    & k_1^i k_2^j I_{ij}(\boldsymbol{r}_1 - \boldsymbol{r}_2)
    \nonumber \\
    &=
    (\boldsymbol{k}_1 \cdot \boldsymbol{k}_2)\!
    \left[\sigma_0^2\!\big(r_{\rm BAO}|\hat{r}_1 - \hat{r}_2|\big)
    + 2\sigma_2^2\!\big(r_{\rm BAO}|\hat{r}_1 - \hat{r}_2|\big)\right]
    \nonumber \\
    &\quad - \frac{3}{2}
    \left[k_1 k_2 + (\boldsymbol{k}_1 \cdot \boldsymbol{k}_2)\right]
    \sigma_2^2\!\big(r_{\rm BAO}|\hat{r}_1 - \hat{r}_2|\big).
\end{align}
Here we have used the approximation
\begin{equation}
    k_1^i k_2^j \,\hat{r}_{12,i}\,\hat{r}_{12,j}
    \approx -\frac{1}{2}
    \left[k_1 k_2 - (\boldsymbol{k}_1 \cdot \boldsymbol{k}_2)\right],
\end{equation}
together with
\begin{equation}
    \hat{r}_{12}
    \equiv \frac{\boldsymbol{r}_1 - \boldsymbol{r}_2}{|\boldsymbol{r}_1 - \boldsymbol{r}_2|}
    \approx
    \frac{\hat{k}_1 - \hat{k}_2}
         {|\hat{k}_1 - \hat{k}_2|}.
\end{equation}

To fix the residual angular dependence through $|\hat{r}_1 - \hat{r}_2|$, we further take an angular average over $\mu = \hat{r}_1 \cdot \hat{r}_2$, yielding the estimate
\begin{equation}
    |\hat{r}_1 - \hat{r}_2|
    \approx \frac{1}{2}\int_{-1}^{1} d\mu\, |\hat{r}_1 - \hat{r}_2|
    = \frac{4}{3}.
\end{equation}

Substituting the above results into Eq.~\eqref{eq:Gamma_lin}, we obtain the net linear displacement-mapping exponent,
\begin{align}
    & - \overline{\Gamma}^{\rm (lin)}(\boldsymbol{k}_1, \boldsymbol{k}_2)
    + \Gamma^{\rm (lin)}(\boldsymbol{k}_1, \boldsymbol{k}_2;\, \boldsymbol{r}_1, \boldsymbol{r}_2)
    \nonumber \\
    &= - \frac{1}{2}\left( k_1^2 + k_2^2 + k_3^2 \right)\, \sigma_{\rm BAO}^2
       - (\boldsymbol{k}_1 \cdot \boldsymbol{k}_2)\, \Delta\sigma_{\rm BAO}^2
       \nonumber \\
    &\quad + \frac{3}{2}
    \left[k_1 k_2 + (\boldsymbol{k}_1 \cdot \boldsymbol{k}_2)\right]
    \sigma_2^2\!\left(r'_{\rm BAO}\right),
\end{align}
where $r'_{\rm BAO} = (4/3)\,r_{\rm BAO}$, and the BAO damping parameter specific to the $\text{w--w}$ contribution is defined as
\begin{align}
    \Delta\sigma_{\rm BAO}^2
    &= \sigma_0^2\!\left(r'_{\rm BAO}\right) - \sigma_0^2(r_{\rm BAO})
    \nonumber \\
    &\quad + 2
    \left[\sigma_2^2\!\left(r'_{\rm BAO}\right) - \sigma_2^2(r_{\rm BAO})\right].
\end{align}

Finally, the $\text{w--w}$ contribution to the ULPT bispectrum can be written as
\begin{align}
    B_{\text{w--w}}(\boldsymbol{k}_1,\boldsymbol{k}_2)
    &= e^{-\tfrac{1}{2}(k_1^2 + k_2^2 + k_3^2)\sigma_{\rm BAO}^2}\,
    \mathcal{D}_{\text{w--w}}(\boldsymbol{k}_1,\boldsymbol{k}_2)
    \nonumber \\
    &\quad \times
    B_{\text{J, w--w}}^{\text{(tree+1\mbox{-}loop)}}(\boldsymbol{k}_1,\boldsymbol{k}_2),
\end{align}
where the damping function specific to the $\text{w--w}$ contribution is given by
\begin{equation}
    \mathcal{D}_{\text{w--w}}(\boldsymbol{k}_1,\boldsymbol{k}_2)
    =
    e^{- (\boldsymbol{k}_1 \cdot \boldsymbol{k}_2)\, \Delta\sigma_{\rm BAO}^2
    + \frac{3}{2}\left(k_1 k_2 + \boldsymbol{k}_1 \cdot \boldsymbol{k}_2\right)
    \sigma_2^2\!\left(r'_{\rm BAO}\right)}.
    \label{eq:Dww}
\end{equation}

\subsubsection{Decomposition of the source bispectrum into wiggle and no-wiggle components}
\label{sec:IR_source}

To complete the construction of the IR-resummed bispectrum model within the ULPT framework, the remaining step is to specify the source bispectrum. As shown in Eq.~\eqref{eq:B_SPT_J_tree}, the tree-level source bispectrum coincides with the SPT bispectrum. By applying the wiggle--no-wiggle decomposition $P_{\rm lin} = P_{\rm w} + P_{\rm nw}$ to each linear power spectrum appearing in the tree-level expression, we obtain
\begin{align}
    B^{(\rm tree)}_{{\rm J},\,\text{w--w}}(\boldsymbol{k}_1,\boldsymbol{k}_2)
    &= 2\, F_2(\boldsymbol{k}_1,\boldsymbol{k}_2)\,P_{\rm w}(k_1)\,P_{\rm w}(k_2), \nonumber \\
    B^{(\rm tree)}_{{\rm J},\,\text{w--nw}}(\boldsymbol{k}_1,\boldsymbol{k}_2)
    &= 2\, F_2(\boldsymbol{k}_1,\boldsymbol{k}_2)\,P_{\rm w}(k_1)\,P_{\rm nw}(k_2), \nonumber \\
    B^{(\rm tree)}_{{\rm J},\,\text{nw--w}}(\boldsymbol{k}_1,\boldsymbol{k}_2)
    &= 2\, F_2(\boldsymbol{k}_1,\boldsymbol{k}_2)\,P_{\rm nw}(k_1)\,P_{\rm w}(k_2), \nonumber \\
    B^{(\rm tree)}_{{\rm J},\,\text{nw--nw}}(\boldsymbol{k}_1,\boldsymbol{k}_2)
    &= 2\, F_2(\boldsymbol{k}_1,\boldsymbol{k}_2)\,P_{\rm nw}(k_1)\,P_{\rm nw}(k_2).
\end{align}

To evaluate the one-loop corrections to the source bispectrum, we adopt the simplifying assumption that BAO features arising inside the mode-coupling integrals of the source bispectrum are subdominant and can therefore be neglected. It is important to emphasize that, while we retain the BAO signal generated through the mode couplings induced by the displacement-mapping factor, we neglect BAO oscillations that reside within the mode-coupling integrals of the source bispectrum itself. For the power spectrum, an analogous approximation has been thoroughly validated in Ref.~\cite{Sugiyama:2025myq}, and we assume that the same reasoning applies to the bispectrum as well. Consequently, all mode-coupling terms in the source bispectrum are evaluated using the no-wiggle linear power spectrum.

Under this assumption, the BAO dependence of the source bispectrum is entirely encoded in the terms that depend explicitly on the linear power spectrum, i.e., in the factors that lie outside the mode-coupling integrals. Substituting $P_{\rm lin} = P_{\rm w} + P_{\rm nw}$ into Eq.~\eqref{eq:B_J_1loop}, the one-loop source bispectrum can be decomposed as
\begin{align}
    B^{(1\text{-loop})}_{{\rm J},\,\text{w--w}}(\boldsymbol{k}_1,\boldsymbol{k}_2)
    &= B^{(1\text{-loop})}_{{\rm J},\,{\rm GG},\,{\rm nw}}(\boldsymbol{k}_1,\boldsymbol{k}_2)\,
       P_{\rm w}(k_1)\,P_{\rm w}(k_2), \nonumber \\
    B^{(1\text{-loop})}_{{\rm J},\,\text{w--nw}}(\boldsymbol{k}_1,\boldsymbol{k}_2)
    &= \left[
       B^{(1\text{-loop})}_{{\rm J},\,{\rm GG},\,{\rm nw}}(\boldsymbol{k}_1,\boldsymbol{k}_2)
       + \frac{B^{(1\text{-loop})}_{{\rm J},\,{\rm GM},\,{\rm nw}}(\boldsymbol{k}_1,\boldsymbol{k}_2)}{P_{\rm nw}(k_2)}
       \right] \nonumber\\
    &\quad \times
       P_{\rm w}(k_1)\,P_{\rm nw}(k_2), \nonumber \\
    B^{(1\text{-loop})}_{{\rm J},\,\text{nw--w}}(\boldsymbol{k}_1,\boldsymbol{k}_2)
    &= \left[
       B^{(1\text{-loop})}_{{\rm J},\,{\rm GG},\,{\rm nw}}(\boldsymbol{k}_1,\boldsymbol{k}_2)
       + \frac{B^{(1\text{-loop})}_{{\rm J},\,{\rm GM},\,{\rm nw}}(\boldsymbol{k}_1,\boldsymbol{k}_2)}{P_{\rm nw}(k_1)}
       \right] \nonumber \\
    &\quad \times
       P_{\rm nw}(k_1)\,P_{\rm w}(k_2).
    \label{eq:B_J_1loop_decomp}
\end{align}
Here, $B^{(1\text{-loop})}_{{\rm J},\,{\rm GG},\,{\rm nw}}$, $B^{(1\text{-loop})}_{{\rm J},\,{\rm GM},\,{\rm nw}}$, and $B^{(1\text{-loop})}_{{\rm J},\,{\rm MG},\,{\rm nw}}$ denote the one-loop components defined in Eq.~\eqref{eq:B_J_X_1loop}, evaluated with the linear power spectrum $P_{\rm lin}$ replaced by its no-wiggle counterpart $P_{\rm nw}$.

\subsubsection{Final expression for the IR-resummed bispectrum in ULPT}

Combining the results obtained in Secs.~\ref{sec:decom}--\ref{sec:IR_source}, the IR-resummed bispectrum at tree level in ULPT takes the form
\begin{align}
    & B_{\rm IR\mbox{-}resum}(\boldsymbol{k}_1,\boldsymbol{k}_2)
    \nonumber \\
    &= 2\,F_2(\boldsymbol{k}_1,\boldsymbol{k}_2)
    \Big[
        P_{\rm nw}(k_1)P_{\rm nw}(k_2)
        \nonumber \\
        & \quad + e^{-k_1^2\sigma_{\rm BAO}^2}P_{\rm w}(k_1)P_{\rm nw}(k_2)
        + e^{-k_2^2\sigma_{\rm BAO}^2}P_{\rm nw}(k_1)P_{\rm w}(k_2)
        \nonumber \\
        & \quad + e^{-\tfrac{1}{2}(k_1^2+k_2^2+k_3^2)\sigma_{\rm BAO}^2}\,
        {\cal D}_{\text{w--w}}(\boldsymbol{k}_1,\boldsymbol{k}_2)\,
        P_{\rm w}(k_1)P_{\rm w}(k_2)
    \Big],
    \label{eq:IR_resum_ULPT}
\end{align}
which can be directly compared with the previously proposed IR-resummed models, Eqs.~\eqref{eq:bispec_IR_resum_2} and \eqref{eq:bispec_IR_resum_nao}.

At ${\cal O}(P_{\rm w})$, the ULPT expression coincides with Eq.~\eqref{eq:bispec_IR_resum_2}, and the exponential suppression of the BAO feature is governed solely by the parameter $\sigma_{\rm BAO}^2$. This result is also fully consistent with the IR-resummed bispectrum obtained within the TSPT framework. It therefore demonstrates that ULPT correctly reproduces the standard leading-order BAO damping, including the wiggle contribution embedded in the mode-coupling structure.

At ${\cal O}(P_{\rm w}^2)$, however, the ULPT prediction exhibits a qualitatively new exponential structure,
\[
e^{-\tfrac{1}{2}(k_1^2+k_2^2+k_3^2)\sigma_{\rm BAO}^2}\,
{\cal D}_{\text{w--w}}(\boldsymbol{k}_1,\boldsymbol{k}_2),
\]
which represents a key difference from the existing IR-resummed models. The first factor describes the conventional BAO damping induced by large-scale bulk flows and is governed solely by the universal parameter $\sigma_{\rm BAO}^2$, as in the standard IR-resummation framework. In contrast, the additional exponential factor ${\cal D}_{\text{w--w}}$ is unique to the $\text{w--w}$ contribution and encodes a genuinely three-point configuration-dependent, nonperturbative nonlinear modulation of the BAO signal that has no counterpart in previously proposed models.

This additional factor reflects the nontrivial correlation of the relative displacements associated with the two BAO separations and depends explicitly on the mutual orientation of $\boldsymbol{k}_1$ and $\boldsymbol{k}_2$. The emergence of ${\cal D}_{\text{w--w}}$ therefore constitutes one of the central new predictions of ULPT for the bispectrum and directly demonstrates how the unified Lagrangian structure fixes the IR behavior of higher-order clustering statistics beyond the power spectrum.

For a fiducial $\Lambda$CDM cosmology consistent with Planck 2015 at $z=0$, we find the numerical values
\begin{align}
    \bar{\sigma}^2 & = 35.774\,(h^{-1}\,{\rm Mpc})^2, \nonumber \\
    \sigma_{\rm BAO}^2 & = 36.303\,(h^{-1}\,{\rm Mpc})^2, \nonumber \\
    \Delta \sigma_{\rm BAO}^2 & = -1.042\,(h^{-1}\,{\rm Mpc})^2, \nonumber \\
    \sigma_2^2(r'_{\rm BAO}) & = -2.324\,(h^{-1}\,{\rm Mpc})^2,
\end{align}
where $r'_{\rm BAO} = \tfrac{4}{3}\,r_{\rm BAO}$. The difference between $\bar{\sigma}^2$ and $\sigma_{\rm BAO}^2$ is only $1.45\%$, while $\Delta\sigma_{\rm BAO}^2$ and $\sigma_2^2(r'_{\rm BAO})$ amount to merely $2.9\%$ and $6.5\%$ of $\bar{\sigma}^2$, respectively. Hence, for $\Lambda$CDM-like cosmologies, the approximations
\begin{align}
    \bar{\sigma}^2 & \approx \sigma_{\rm BAO}^2, \nonumber \\
    \Delta\sigma_{\rm BAO}^2 & \approx \sigma_2^2(r'_{\rm BAO}) \approx 0
\end{align}
hold with excellent accuracy.

Under these approximations, the ULPT-based IR-resummed model in Eq.~\eqref{eq:IR_resum_ULPT} reduces to the previously proposed expression Eq.~\eqref{eq:bispec_IR_resum_nao}, which corresponds to the case in which all BAO contributions inside the mode-coupling integrals are neglected. This demonstrates that, as in the power-spectrum case, BAO features arising from the mode-coupling integrals of the bispectrum are indeed subdominant for $\Lambda$CDM-like cosmologies.

Including the one-loop correction, the full IR-resummed bispectrum in ULPT is given by
\begin{widetext}
\begin{align}
    B_{\rm IR\mbox{-}resum}(\boldsymbol{k}_1,\boldsymbol{k}_2)
    &= e^{-\tfrac{1}{2}(k_1^2 + k_2^2+k_3^2)\sigma_{\rm BAO}^2}\,
    {\cal D}_{\text{w--w}}(\boldsymbol{k}_1,\boldsymbol{k}_2)\,
        B_{\rm J,GG,nw}^{\rm (tree + 1\text{-loop})}(\boldsymbol{k}_1,\boldsymbol{k}_2)\,
        P_{\rm w}(k_1)\, P_{\rm w}(k_2)
    \nonumber \\
    &\quad + e^{-k_1^2\sigma_{\rm BAO}^2}
    \left[
    B_{\rm J,GG,nw}^{\rm (tree + 1\text{-loop})}(\boldsymbol{k}_1,\boldsymbol{k}_2)
    + \frac{B_{\rm J,GM,nw}^{\rm (1\text{-loop})}(\boldsymbol{k}_1,\boldsymbol{k}_2)}{P_{\rm nw}(k_2)}
    \right]
    P_{\rm w}(k_1)\, P_{\rm nw}(k_2)
    \nonumber \\
    &\quad + e^{-k_2^2\sigma_{\rm BAO}^2}
    \left[
    B_{\rm J,GG,nw}^{\rm (tree + 1\text{-loop})}(\boldsymbol{k}_1,\boldsymbol{k}_2)
    + \frac{B_{\rm J,MG,nw}^{\rm (1\text{-loop})}(\boldsymbol{k}_1,\boldsymbol{k}_2)}{P_{\rm nw}(k_1)}
    \right]
    P_{\rm nw}(k_1)\, P_{\rm w}(k_2)
    \nonumber \\
    &\quad + B^{\rm (tree + 1\text{-loop})}_{\rm SPT,nw}(\boldsymbol{k}_1,\boldsymbol{k}_2).
\end{align}
\end{widetext}

\subsection{Post-reconstruction IR-resummed bispectrum in ULPT}
\label{sec:ULPT_bispec_IR_recon}

Within the ULPT framework, the mathematical structure of the density fluctuations remains unchanged before and after reconstruction. As shown in Eq.~\eqref{eq:delta_rec}, the reconstruction procedure modifies only the displacement field, while the Jacobian deviation remains invariant. Consequently, the IR-resummed bispectrum model after reconstruction preserves the same analytic structure as in the pre-reconstruction case, with appropriate modifications applied solely to the displacement field and the nonlinear kernel functions.

The source bispectrum after reconstruction is evaluated using the nonlinear kernel defined for the reconstructed density field. At tree level, the second-order kernel $F_2(\boldsymbol{k}_1,\boldsymbol{k}_2)$ is replaced by its reconstructed counterpart $F_{{\rm rec},2}(\boldsymbol{k}_1,\boldsymbol{k}_2)$, given in Eq.~\eqref{eq:F2_rec}.

The displacement-mapping factor must also be evaluated using the reconstructed displacement field defined in Eq.~\eqref{eq:Psi_rec}. We define the post-reconstruction linear-order displacement correlation tensor as
\begin{align}
    I_{{\rm rec},ij}(\boldsymbol{r})
    = \left\langle \Psi^{(1)}_{{\rm rec},i}(\boldsymbol{q}) \,
      \Psi^{(1)}_{{\rm rec},j}(\boldsymbol{q}') \right\rangle_{\rm c},
    \label{eq:Iij_def_rec}
\end{align}
where $\boldsymbol{r} = \boldsymbol{q} - \boldsymbol{q}'$, and the post-reconstruction linear displacement vector is given by
\begin{equation}
    \boldsymbol{\Psi}^{(1)}_{\rm rec}(\boldsymbol{q})
    = i \int \frac{d^3k}{(2\pi)^3} e^{i\boldsymbol{k}\cdot\boldsymbol{q}}
      \frac{\boldsymbol{k}}{k^2} \left[ 1 - W_{\rm G}(kR) \right]
      \tilde{\delta}^{(1)}(\boldsymbol{k}).
\end{equation}

The reconstructed displacement correlation tensor can be decomposed as
\begin{equation}
    I_{{\rm rec},ij}(\boldsymbol{r})
    = \delta_{ij}\,\sigma_{{\rm rec},0}^2(r)
      + 2\left(\frac{3\hat{r}_i \hat{r}_j - \delta_{ij}}{2}\right)
        \sigma_{{\rm rec},2}^2(r),
\end{equation}
with
\begin{equation}
    \sigma_{{\rm rec},\ell}^2(r)
    = \frac{1}{3} i^{\ell}
      \int \frac{dp}{2\pi^2} j_{\ell}(pr)
      \left[ 1 - W_{\rm G}(pR) \right]^2 P_{\rm lin}(p).
\end{equation}
In the limit $r \to 0$, this reduces to
\begin{equation}
    \bar{\sigma}_{\rm rec}^2
    = \sigma_{{\rm rec},0}^2(0)
    = \frac{1}{3} \int \frac{dp}{2\pi^2}
      \left[ 1 - W_{\rm G}(pR) \right]^2 P_{\rm lin}(p).
\end{equation}

Using $I_{{\rm rec},ij}$, the linear contributions to the displacement-mapping exponent are given by
\begin{align}
    \overline{\Gamma}_{\rm rec}^{\rm (lin)}(\boldsymbol{k}_1,\boldsymbol{k}_2)
    & = \frac{1}{2}\left( k_1^2 + k_2^2 + k_3^2 \right)\bar{\sigma}_{\rm rec}^2 , \nonumber \\
    \Gamma_{\rm rec}^{\rm (lin)}(\boldsymbol{k}_1,\boldsymbol{k}_2,\boldsymbol{r}_1,\boldsymbol{r}_2)
    & = - k_1^i k_3^j I_{{\rm rec}, ij}(\boldsymbol{r}_1)
        - k_2^i k_3^j I_{{\rm rec}, ij}(\boldsymbol{r}_2) \nonumber \\
        &\quad - k_1^i k_2^j I_{{\rm rec}, ij}(\boldsymbol{r}_{1}-\boldsymbol{r}_2).
    \label{eq:Gamma_rec_lin}
\end{align}

Accordingly, the BAO damping parameter after reconstruction is given by
\begin{align}
    \sigma_{\rm BAO,rec}^2
    & = \bar{\sigma}_{\rm rec}^2
      - \sigma_{{\rm rec},0}^2(r_{\rm BAO})
      - 2\sigma_{{\rm rec},2}^2(r_{\rm BAO}),
    \label{eq:BAO_rec_param}
\end{align}
which is consistent with the definition in Eq.~\eqref{eq:sigma_BAO_recon}. The exponential function specific to the $\text{w--w}$ contribution is then given by
\begin{equation}
    \mathcal{D}_{\text{rec,\,w--w}}(\boldsymbol{k}_1,\boldsymbol{k}_2)
    = e^{
        -(\boldsymbol{k}_1 \cdot \boldsymbol{k}_2)\,
        \Delta\sigma_{\rm BAO,rec}^2
        + \frac{3}{2}\left(k_1 k_2 + \boldsymbol{k}_1 \cdot \boldsymbol{k}_2\right)
        \sigma_{{\rm rec},2}^2(r'_{\rm BAO})},
    \label{eq:Dww_rec}
\end{equation}
where
\begin{align}
    \Delta\sigma_{\rm BAO,rec}^2
    &= - \left[\sigma_{{\rm rec},0}^2(r_{\rm BAO})
       + 2\sigma_{{\rm rec},2}^2(r_{\rm BAO})\right] \nonumber \\
    &\quad + \sigma_{{\rm rec},0}^2(r'_{\rm BAO})
       + 2\sigma_{{\rm rec},2}^2(r'_{\rm BAO}),
\end{align}
with $r'_{\rm BAO} = \tfrac{4}{3}r_{\rm BAO}$.

With these definitions, the full post-reconstruction IR-resummed bispectrum at tree level is given by
\begin{align}
    & B_{\rm IR\mbox{-}resum,\,rec}(\boldsymbol{k}_1,\boldsymbol{k}_2) \nonumber \\
    &= 2\,F_{{\rm rec},2}(\boldsymbol{k}_1,\boldsymbol{k}_2)
       \Big[P_{\rm nw}(k_1)P_{\rm nw}(k_2) \nonumber \\
    &\quad + e^{-k_1^2\sigma_{\rm BAO,rec}^2}
       P_{\rm w}(k_1)P_{\rm nw}(k_2)
       + e^{-k_2^2\sigma_{\rm BAO,rec}^2}
       P_{\rm nw}(k_1)P_{\rm w}(k_2) \nonumber \\
    &\quad + e^{-\tfrac{1}{2}(k_1^2+k_2^2+k_3^2)
        \sigma_{\rm BAO,rec}^2}
       \mathcal{D}_{\text{rec,\,w--w}}(\boldsymbol{k}_1,\boldsymbol{k}_2)
       P_{\rm w}(k_1)P_{\rm w}(k_2)\Big].
    \label{eq:IR_resum_ULPT_rec}
\end{align}

For the fiducial $\Lambda$CDM cosmology at redshift $z=0$ and smoothing scale $R=15\,h^{-1}{\rm Mpc}$, the damping parameters take the numerical values
\begin{align}
    \bar{\sigma}_{\rm rec}^2 &= 12.471\,(h^{-1}\,{\rm Mpc})^2, \nonumber \\
    \sigma_{\rm BAO,rec}^2 &= 12.448\,(h^{-1}\,{\rm Mpc})^2, \nonumber \\
    \Delta\sigma_{\rm BAO,rec}^2 &= -0.036\,(h^{-1}\,{\rm Mpc})^2, \nonumber \\
    \sigma_{{\rm rec},2}^2(r'_{\rm BAO}) &= -0.005\,(h^{-1}\,{\rm Mpc})^2.
    \label{eq:sigma_rec_values}
\end{align}
The relative difference between $\bar{\sigma}_{\rm rec}^2$ and $\sigma_{\rm BAO,rec}^2$ is only $0.184\%$, while $\Delta\sigma_{\rm BAO,rec}^2$ and $\sigma_{{\rm rec},2}^2(r'_{\rm BAO})$ amount to merely $0.289\%$ and $0.04\%$ of $\bar{\sigma}_{\rm rec}^2$, respectively. These results demonstrate that the approximation
\begin{align}
    \bar{\sigma}_{\rm rec}^2 &\approx \sigma_{\rm BAO,rec}^2, \nonumber \\
    \Delta\sigma_{\rm BAO,rec}^2 &\approx
      \sigma_{{\rm rec},2}^2(r'_{\rm BAO}) \approx 0
\end{align}
holds with significantly higher accuracy after reconstruction than before. This validates the approximation of neglecting BAO-induced features in the mode-coupling integrals, which underlies the simplified IR-resummed bispectrum model given in Eq.~\eqref{eq:bispec_IR_resum_nao_rec} proposed in Ref.~\cite{Sugiyama:2024qsw}.

\section{Cross bispectrum between pre- and post-reconstruction density fields}
\label{sec:cross_bispectrum}

In this section, we compute the cross bispectrum within the ULPT framework, focusing on correlations between density fluctuations evaluated before and after reconstruction. When different displacement fields are involved, nonperturbative IR effects no longer cancel and instead generate a universal exponential suppression factor in the cross bispectrum. We explicitly demonstrate how this damping factor arises from the displacement-mapping sector and construct the corresponding IR-resummed model, which accurately captures the nonlinear suppression of the BAO signal in the cross correlations.

\subsection{General structure}

The cross bispectrum involving different density fields is defined as
\begin{equation}
    \langle \tilde{\delta}_{\rm a}(\boldsymbol{k}_1)
    \tilde{\delta}_{\rm b}(\boldsymbol{k}_2)
    \tilde{\delta}_{\rm c}(\boldsymbol{k}_3)  \rangle
    = (2\pi)^3\delta_{\rm D}(\boldsymbol{k}_{123})\,
    B_{\rm (a,b,c)}(\boldsymbol{k}_1,\boldsymbol{k}_2,\boldsymbol{k}_3),
\end{equation}
where $a$, $b$, and $c$ denote either \emph{pre} or \emph{post}, corresponding to the density fields before and after reconstruction. For example, the pre-reconstruction auto bispectrum is denoted by $B_{\rm (pre,pre,pre)}$, while the post-reconstruction auto bispectrum is $B_{\rm (post,post,post)}$.

Within the ULPT framework, the general expression for the cross bispectrum can be written as
\begin{align}
    & B_{\alpha\beta, {\rm (a,b,c)}}(\boldsymbol{k}_{\alpha},\boldsymbol{k}_{\beta})
    \nonumber \\
    &=
    e^{-\overline{\Gamma}_{\alpha\beta, {\rm (a,b,c)}}(\boldsymbol{k}_{\alpha},\boldsymbol{k}_{\beta})}
    \int d^3r_{\alpha}\, e^{-i \boldsymbol{k}_{\alpha} \cdot \boldsymbol{r}_{\alpha}}
    \int d^3r_{\beta}\, e^{-i \boldsymbol{k}_{\beta} \cdot \boldsymbol{r}_{\beta}} \nonumber \\
    &\quad \times
    e^{\Gamma_{\alpha\beta, {\rm (a,b,c)}}(\boldsymbol{k}_{\alpha},\boldsymbol{k}_{\beta},\boldsymbol{r}_{\alpha},\boldsymbol{r}_{\beta})}
    \zeta_{{\rm J}_{\alpha\beta}, {\rm (a,b,c)}}(\boldsymbol{r}_{\alpha},\boldsymbol{r}_{\beta}),
    \label{eq:B_ULPT_general_ij_abc}
\end{align}
where $(\alpha,\beta)$ runs over the three cyclic combinations $(1,2)$, $(2,3)$, and $(3,1)$. The full cross bispectrum is obtained as the sum of these three contributions, as in Eq.~\eqref{eq:B12_23_31}. For the auto bispectrum, these terms are related by cyclic symmetry, so that it is sufficient to evaluate only the $(\alpha,\beta)=(1,2)$ contribution. In the cross case, however, this symmetry is broken, and all three terms must be evaluated explicitly.

The exponent of the displacement-mapping factor is given by
\begin{align}
    \sum_{m=2}^{\infty}\frac{1}{m!}\langle X_{\rm (a,b,c)}^m \rangle_{\rm c}
    &= - \overline{\Gamma}_{\alpha\beta, {\rm (a,b,c)}}(\boldsymbol{k}_{\alpha},\boldsymbol{k}_{\beta})
    \nonumber \\
    &\quad +
    \Gamma_{\alpha\beta, {\rm (a,b,c)}}(\boldsymbol{k}_{\alpha},\boldsymbol{k}_{\beta},
    \boldsymbol{r}_{\alpha},\boldsymbol{r}_{\beta}),
\end{align}
with
\begin{equation}
    X_{\rm (a,b,c)} = -i\boldsymbol{k}_1\cdot\boldsymbol{\Psi}_{\rm a}(\boldsymbol{q}_1)
                 -i\boldsymbol{k}_2\cdot\boldsymbol{\Psi}_{\rm b}(\boldsymbol{q}_2)
                 -i\boldsymbol{k}_3\cdot\boldsymbol{\Psi}_{\rm c}(\boldsymbol{q}_3),
\end{equation}
where $\boldsymbol{\Psi}_{\rm a}$, $\boldsymbol{\Psi}_{\rm b}$, and $\boldsymbol{\Psi}_{\rm c}$ denote the displacement fields associated with the density fields $\delta_{\rm a}$, $\delta_{\rm b}$, and $\delta_{\rm c}$, respectively.

In contrast to the auto-bispectrum case, the displacement-mapping contribution does \emph{not} cancel in the IR limit for the cross bispectrum,
\begin{equation}
    \overline{\Gamma}_{\alpha\beta, {\rm (a,b,c)}}(\boldsymbol{k}_{\alpha},\boldsymbol{k}_{\beta})
    \neq
    \Gamma_{\alpha\beta, {\rm (a,b,c)}}(\boldsymbol{k}_{\alpha},\boldsymbol{k}_{\beta},
    \boldsymbol{r}_{\alpha}=\boldsymbol{0},\boldsymbol{r}_{\beta}=\boldsymbol{0}),
\end{equation}
in sharp contrast with the auto-bispectrum relation in Eq.~\eqref{eq:Gamma_relation}. This inequality expresses the essential origin of the IR effect in the cross case: the long-wavelength displacement contributions no longer cancel but combine into a nonvanishing exponential factor. As a result, the cross bispectrum acquires a universal exponential suppression that multiplies its entire shape.

The cross source three-point correlation function and the corresponding source bispectrum are related by a double Fourier transform,
\begin{align}
    & \zeta_{{\rm J}_{\alpha\beta}, {\rm (a,b,c)}}(\boldsymbol{r}_{\alpha},\boldsymbol{r}_{\beta}) \nonumber \\
    & =
    \int \frac{d^3k_{\alpha}}{(2\pi)^3} e^{i\boldsymbol{k}_{\alpha}\cdot\boldsymbol{r}_{\alpha}}
    \int \frac{d^3k_{\beta}}{(2\pi)^3} e^{i\boldsymbol{k}_{\beta}\cdot\boldsymbol{r}_{\beta}}\,
    B_{{\rm J}_{\alpha\beta}, {\rm (a,b,c)}}(\boldsymbol{k}_{\alpha},\boldsymbol{k}_{\beta}).
\end{align}

In the following, we focus on two representative classes of cross bispectra:
(i) $(a,b,c)=({\rm pre,pre,post})$, discussed in Sec.~\ref{sec:preprepost}, which involves two pre-reconstruction fields and one post-reconstruction field; and
(ii) $(a,b,c)=({\rm pre,post,post})$, discussed in Sec.~\ref{sec:prepostpost}, which involves one pre-reconstruction field and two post-reconstruction fields.

\subsection{Displacement-mapping factor for the cross bispectrum}

To evaluate the displacement-mapping contribution to the cross bispectrum, we begin by defining the linear-order cross displacement correlation tensor,
\begin{equation}
    I_{{\rm cross},ij}(\boldsymbol{r})
    = \left\langle \Psi^{(1)}_{{\rm pre},i}(\boldsymbol{q})\,\Psi^{(1)}_{{\rm post},j}(\boldsymbol{q}') \right\rangle
    = \left\langle \Psi^{(1)}_{{\rm post},i}(\boldsymbol{q})\,\Psi^{(1)}_{{\rm pre},j}(\boldsymbol{q}') \right\rangle,
    \label{eq:I_cross_origin}
\end{equation}
where $\boldsymbol{r}=\boldsymbol{q}-\boldsymbol{q}'$. This tensor characterizes the statistical correlation between the pre- and post-reconstruction displacement fields. By statistical isotropy, it can be decomposed as
\begin{equation}
    I_{{\rm cross},ij}(\boldsymbol{r})
    = \delta_{ij}\,\sigma_{{\rm cross},0}^2(r)
    + \left(\frac{3\hat{r}_i \hat{r}_j - \delta_{ij}}{2}\right)
    2\sigma_{{\rm cross},2}^2(r),
    \label{eq:Iij_def_cross}
\end{equation}
where the scalar functions $\sigma_{{\rm cross},\ell}^2(r)$ are defined by
\begin{align}
    \sigma_{{\rm cross},\ell}^2(r)
    = \frac{1}{3}\, i^{\ell}
    \int \frac{dp}{2\pi^2}\, j_{\ell}(pr)\,
    \bigl[ 1 - W_{\rm G}(pR) \bigr]\,
    P_{\rm lin}(p),
    \label{eq:sigma_def_cross}
\end{align}
with $W_{\rm G}(pR)=\exp(-p^2R^2/2)$ being the Gaussian smoothing kernel used in the reconstruction procedure.

For later use, we also introduce the cross-correlation tensor between the reconstruction shift vector $\boldsymbol{t}(\boldsymbol{q})$ [defined in Eq.~\eqref{eq:t}] and the pre-reconstruction displacement field,
\begin{equation}
    I_{t\Psi,ij}(\boldsymbol{r})
    = \left\langle \Psi_{{\rm pre},i}(\boldsymbol{q})\,t_{j}(\boldsymbol{q}') \right\rangle
    = \left\langle t_{i}(\boldsymbol{q})\,\Psi_{{\rm pre},j}(\boldsymbol{q}') \right\rangle.
    \label{eq:Iij_tPsi}
\end{equation}
This tensor admits the same tensorial decomposition as Eq.~\eqref{eq:Iij_def_cross},
\begin{equation}
    I_{t\Psi,ij}(\boldsymbol{r})
    = \delta_{ij}\,\sigma_{t\Psi,0}^2(r)
    + \left(\frac{3\hat{r}_i \hat{r}_j - \delta_{ij}}{2}\right)
    2\sigma_{t\Psi,2}^2(r),
    \label{eq:Iij_def_tp}
\end{equation}
where
\begin{equation}
    \sigma_{t\Psi,\ell}^2(r)
    = \frac{1}{3}\, i^{\ell}
    \int \frac{dp}{2\pi^2}\, j_{\ell}(pr)\,
    \bigl[- W_{\rm G}(pR)\bigr]\,
    P_{\rm lin}(p),
\end{equation}
and the corresponding one-point variance is
\begin{equation}
    \bar{\sigma}_{t\Psi}^2
    = \frac{1}{3}
    \int \frac{dp}{2\pi^2}\,
    \bigl[- W_{\rm G}(pR)\bigr]\,
    P_{\rm lin}(p).
\end{equation}

For completeness, we also define the auto-correlation tensor of the reconstruction shift vector,
\begin{align}
    I_{tt,ij}(\boldsymbol{r})
    &= \left\langle t_{i}(\boldsymbol{q})\,t_{j}(\boldsymbol{q}') \right\rangle \nonumber \\
    &= \delta_{ij}\,\sigma_{tt,0}^2(r)
    + \left(\frac{3\hat{r}_i \hat{r}_j - \delta_{ij}}{2}\right)
    2\sigma_{tt,2}^2(r),
    \label{eq:Iij_tt}
\end{align}
with
\begin{align}
    \sigma_{tt,\ell}^2(r)
    &= \frac{1}{3}\, i^{\ell}
    \int \frac{dp}{2\pi^2}\, j_{\ell}(pr)\,
    \bigl[- W_{\rm G}(pR)\bigr]^2\,
    P_{\rm lin}(p), \nonumber \\
    \bar{\sigma}_{tt}^2
    &= \frac{1}{3}
    \int \frac{dp}{2\pi^2}\,
    \bigl[- W_{\rm G}(pR)\bigr]^2\,
    P_{\rm lin}(p).
    \label{eq:sigma_tt}
\end{align}

The displacement correlation tensors defined above
[Eqs.~\eqref{eq:I_cross_origin}, \eqref{eq:Iij_tPsi}, and \eqref{eq:Iij_tt}]
satisfy the following relations with the auto displacement correlation tensors for the pre- and post-reconstruction cases, defined in Eqs.~\eqref{eq:Iij_def_start} and \eqref{eq:Iij_def_rec}:
\begin{equation}
    \boldsymbol{I}_{\rm cross} = \boldsymbol{I} + \boldsymbol{I}_{t\Psi},
    \label{eq:I_cross_total}
\end{equation}
and
\begin{equation}
    \boldsymbol{I}_{\rm rec} = \boldsymbol{I} + 2\,\boldsymbol{I}_{t\Psi} + \boldsymbol{I}_{tt},
    \label{eq:I_rec_total}
\end{equation}
which immediately imply the corresponding relations for the multipole components,
\begin{align}
    \sigma^2_{{\rm cross},\ell}(r) &= \sigma^2_{\ell}(r)
    + \sigma^2_{t\Psi,\ell}(r), \nonumber \\
    \sigma^2_{{\rm rec},\ell}(r) &= \sigma^2_{\ell}(r)
    + 2\,\sigma^2_{t\Psi,\ell}(r)
    + \sigma^2_{tt,\ell}(r).
\end{align}

\subsection{IR-resummed model for the pre--pre--post configuration}
\label{sec:preprepost}

In this subsection, we consider the cross bispectrum $B_{\rm (pre,pre,post)}(\boldsymbol{k}_1,\boldsymbol{k}_2,\boldsymbol{k}_3)$, in which the density fluctuations associated with $\boldsymbol{k}_1$ and $\boldsymbol{k}_2$ are drawn from the pre-reconstruction field, while that associated with $\boldsymbol{k}_3$ is taken from the post-reconstruction field. We refer to this configuration as the \emph{pre--pre--post} case and derive the
corresponding IR-resummed model.

\subsubsection{Source bispectrum}

At tree level, the source bispectrum coincides with the standard SPT solution. Accordingly, the tree-level cross source bispectrum is given by
\begin{align}
    B^{(\rm tree)}_{{\rm J}_{12},{\rm (pre,pre,post)}}(\boldsymbol{k}_1,\boldsymbol{k}_2)
    &=
    2F_{{\rm rec},2}(\boldsymbol{k}_1,\boldsymbol{k}_2)\,
    P_{\rm lin}(k_1)\,P_{\rm lin}(k_2), \nonumber \\
    B^{(\rm tree)}_{{\rm J}_{23},{\rm (pre,pre,post)}}(\boldsymbol{k}_2,\boldsymbol{k}_3)
    &=
    2F_{2}(\boldsymbol{k}_2,\boldsymbol{k}_3)\,
    P_{\rm lin}(k_2)\,P_{\rm lin}(k_3), \nonumber \\
    B^{(\rm tree)}_{{\rm J}_{31},{\rm (pre,pre,post)}}(\boldsymbol{k}_3,\boldsymbol{k}_1)
    &=
    2F_{2}(\boldsymbol{k}_3,\boldsymbol{k}_1)\,
    P_{\rm lin}(k_3)\,P_{\rm lin}(k_1),
    \label{eq:BJ_preprepost}
\end{align}
where $F_{2}$ and $F_{{\rm rec},2}$ denote the second-order kernel functions for the pre- and post-reconstruction fields, respectively. Their relation is given in Eq.~\eqref{eq:F2_rec}.

\subsubsection{IR limit}
\label{sec:IR_limit_pre_pre_post}

For the mixed configuration in which two fields are evaluated before reconstruction and one after (the pre--pre--post case), the linear-order displacement-mapping exponent takes the form
\begin{align}
    \overline{\Gamma}_{\alpha\beta,{\rm (pre,pre,post)}}^{\rm (lin)}
    (\boldsymbol{k}_{\alpha},\boldsymbol{k}_{\beta})
    =
    \frac{1}{2}\left(
    k_1^2\bar{\sigma}^2
    + k_2^2\bar{\sigma}^2
    + k_3^2\bar{\sigma}_{\rm rec}^2
    \right),
\end{align}
with $(\alpha,\beta)=(1,2),(2,3),(3,1)$. The scale-dependent parts entering $\Gamma_{\alpha\beta}$ are given by
\begin{align}
    & \Gamma_{12,{\rm (pre,pre,post)}}^{\rm (lin)}(\boldsymbol{k}_1,\boldsymbol{k}_2,\boldsymbol{r}_1,\boldsymbol{r}_2)
    \nonumber \\
    &=
    - k_1^i k_3^j I_{{\rm cross},ij}(\boldsymbol{r}_1)
    - k_2^i k_3^j I_{{\rm cross},ij}(\boldsymbol{r}_2)
    - k_1^i k_2^j I_{ij}(\boldsymbol{r}_1-\boldsymbol{r}_2), \nonumber \\
    & \Gamma_{23,{\rm (pre,pre,post)}}^{\rm (lin)}(\boldsymbol{k}_2,\boldsymbol{k}_3,\boldsymbol{r}_2,\boldsymbol{r}_3)
    \nonumber \\
    &=
    - k_2^i k_1^j I_{ij}(\boldsymbol{r}_2)
    - k_3^i k_1^j I_{{\rm cross},ij}(\boldsymbol{r}_3)
    - k_2^i k_3^j I_{{\rm cross},ij}(\boldsymbol{r}_2-\boldsymbol{r}_3), \nonumber \\
    & \Gamma_{31,{\rm (pre,pre,post)}}^{\rm (lin)}(\boldsymbol{k}_3,\boldsymbol{k}_1,\boldsymbol{r}_3,\boldsymbol{r}_1)
    \nonumber \\
    &=
    - k_3^i k_2^j I_{{\rm cross},ij}(\boldsymbol{r}_3)
    - k_1^i k_2^j I_{ij}(\boldsymbol{r}_1)
    - k_3^i k_1^j I_{{\rm cross},ij}(\boldsymbol{r}_3-\boldsymbol{r}_1),
    \label{eq:Gamma_pre_pre_post}
\end{align}
where $\bar{\sigma}$, $\bar{\sigma}_{\rm rec}$, $I_{ij}$, and $I_{{\rm cross},ij}$ are defined in Eqs.~\eqref{eq:sigma_bar}, \eqref{eq:sigma_bar_rec}, \eqref{eq:Iij_def_start}, and \eqref{eq:I_cross_origin}, respectively.

Taking the IR limit $\boldsymbol{r}_{\alpha},\boldsymbol{r}_{\beta}\to \boldsymbol{0}$, we obtain
\begin{align}
    & -\overline{\Gamma}_{\alpha\beta,{\rm (pre,pre,post)}}^{\rm (lin)}
    (\boldsymbol{k}_{\alpha},\boldsymbol{k}_{\beta})
    + \Gamma_{\alpha\beta,{\rm (pre,pre,post)}}^{\rm (lin)}
    (\boldsymbol{k}_{\alpha},\boldsymbol{k}_{\beta},\boldsymbol{0},\boldsymbol{0}) \nonumber \\
    & = -\frac{1}{2}k_3^2\,\bar{\sigma}_{tt}^2 ,
    \label{eq:DM_cross_preprepost_final}
\end{align}
which is independent of the choice of $(\alpha,\beta)$. Here $\bar{\sigma}_{tt}^2$, defined in Eq.~\eqref{eq:sigma_tt}, is the one-point variance of the reconstruction shift field.

Substituting Eqs.~\eqref{eq:BJ_preprepost} and \eqref{eq:DM_cross_preprepost_final} into the general ULPT cross-bispectrum formula \eqref{eq:B_ULPT_general_ij_abc}, the tree-level cross bispectrum in the IR limit becomes
\begin{align}
    & B_{\rm (pre,pre,post)}(\boldsymbol{k}_1,\boldsymbol{k}_2,\boldsymbol{k}_3) \nonumber \\
    &=
    e^{-\frac{1}{2}k_3^2\bar{\sigma}_{tt}^2}
    \Big[
    2F_{{\rm rec},2}(\boldsymbol{k}_1,\boldsymbol{k}_2)\,
    P_{\rm lin}(k_1)P_{\rm lin}(k_2) \nonumber \\
    &\qquad\qquad\quad
    + 2F_{2}(\boldsymbol{k}_2,\boldsymbol{k}_3)\,
    P_{\rm lin}(k_2)P_{\rm lin}(k_3) \nonumber \\
    &\qquad\qquad\quad
    + 2F_{2}(\boldsymbol{k}_3,\boldsymbol{k}_1)\,
    P_{\rm lin}(k_3)P_{\rm lin}(k_1)
    \Big].
\end{align}

This expression shows that the pre--pre--post cross bispectrum acquires a universal exponential suppression factor,
\[
    \exp\!\left[-\frac{1}{2}k_3^2\bar{\sigma}_{tt}^2\right],
\]
which multiplies the entire bispectrum shape. The characteristic damping scale is set by the wavenumber $k_3$ associated with the post-reconstruction density field, reflecting the fact that only the post-reconstruction field retains the smoothing induced by the reconstruction shift. An intuitive physical derivation of this behavior in the IR limit is presented in Appendix~\ref{sec:appendix_universal}.

\subsubsection{IR-resummed model}
\label{sec:IR_resum_pre_pre_post}

To derive the IR-resummed model for the pre--pre--post cross bispectrum, we closely follow the procedure outlined in Sec.~\ref{sec:ULPT_bispec_IR}. In particular, we apply the same set of replacements to the linear displacement-mapping exponent \(\Gamma_{\alpha\beta, {\rm (pre,pre,post)}}^{(\rm lin)} (\boldsymbol{k}_{\alpha},\boldsymbol{k}_{\beta}, \boldsymbol{r}_{\alpha},\boldsymbol{r}_{\beta})\) in Eq.~\eqref{eq:Gamma_pre_pre_post}, evaluated for \((\alpha,\beta)=(1,2),(2,3),(3,1)\).

For the nw--nw contribution, we take the limit \(\boldsymbol{r}_{\alpha}\to \boldsymbol{0}\) and \(\boldsymbol{r}_{\beta}\to \boldsymbol{0}\). For the w--nw contribution, we set \(r_{\alpha}=r_{\rm BAO}\), \(\hat{r}_{\alpha}=\hat{k}_{\alpha}\), and \(\boldsymbol{r}_{\beta}=\boldsymbol{0}\). For the nw--w case, we set \(r_{\beta}=r_{\rm BAO}\), \(\hat{r}_{\beta}=\hat{k}_{\beta}\), and \(\boldsymbol{r}_{\alpha}=\boldsymbol{0}\). For the w--w contribution, we choose \(r_{\alpha}=r_{\rm BAO}\), \(\hat{r}_{\alpha}=\hat{k}_{\alpha}\), \(r_{\beta}=r_{\rm BAO}\), \(\hat{r}_{\beta}=\hat{k}_{\beta}\), and additionally adopt the approximation \(|\hat{r}_{\alpha}-\hat{r}_{\beta}|\approx 4/3\).

Applying the above replacements, the IR-resummed expression for the \(B_{12}\) component of the pre--pre--post cross bispectrum becomes
\begin{align}
    & B_{12, {\rm (pre,pre,post)}, {\rm IR\text{-}resum}}(\boldsymbol{k}_1,\boldsymbol{k}_2)
    \nonumber \\
    &= 2\,F_{{\rm rec},2}(\boldsymbol{k}_1,\boldsymbol{k}_2)
    e^{-\tfrac{1}{2}k_3^2\bar{\sigma}_{tt}^2}
    \Big[
        P_{\rm nw}(k_1)P_{\rm nw}(k_2)
        \nonumber \\
        &\quad
        + e^{-k_1^2\sigma_{\rm BAO,cross}^2}
          e^{-(\boldsymbol{k}_1\cdot\boldsymbol{k}_2)\sigma_{{\rm BAO},t\Psi}^2}
          P_{\rm w}(k_1)P_{\rm nw}(k_2)
        \nonumber \\
        &\quad
        + e^{-k_2^2\sigma_{\rm BAO,cross}^2}
          e^{-(\boldsymbol{k}_1\cdot\boldsymbol{k}_2)\sigma_{{\rm BAO},t\Psi}^2}
          P_{\rm nw}(k_1)P_{\rm w}(k_2)
        \nonumber \\
        &\quad
        + e^{-\tfrac{1}{2}(k_1^2+k_2^2+k_3^2)\sigma_{\rm BAO,cross}^2}
          e^{-(\boldsymbol{k}_1\cdot\boldsymbol{k}_2)\sigma_{{\rm BAO},t\Psi}^2}
        \nonumber \\
        &\qquad\qquad
          \times
          \mathcal{D}_{{\rm w\text{-}w}}(\boldsymbol{k}_1,\boldsymbol{k}_2)
          P_{\rm w}(k_1)P_{\rm w}(k_2)
      \Big],
\end{align}
where the damping scales appearing in the above expression are given by
\begin{align}
    \sigma_{\rm BAO,cross}^2
    &= \bar{\sigma}_{\rm cross}^2
       - \sigma_{{\rm cross},0}^2(r_{\rm BAO})
       - 2\sigma_{{\rm cross},2}^2(r_{\rm BAO})
    \nonumber \\
    &= \frac{1}{3}\int\frac{dp}{2\pi^2}
       \left[1 - j_0(pr_{\rm BAO}) + 2j_2(pr_{\rm BAO})\right] \nonumber \\
     & \quad \times  \left[1 - W_{\rm G}(pR)\right] P_{\rm lin}(p),
       \label{eq:sigma_BAO_cross}
\end{align}
and
\begin{align}
    \sigma_{{\rm BAO},t\Psi}^2
    &= \bar{\sigma}_{t\Psi}^2
       - \sigma_{t\Psi,0}^2(r_{\rm BAO})
       - 2\sigma_{t\Psi,2}^2(r_{\rm BAO})
    \nonumber \\
    &= \frac{1}{3}\int\frac{dp}{2\pi^2}
       \left[1 - j_0(pr_{\rm BAO}) + 2j_2(pr_{\rm BAO})\right] \nonumber \\
       & \quad \times  \left[-W_{\rm G}(pR)\right] P_{\rm lin}(p).
       \label{eq:sigma_BAO_tPsi}
\end{align}
The w--w specific exponential function for \(B_{12}\) is identical to that appearing in the auto bispectrum, $\mathcal{D}_{{\rm w\text{-}w}}$ given by Eq.~\eqref{eq:Dww}.

The remaining components follow analogously. The \(B_{23}\) contribution is
\begin{align}
    & B_{23,{\rm (pre,pre,post)}, {\rm IR\text{-}resum}}(\boldsymbol{k}_2,\boldsymbol{k}_3)
    \nonumber \\
    &= 2\,F_{2}(\boldsymbol{k}_2,\boldsymbol{k}_3)
       e^{-\tfrac{1}{2}k_3^2\bar{\sigma}_{tt}^2}
    \Big[
        P_{\rm nw}(k_2)P_{\rm nw}(k_3)
        \nonumber \\
        &\quad
        + e^{-k_2^2\sigma_{\rm BAO,cross}^2}
          e^{-(\boldsymbol{k}_1\cdot\boldsymbol{k}_2)\sigma_{{\rm BAO},t\Psi}^2}
          P_{\rm w}(k_2)P_{\rm nw}(k_3)
        \nonumber \\
        &\quad
        + e^{-k_3^2\sigma_{\rm BAO,cross}^2}
          P_{\rm nw}(k_2)P_{\rm w}(k_3)
        \nonumber \\
        &\quad
        + e^{-\tfrac{1}{2}(k_1^2+k_2^2+k_3^2)\sigma_{\rm BAO,cross}^2}
          e^{-(\boldsymbol{k}_1\cdot\boldsymbol{k}_2)\sigma_{{\rm BAO},t\Psi}^2}
        \nonumber \\
        &\qquad\qquad
          \times
          \mathcal{D}_{{\rm cross}, {\rm w\text{-}w}}(\boldsymbol{k}_2,\boldsymbol{k}_3)
          P_{\rm w}(k_2)P_{\rm w}(k_3)
      \Big].
\end{align}

The \(B_{31}\) contribution is
\begin{align}
    & B_{31,{\rm (pre,pre,post)}, {\rm IR\text{-}resum}}(\boldsymbol{k}_3,\boldsymbol{k}_1)
    \nonumber \\
    &= 2\,F_{2}(\boldsymbol{k}_3,\boldsymbol{k}_1)
       e^{-\tfrac{1}{2}k_3^2\bar{\sigma}_{tt}^2}
    \Big[
        P_{\rm nw}(k_3)P_{\rm nw}(k_1)
        \nonumber \\
        &\quad
        + e^{-k_3^2\sigma_{\rm BAO,cross}^2}
          P_{\rm w}(k_3)P_{\rm nw}(k_1)
        \nonumber \\
        &\quad
        + e^{-k_1^2\sigma_{\rm BAO,cross}^2}
          e^{-(\boldsymbol{k}_1\cdot\boldsymbol{k}_2)\sigma_{{\rm BAO},t\Psi}^2}
          P_{\rm nw}(k_3)P_{\rm w}(k_1)
        \nonumber \\
        &\quad
        + e^{-\tfrac{1}{2}(k_1^2+k_2^2+k_3^2)\sigma_{\rm BAO,cross}^2}
          e^{-(\boldsymbol{k}_1\cdot\boldsymbol{k}_2)\sigma_{{\rm BAO},t\Psi}^2}
        \nonumber \\
        &\qquad\qquad
          \times
          \mathcal{D}_{{\rm cross}, {\rm w\text{-}w}}(\boldsymbol{k}_3,\boldsymbol{k}_1)
          P_{\rm w}(k_3)P_{\rm w}(k_1)
    \Big].
\end{align}
The w--w function for these two components is
\begin{align}
    \mathcal{D}_{{\rm cross}, {\rm w\text{-}w}}(\boldsymbol{k}_2,\boldsymbol{k}_3)
    &=
    \exp\!\Big[
        -( \boldsymbol{k}_2\!\cdot\!\boldsymbol{k}_3 )\,
        \Delta\sigma_{\rm BAO,cross}^2
        \nonumber \\
        &\quad
        + \tfrac{3}{2}\left(k_2 k_3 + \boldsymbol{k}_2\!\cdot\!\boldsymbol{k}_3\right)
        \sigma_{{\rm cross},2}^2(r'_{\rm BAO})
    \Big],
    \label{eq:Dww_preprepost}
\end{align}
with
\begin{align}
    \Delta\sigma_{\rm BAO,cross}^2
    &=
      \sigma_{{\rm cross},0}^2(r'_{\rm BAO})
      - \sigma_{{\rm cross},0}^2(r_{\rm BAO})
    \nonumber \\
    &\quad
    + 2\left[
        \sigma_{{\rm cross},2}^2(r'_{\rm BAO})
        - \sigma_{{\rm cross},2}^2(r_{\rm BAO})
    \right],
\end{align}
and \(r'_{\rm BAO} = (4/3)r_{\rm BAO}\).

The full pre--pre--post cross bispectrum is finally obtained by summing the three components: \(B_{12} + B_{23} + B_{31}\).

We now discuss the behavior of the IR-resummed model for the pre--pre--post cross bispectrum. As already shown in the IR limit, the overall shape of the bispectrum is controlled by a universal exponential damping factor, \(\exp\!\left[-\tfrac{1}{2}k_3^2\bar{\sigma}^2_{tt}\right]\), which multiplies the entire expression. Here, \(k_3\) is the wavenumber associated with the density field evaluated after reconstruction in the pre--pre--post configuration.

The three contributions \(B_{12}\), \(B_{23}\), and \(B_{31}\) exhibit a clear asymmetry in the appearance of the second-order kernels. The \(B_{12}\) term is governed by the nonlinear kernel \(F_{{\rm rec},2}\), which explicitly incorporates the effect of reconstruction, whereas the remaining two terms, \(B_{23}\) and \(B_{31}\), are described by the standard pre-reconstruction kernel \(F_{2}\). This structure reflects the fact that only one of the three density fields is evaluated after reconstruction.

The nonlinear modification of the BAO signal is primarily encoded in the parameter \(\sigma^2_{\rm BAO,cross}\), which appears in the factors \(\exp(-k_i^2\sigma^2_{\rm BAO,cross})\) for $i=1,2,3$. This closely parallels the auto bispectrum case, where the BAO damping is parameterized by \(\sigma^2_{\rm BAO}\), and the formal structure of the model is therefore very similar. A key difference from the auto bispectrum, however, is the appearance of an additional exponential factor, \(\exp\!\left[-(\boldsymbol{k}_1\cdot\boldsymbol{k}_2)\sigma^2_{{\rm BAO},t\Psi}\right]\), which describes further nonlinear effects on the BAO signal. The dependence on \(\boldsymbol{k}_1\cdot\boldsymbol{k}_2\) implies that this term is controlled by the scales of the two pre-reconstruction density fields. Consequently, this factor appears in all three components, \(B_{12}\), \(B_{23}\), and \(B_{31}\), and in each of their w--nw, nw--w, and w--w contributions. The only exceptions are the nw--w term of \(B_{23}\) and the w--nw term of \(B_{31}\), where this factor is absent because in those cases it is the post-reconstruction mode \(k_3\) that governs the BAO damping.

Moreover, for the $B_{23}$ and $B_{31}$ components, the w--w contributions involve the exponential factor $\mathcal{D}_{{\rm cross},{\rm w\text{-}w}}$, defined in Eq.~\eqref{eq:Dww_preprepost}. This factor is the direct counterpart of the w--w exponential term appearing in the auto-bispectrum, but evaluated with the parameters appropriate to the cross configuration. Although it shares the same functional dependence on the wavenumbers, the different parameter values lead to a quantitatively distinct BAO modulation.

To quantify the structure of the exponential factors in the IR-resummed model, we list the numerical values of the relevant parameters for our fiducial $\Lambda$CDM cosmology at redshift \(z = 0\) and smoothing scale \(R=15\,h^{-1}\,\mathrm{Mpc}\):
\begin{align}
    \bar{\sigma}^2_{\rm cross} & = 15.981\,(h^{-1}\,\mathrm{Mpc})^2, \nonumber \\
    \sigma^2_{\rm BAO,cross} & = 16.276\,(h^{-1}\,\mathrm{Mpc})^2, \nonumber \\
    \bar{\sigma}^2_{t\Psi} & = -19.793\,(h^{-1}\,\mathrm{Mpc})^2, \nonumber \\
    \sigma^2_{{\rm BAO},t\Psi} & = -20.027\,(h^{-1}\,\mathrm{Mpc})^2, \nonumber \\
    \Delta \sigma^2_{\rm BAO,cross} & = 0.081\,(h^{-1}\,\mathrm{Mpc})^2, \nonumber \\
    \sigma^2_{{\rm cross},2}(r'_{\rm BAO}) & = -0.102\,(h^{-1}\,\mathrm{Mpc})^2.
    \label{eq:sigma_cross_values}
\end{align}
The relative difference between \(\bar{\sigma}^2_{\rm cross}\) and \(\sigma^2_{\rm BAO,cross}\) is about \(1.8\%\), while that between \(\bar{\sigma}^2_{t\Psi}\) and \(\sigma^2_{{\rm BAO},t\Psi}\) is about \(1.2\%\). Furthermore, \(\Delta \sigma^2_{\rm BAO,cross}\) and \(\sigma^2_{{\rm cross},2}(r'_{\rm BAO})\) are only \(0.5\%\) and \(0.6\%\), respectively, of \(\bar{\sigma}^2_{\rm cross}\). These results confirm that the following numerical approximation is well justified:
\begin{align}
    \bar{\sigma}^2_{\rm cross} & \approx \sigma^2_{\rm BAO,cross}, \nonumber \\
    \bar{\sigma}^2_{t\Psi} & \approx \sigma^2_{{\rm BAO},t\Psi}, \nonumber \\
    \Delta \sigma^2_{\rm BAO,cross}
    & \approx \sigma^2_{{\rm cross},2}(r'_{\rm BAO}) \approx 0.
\end{align}
In physical terms, this approximation corresponds to neglecting the BAO contribution of the mode-coupling terms. Under this approximation, the IR-resummed model for the pre--pre--post cross bispectrum takes a particularly simple form.

The \(B_{12}\) component reduces to
\begin{align}
    & B_{12, {\rm (pre,pre,post)}, {\rm IR\text{-}resum}}(\boldsymbol{k}_1,\boldsymbol{k}_2)
    \nonumber \\
    &= 2\,F_{{\rm rec},2}(\boldsymbol{k}_1,\boldsymbol{k}_2)
    e^{-\tfrac{1}{2}k_3^2\bar{\sigma}_{tt}^2}
    \Big[
        P_{\rm nw}(k_1)P_{\rm nw}(k_2)
        \nonumber \\
        &\quad
        + e^{-k_1^2\bar{\sigma}_{\rm cross}^2}
          e^{-(\boldsymbol{k}_1\cdot\boldsymbol{k}_2)\bar{\sigma}^2_{t\Psi}}
          P_{\rm w}(k_1)P_{\rm nw}(k_2)
        \nonumber \\
        &\quad
        + e^{-k_2^2\bar{\sigma}_{\rm cross}^2}
          e^{-(\boldsymbol{k}_1\cdot\boldsymbol{k}_2)\bar{\sigma}^2_{t\Psi}}
          P_{\rm nw}(k_1)P_{\rm w}(k_2)
        \nonumber \\
        &\quad
        + e^{-\tfrac{1}{2}(k_1^2+k_2^2+k_3^2)\bar{\sigma}_{\rm cross}^2}
          e^{-(\boldsymbol{k}_1\cdot\boldsymbol{k}_2)\bar{\sigma}^2_{t\Psi}}
          P_{\rm w}(k_1)P_{\rm w}(k_2)
    \Big],
    \label{eq:IR_resum_preprepost_12_approx}
\end{align}
while the \(B_{23}\) component becomes
\begin{align}
    & B_{23, {\rm (pre,pre,post)}, {\rm IR\text{-}resum}}(\boldsymbol{k}_2,\boldsymbol{k}_3)
    \nonumber \\
    &= 2\,F_{2}(\boldsymbol{k}_2,\boldsymbol{k}_3)
    e^{-\tfrac{1}{2}k_3^2\bar{\sigma}_{tt}^2}
    \Big[
        P_{\rm nw}(k_2)P_{\rm nw}(k_3)
        \nonumber \\
        &\quad
        + e^{-k_2^2\bar{\sigma}_{\rm cross}^2}
          e^{-(\boldsymbol{k}_1\cdot\boldsymbol{k}_2)\bar{\sigma}^2_{t\Psi}}
          P_{\rm w}(k_2)P_{\rm nw}(k_3)
        \nonumber \\
        &\quad
        + e^{-k_3^2\bar{\sigma}_{\rm cross}^2}
          P_{\rm nw}(k_2)P_{\rm w}(k_3)
        \nonumber \\
        &\quad
        + e^{-\tfrac{1}{2}(k_1^2+k_2^2+k_3^2)\bar{\sigma}_{\rm cross}^2}
          e^{-(\boldsymbol{k}_1\cdot\boldsymbol{k}_2)\bar{\sigma}^2_{t\Psi}}
          P_{\rm w}(k_2)P_{\rm w}(k_3)
    \Big],
    \label{eq:IR_resum_preprepost_23_approx}
\end{align}
and the \(B_{31}\) component is
\begin{align}
    & B_{31, {\rm (pre,pre,post)}, {\rm IR\text{-}resum}}(\boldsymbol{k}_3,\boldsymbol{k}_1)
    \nonumber \\
    &= 2\,F_{2}(\boldsymbol{k}_3,\boldsymbol{k}_1)
    e^{-\tfrac{1}{2}k_3^2\bar{\sigma}_{tt}^2}
    \Big[
        P_{\rm nw}(k_3)P_{\rm nw}(k_1)
        \nonumber \\
        &\quad
        + e^{-k_3^2\bar{\sigma}_{\rm cross}^2}
          P_{\rm w}(k_3)P_{\rm nw}(k_1)
        \nonumber \\
        &\quad
        + e^{-k_1^2\bar{\sigma}_{\rm cross}^2}
          e^{-(\boldsymbol{k}_1\cdot\boldsymbol{k}_2)\bar{\sigma}^2_{t\Psi}}
          P_{\rm nw}(k_3)P_{\rm w}(k_1)
        \nonumber \\
        &\quad
        + e^{-\tfrac{1}{2}(k_1^2+k_2^2+k_3^2)\bar{\sigma}_{\rm cross}^2}
          e^{-(\boldsymbol{k}_1\cdot\boldsymbol{k}_2)\bar{\sigma}^2_{t\Psi}}
          P_{\rm w}(k_3)P_{\rm w}(k_1)
    \Big].
    \label{eq:IR_resum_preprepost_31_approx}
\end{align}
In this approximate form, the structure of the cross bispectrum becomes transparent: all nonlinear BAO effects are governed by only two characteristic scales, \(\bar{\sigma}^2_{\rm cross}\) and \(\bar{\sigma}^2_{t\Psi}\), together with the universal damping scale \(\bar{\sigma}^2_{tt}\) associated with the post-reconstruction density mode \(k_3\).

\subsection{IR-resummed model for the pre--post--post configuration}
\label{sec:prepostpost}

In this subsection, we extend the analysis of Sec.~\ref{sec:preprepost} to the cross bispectrum $B_{\rm (pre,post,post)}(\boldsymbol{k}_1,\boldsymbol{k}_2,\boldsymbol{k}_3)$, in which the density fluctuation associated with $\boldsymbol{k}_1$ is taken from the pre-reconstruction field, while those associated with $\boldsymbol{k}_2$ and $\boldsymbol{k}_3$ are taken from the post-reconstruction field. We refer to this arrangement as the \emph{pre--post--post} configuration.

\subsubsection{Source bispectrum}

For the pre--post--post configuration, the tree-level cross source bispectrum is given by
\begin{align}
    B^{(\rm tree)}_{{\rm J}_{12}, {\rm (pre,post,post)}}(\boldsymbol{k}_1,\boldsymbol{k}_2)
    & = 2F_{{\rm rec},2}(\boldsymbol{k}_1,\boldsymbol{k}_2)\,
    P_{\rm lin}(k_1)\,P_{\rm lin}(k_2)\,, \nonumber \\
    B^{(\rm tree)}_{{\rm J}_{23}, {\rm (pre,post,post)}}(\boldsymbol{k}_2,\boldsymbol{k}_3)
    & = 2F_{2}(\boldsymbol{k}_2,\boldsymbol{k}_3)\,
    P_{\rm lin}(k_2)\,P_{\rm lin}(k_3)\,, \nonumber \\
    B^{(\rm tree)}_{{\rm J}_{31}, {\rm (pre,post,post)}}(\boldsymbol{k}_3,\boldsymbol{k}_1)
    & = 2F_{{\rm rec},2}(\boldsymbol{k}_3,\boldsymbol{k}_1)\,
    P_{\rm lin}(k_3)\,P_{\rm lin}(k_1)\,,
    \label{eq:BJ_prepostpost}
\end{align}
where $F_{2}$ and $F_{{\rm rec},2}$ denote the standard and reconstructed second-order kernels, respectively. The $B_{12}$ and $B_{31}$ components are governed by the reconstructed kernel $F_{{\rm rec},2}$, whereas the $B_{23}$ component is described by the pre-reconstruction kernel $F_{2}$.

\subsubsection{IR limit}

For the pre--post--post configuration, the linear-order contribution to the displacement-mapping exponent is given by
\begin{align}
    \overline{\Gamma}_{\alpha\beta, {\rm (pre,post,post)}}^{\rm (lin)}
    (\boldsymbol{k}_{\alpha},\boldsymbol{k}_{\beta})
    = \frac{1}{2}
    \left(
         k_1^2 \bar{\sigma}^2
         + k_2^2 \bar{\sigma}_{\rm rec}^2
         + k_3^2 \bar{\sigma}_{\rm rec}^2
    \right),
\end{align}
for $(\alpha,\beta)=(1,2),(2,3),(3,1)$, where $\bar{\sigma}$ and $\bar{\sigma}_{\rm rec}$ are defined in Eqs.~\eqref{eq:sigma_bar} and \eqref{eq:sigma_bar_rec}, respectively. The configuration-dependent contributions are encoded in
\begin{align}
    & \Gamma_{12, {\rm (pre,post,post)}}^{\rm (lin)}
    (\boldsymbol{k}_1,\boldsymbol{k}_2,\boldsymbol{r}_1,\boldsymbol{r}_2)
    \nonumber \\
    & = - k_1^i k_2^j I_{{\rm cross}, ij}(\boldsymbol{r}_{1}-\boldsymbol{r}_2)
        - k_1^i k_3^j I_{{\rm cross}, ij}(\boldsymbol{r}_{1})
        - k_2^i k_3^j I_{{\rm rec}, ij}(\boldsymbol{r}_2), \\
    & \Gamma_{23, {\rm (pre,post,post)}}^{\rm (lin)}
    (\boldsymbol{k}_2,\boldsymbol{k}_3,\boldsymbol{r}_2,\boldsymbol{r}_3)
    \nonumber \\
    & = - k_1^i k_2^j I_{{\rm cross}, ij}(\boldsymbol{r}_2)
        - k_1^i k_3^j I_{{\rm cross}, ij}(\boldsymbol{r}_3)
        - k_2^i k_3^j I_{{\rm rec}, ij}(\boldsymbol{r}_{2}-\boldsymbol{r}_3), \\
    & \Gamma_{31, {\rm (pre,post,post)}}^{\rm (lin)}
    (\boldsymbol{k}_3,\boldsymbol{k}_1,\boldsymbol{r}_3,\boldsymbol{r}_1)
    \nonumber \\
    & = - k_1^i k_2^j I_{{\rm cross}, ij}(\boldsymbol{r}_1)
        - k_1^i k_3^j I_{{\rm cross}, ij}(\boldsymbol{r}_3-\boldsymbol{r}_1)
        - k_2^i k_3^j I_{{\rm rec}, ij}(\boldsymbol{r}_3),
\end{align}
where $I_{{\rm cross},ij}$ and $I_{{\rm rec},ij}$ are given by Eqs.~\eqref{eq:I_cross_origin} and \eqref{eq:Iij_def_rec}, respectively.

Taking the IR limit $\boldsymbol{r}_{\alpha},\boldsymbol{r}_{\beta}\to \boldsymbol{0}$, we obtain
\begin{align}
    & -\overline{\Gamma}_{\alpha\beta, {\rm (pre,post,post)}}^{\rm (lin)}
      (\boldsymbol{k}_{\alpha},\boldsymbol{k}_{\beta})
    + \Gamma_{\alpha\beta, {\rm (pre,post,post)}}^{\rm (lin)}
    (\boldsymbol{k}_{\alpha},\boldsymbol{k}_{\beta}, \boldsymbol{0}, \boldsymbol{0}) \nonumber \\
    & \quad = -\frac{1}{2}k_1^2\bar{\sigma}_{tt}^2,
    \label{eq:DM_cross_prepostpost}
\end{align}
for all $(\alpha,\beta)=(1,2),(2,3),(3,1)$.

Substituting Eqs.~\eqref{eq:BJ_prepostpost} and \eqref{eq:DM_cross_prepostpost} into the general ULPT cross-bispectrum expression \eqref{eq:B_ULPT_general_ij_abc}, we find that the tree-level pre--post--post cross bispectrum in the IR limit takes the form
\begin{align}
    & B_{\rm (pre,post,post)}(\boldsymbol{k}_1,\boldsymbol{k}_2,\boldsymbol{k}_3) \nonumber \\
    &= e^{-\frac{1}{2}k_1^2\bar{\sigma}_{tt}^2}
    \Big[
        2F_{{\rm rec}, 2}(\boldsymbol{k}_1,\boldsymbol{k}_2)\,
          P_{\rm lin}(k_1)\,P_{\rm lin}(k_2) \nonumber \\
    &\hspace{1.8cm}
        + 2F_{2}(\boldsymbol{k}_2,\boldsymbol{k}_3)\,
          P_{\rm lin}(k_2)\,P_{\rm lin}(k_3) \nonumber \\
    &\hspace{1.8cm}
        + 2F_{{\rm rec}, 2}(\boldsymbol{k}_3,\boldsymbol{k}_1)\,
          P_{\rm lin}(k_3)\,P_{\rm lin}(k_1)
    \Big].
\end{align}

Thus, in the pre--post--post configuration, the entire shape of the cross bispectrum is multiplied by the universal exponential factor $e^{-\frac{1}{2}k_1^2\bar{\sigma}_{tt}^2}$. The characteristic damping scale is therefore set by the wavenumber $k_1$, which is associated with the pre-reconstruction density field in this configuration.

By contrast, in the pre--pre--post configuration analyzed in Sec.~\ref{sec:preprepost}, the damping scale is determined by $k_3$, the wavenumber associated with the post-reconstruction field. More generally, whenever two of the three density fields are taken from either the pre- or post-reconstruction class and the remaining one from the opposite class, the exponential damping factor is governed by the wavenumber of the single field that differs from the other two. For example, in the pre--post--pre configuration, the damping factor is controlled by $k_2$ and is given by $e^{-\frac{1}{2}k_2^2\bar{\sigma}_{tt}^2}$. An intuitive physical interpretation of this universal rule in the IR limit is presented in Appendix~\ref{sec:appendix_universal}.

\subsubsection{IR-resummed model}

In this subsection, we present the IR-resummed expression for the pre--post--post cross bispectrum. The derivation closely follows the procedure described in Sec.~\ref{sec:preprepost}.

For the $B_{12}$ contribution, the IR-resummed bispectrum is
\begin{align}
    & B_{12, {\rm (pre,post,post)}, {\rm IR\text{-}resum}}(\boldsymbol{k}_1,\boldsymbol{k}_2)
    \nonumber \\
    &= 2\,F_{{\rm rec},2}(\boldsymbol{k}_1,\boldsymbol{k}_2)\,
    e^{-\tfrac{1}{2}k_1^2\bar{\sigma}_{tt}^2}
    \Big[
        P_{\rm nw}(k_1)P_{\rm nw}(k_2)
        \nonumber \\
        &\quad
        + e^{-k_1^2\sigma_{\rm BAO,cross}^2}
          P_{\rm w}(k_1)P_{\rm nw}(k_2)
        \nonumber \\
        &\quad
        + e^{-k_2^2\sigma_{\rm BAO,cross}^2}
          e^{\boldsymbol{k}_2\cdot\boldsymbol{k}_3\left(\sigma^2_{{\rm BAO},t\Psi} + \sigma_{{\rm BAO},tt}^2\right)}
          P_{\rm nw}(k_1)P_{\rm w}(k_2)
        \nonumber \\
        &\quad
        + e^{-\tfrac{1}{2}(k_1^2+k_2^2+k_3^2)\sigma_{\rm BAO,cross}^2}
          e^{\boldsymbol{k}_2\cdot\boldsymbol{k}_3\left(\sigma_{{\rm BAO},t\Psi}^2 + \sigma_{{\rm BAO},tt}^2\right)}
        \nonumber \\
        &\qquad\qquad
          \times
          \mathcal{D}_{{\rm cross},{\rm w\text{-}w}}(\boldsymbol{k}_1,\boldsymbol{k}_2)\,
          P_{\rm w}(k_1)P_{\rm w}(k_2)
      \Big].
\end{align}

The $B_{23}$ contribution is
\begin{align}
    & B_{23,{\rm (pre,post,post)}, {\rm IR\text{-}resum}}(\boldsymbol{k}_2,\boldsymbol{k}_3)
    \nonumber \\
    &= 2\,F_{2}(\boldsymbol{k}_2,\boldsymbol{k}_3)\,
       e^{-\tfrac{1}{2}k_1^2\bar{\sigma}_{tt}^2}
    \Big[
        P_{\rm nw}(k_2)P_{\rm nw}(k_3)
        \nonumber \\
        &\quad
        + e^{-k_2^2\sigma_{\rm BAO,cross}^2}
          e^{\boldsymbol{k}_2\cdot\boldsymbol{k}_3\left(\sigma_{{\rm BAO},t\Psi}^2 + \sigma_{{\rm BAO},tt}^2\right)}
          P_{\rm w}(k_2)P_{\rm nw}(k_3)
        \nonumber \\
        &\quad
        + e^{-k_3^2\sigma_{\rm BAO,cross}^2}
          e^{\boldsymbol{k}_2\cdot\boldsymbol{k}_3\left(\sigma_{{\rm BAO},t\Psi}^2 + \sigma_{{\rm BAO},tt}^2\right)}
          P_{\rm nw}(k_2)P_{\rm w}(k_3)
        \nonumber \\
        &\quad
        + e^{-\tfrac{1}{2}(k_1^2+k_2^2+k_3^2)\sigma_{\rm BAO,cross}^2}
          e^{\boldsymbol{k}_2\cdot\boldsymbol{k}_3\left(\sigma_{{\rm BAO},t\Psi}^2 + \sigma_{{\rm BAO},tt}^2\right)}
        \nonumber \\
        &\qquad\qquad
          \times
          \mathcal{D}_{{\rm rec},{\rm w\text{-}w}}(\boldsymbol{k}_2,\boldsymbol{k}_3)\,
          P_{\rm w}(k_2)P_{\rm w}(k_3)
    \Big].
\end{align}

The $B_{31}$ contribution is
\begin{align}
    & B_{31,{\rm (pre,post,post)}, {\rm IR\text{-}resum}}(\boldsymbol{k}_3,\boldsymbol{k}_1)
    \nonumber \\
    &= 2\,F_{{\rm rec},2}(\boldsymbol{k}_3,\boldsymbol{k}_1)\,
       e^{-\tfrac{1}{2}k_1^2\bar{\sigma}_{tt}^2}
    \Big[
        P_{\rm nw}(k_3)P_{\rm nw}(k_1)
        \nonumber \\
        &\quad
        + e^{-k_3^2\sigma_{\rm BAO,cross}^2}
          e^{\boldsymbol{k}_2\cdot\boldsymbol{k}_3\left(\sigma_{{\rm BAO},t\Psi}^2 + \sigma_{{\rm BAO},tt}^2\right)}
          P_{\rm w}(k_3)P_{\rm nw}(k_1)
        \nonumber \\
        &\quad
        + e^{-k_1^2\sigma_{\rm BAO,cross}^2}
          P_{\rm nw}(k_3)P_{\rm w}(k_1)
        \nonumber \\
        &\quad
        + e^{-\tfrac{1}{2}(k_1^2+k_2^2+k_3^2)\sigma_{\rm BAO,cross}^2}
          e^{\boldsymbol{k}_2\cdot\boldsymbol{k}_3\left(\sigma_{{\rm BAO},t\Psi}^2 + \sigma_{{\rm BAO},tt}^2\right)}
        \nonumber \\
        &\qquad\qquad
          \times
          \mathcal{D}_{{\rm cross},{\rm w\text{-}w}}(\boldsymbol{k}_3,\boldsymbol{k}_1)\,
          P_{\rm w}(k_3)P_{\rm w}(k_1)
    \Big].
\end{align}
Here, $\bar{\sigma}_{tt}^2$, $\sigma_{\rm BAO,cross}^2$, and $\sigma_{{\rm BAO},t\Psi}^2$ are defined in Eqs.~\eqref{eq:sigma_tt}, \eqref{eq:sigma_BAO_cross}, and \eqref{eq:sigma_BAO_tPsi}, respectively. The additional BAO parameter $\sigma_{{\rm BAO},tt}^2$ is given by
\begin{align}
    \sigma_{{\rm BAO},tt}^2
    &= \bar{\sigma}_{tt}^2
       - \sigma_{tt,0}^2(r_{\rm BAO})
       - 2\sigma_{tt,2}^2(r_{\rm BAO})
    \nonumber \\
    &= \frac{1}{3}\int\frac{dp}{2\pi^2}
       \left[1 - j_0(pr_{\rm BAO}) + 2j_2(pr_{\rm BAO})\right]
       \nonumber \\
       & \quad \times
       \left[-W_{\rm G}(pR)\right]^2 P_{\rm lin}(p)\,.
\end{align}
The w--w functions $\mathcal{D}_{{\rm cross},{\rm w\text{-}w}}$ and $\mathcal{D}_{{\rm rec},{\rm w\text{-}w}}$ are defined in Eqs.~\eqref{eq:Dww_preprepost} and \eqref{eq:Dww_rec}, respectively.

These expressions show that, as already demonstrated in the IR limit, the entire pre--post--post cross bispectrum is multiplied by the universal damping factor $\exp[-\tfrac{1}{2}k_1^2\bar{\sigma}_{tt}^2]$, where $k_1$ is the wavenumber associated with the pre-reconstruction density field. Furthermore, the $B_{12}$ and $B_{31}$ components are governed by the reconstructed non-linear kernel $F_{{\rm rec},2}$, while the $B_{23}$ component is described by the standard pre-reconstruction kernel $F_{2}$.

The nonlinear modification of the BAO features is primarily governed by $\sigma_{\rm BAO,cross}^2$, in close analogy with the pre--pre--post case. A distinctive feature of the pre--post--post configuration is the presence of an additional exponential factor,
\[
e^{\boldsymbol{k}_2\cdot\boldsymbol{k}_3\left(\sigma_{{\rm BAO},t\Psi}^2
+\sigma_{{\rm BAO},tt}^2\right)},
\]
which encodes further nonlinear BAO effects associated with the pair of post-reconstruction density fields corresponding to $\boldsymbol{k}_2$ and $\boldsymbol{k}_3$. Since this factor depends only on $\boldsymbol{k}_2\!\cdot\!\boldsymbol{k}_3$, it modulates precisely those contributions in which the BAO wiggle resides on either one or both of the post-reconstruction modes, $k_2$ or $k_3$. In contrast, it is absent in the $B_{12}$ w--nw term and the $B_{31}$ nw--w term, where the BAO damping is instead governed by the pre-reconstruction mode $k_1$. This pattern is directly analogous to that found in the pre--pre--post case, where the additional exponential factor is associated with the pair of pre-reconstruction modes and does not appear when the BAO damping is controlled by the post-reconstruction field.

At $z=0$, for the fiducial cosmology adopted in this paper and for a Gaussian reconstruction smoothing scale of $R=15\,h^{-1}\mathrm{Mpc}$, the numerical values of the relevant parameters entering the pre--post--post cross bispectrum are summarized in Eqs.~\eqref{eq:sigma_rec_values} and \eqref{eq:sigma_cross_values}. The remaining parameters satisfy
\begin{align}
    \bar{\sigma}_{tt}^2 & = 16.283\,(h^{-1}\,\mathrm{Mpc})^2, \nonumber \\
    \sigma_{{\rm BAO}, tt}^2 & = 16.199\,(h^{-1}\,\mathrm{Mpc})^2\,,
\end{align}
so that
\begin{equation}
    \bar{\sigma}_{tt}^2 \approx \sigma_{{\rm BAO}, tt}^2\,.
\end{equation}
As in the other IR-resummed models discussed in this paper, this numerical proximity justifies the approximation in which BAO contributions in the mode-coupling sector are neglected.

Under this approximation, the pre--post--post cross bispectrum takes a simplified form. For the $B_{12}$ contribution, we obtain
\begin{align}
    & B_{12, {\rm (pre,post,post)}, {\rm IR\text{-}resum}}(\boldsymbol{k}_1,\boldsymbol{k}_2)
    \nonumber \\
    &= 2\,F_{{\rm rec},2}(\boldsymbol{k}_1,\boldsymbol{k}_2)\,
    e^{-\tfrac{1}{2}k_1^2\bar{\sigma}_{tt}^2}
    \Big[
        P_{\rm nw}(k_1)P_{\rm nw}(k_2)
        \nonumber \\
        &\quad
        + e^{-k_1^2\bar{\sigma}_{\rm cross}^2}
          P_{\rm w}(k_1)P_{\rm nw}(k_2)
        \nonumber \\
        &\quad
        + e^{-k_2^2 \bar{\sigma}_{\rm cross}^2}
          e^{\boldsymbol{k}_2\cdot\boldsymbol{k}_3\left(\bar{\sigma}^2_{t\Psi} +\bar{\sigma}_{tt}^2\right)}
          P_{\rm nw}(k_1)P_{\rm w}(k_2)
        \nonumber \\
        &\quad
        + e^{-\tfrac{1}{2}(k_1^2+k_2^2+k_3^2)\bar{\sigma}_{\rm cross}^2}
          e^{\boldsymbol{k}_2\cdot\boldsymbol{k}_3\left(\bar{\sigma}_{t\Psi}^2 +\bar{\sigma}_{tt}^2\right)}
          P_{\rm w}(k_1)P_{\rm w}(k_2)
      \Big].
\end{align}

For the $B_{23}$ contribution, the simplified expression is
\begin{align}
    & B_{23,{\rm (pre,post,post)}, {\rm IR\text{-}resum}}(\boldsymbol{k}_2,\boldsymbol{k}_3)
    \nonumber \\
    &= 2\,F_{2}(\boldsymbol{k}_2,\boldsymbol{k}_3)\,
       e^{-\tfrac{1}{2}k_1^2\bar{\sigma}_{tt}^2}
    \Big[
        P_{\rm nw}(k_2)P_{\rm nw}(k_3)
        \nonumber \\
        &\quad
        + e^{-k_2^2\bar{\sigma}_{\rm cross}^2}
          e^{\boldsymbol{k}_2\cdot\boldsymbol{k}_3\left(\bar{\sigma}_{t\Psi}^2 + \bar{\sigma}_{tt}^2\right)}
          P_{\rm w}(k_2)P_{\rm nw}(k_3)
        \nonumber \\
        &\quad
        + e^{-k_3^2\bar{\sigma}_{\rm cross}^2}
          e^{\boldsymbol{k}_2\cdot\boldsymbol{k}_3\left(\bar{\sigma}_{t\Psi}^2 + \bar{\sigma}_{tt}^2\right)}
          P_{\rm nw}(k_2)P_{\rm w}(k_3)
        \nonumber \\
        &\quad
        + e^{-\tfrac{1}{2}(k_1^2+k_2^2+k_3^2)\bar{\sigma}_{\rm cross}^2}
          e^{\boldsymbol{k}_2\cdot\boldsymbol{k}_3\left(\bar{\sigma}_{t\Psi}^2 + \bar{\sigma}_{tt}^2\right)}
          P_{\rm w}(k_2)P_{\rm w}(k_3)
    \Big].
\end{align}

The $B_{31}$ contribution becomes
\begin{align}
    & B_{31,{\rm (pre,post,post)}, {\rm IR\text{-}resum}}(\boldsymbol{k}_3,\boldsymbol{k}_1)
    \nonumber \\
    &= 2\,F_{{\rm rec},2}(\boldsymbol{k}_3,\boldsymbol{k}_1)\,
       e^{-\tfrac{1}{2}k_1^2\bar{\sigma}_{tt}^2}
    \Big[
        P_{\rm nw}(k_3)P_{\rm nw}(k_1)
        \nonumber \\
        &\quad
        + e^{-k_3^2\bar{\sigma}_{\rm cross}^2}
          e^{\boldsymbol{k}_2\cdot\boldsymbol{k}_3\left(\bar{\sigma}_{t\Psi}^2 + \bar{\sigma}_{tt}^2\right)}
          P_{\rm w}(k_3)P_{\rm nw}(k_1)
        \nonumber \\
        &\quad
        + e^{-k_1^2\bar{\sigma}_{\rm cross}^2}
          P_{\rm nw}(k_3)P_{\rm w}(k_1)
        \nonumber \\
        &\quad
        + e^{-\tfrac{1}{2}(k_1^2+k_2^2+k_3^2)\bar{\sigma}_{\rm cross}^2}
          e^{\boldsymbol{k}_2\cdot\boldsymbol{k}_3\left(\bar{\sigma}_{t\Psi}^2 + \bar{\sigma}_{tt}^2\right)}
          P_{\rm w}(k_3)P_{\rm w}(k_1)
    \Big].
\end{align}
In this simplified description, the IR-resummed model for the pre--post--post cross bispectrum is fully characterized by the three parameters $\bar{\sigma}^2_{tt}$, $\bar{\sigma}^2_{\rm cross}$, and $\bar{\sigma}^2_{t\Psi}$.

\section{Future Prospects}
\label{sec:future}

The analytic IR-resummed expressions derived in this work rely on several controlled approximations introduced to factor the displacement-mapping exponent out of the ULPT convolution integrals. These include fixing angular configurations such as $\hat{r}_{1}=\hat{k}_{1}$ and adopting characteristic values such as $|\hat{r}_{1}-\hat{r}_{2}|\simeq 4/3$ in the wiggle--wiggle contribution. While these approximations are physically motivated by the structure of ULPT and by the analogous treatment of the power spectrum, their quantitative accuracy for the bispectrum remains to be validated.

A natural direction for future work is therefore the direct numerical evaluation of the \emph{full} ULPT convolution integrals associated with the displacement-mapping factor, without extracting the exponential outside the integral. Such a computation would eliminate the need for fixed-angle assumptions and would automatically reproduce all nonlinear BAO suppression effects and cross-bispectrum exponential factors derived analytically in this paper. Furthermore, experience from the power-spectrum analysis suggests that a precise numerical evaluation of the displacement convolution is not only important for an accurate description of BAO damping but is also likely to lead to significant improvements in modeling the full bispectrum shape \cite{Sugiyama:2025myq}.

\section{Conclusion}
\label{sec:conclusion}

In this work, we have developed a unified analytic framework for modeling the real-space dark matter bispectrum within Unified Lagrangian Perturbation Theory (ULPT). By reorganizing standard Lagrangian perturbation theory into the Jacobian deviation and the displacement-mapping effect, we obtained a consistent description of all auto and cross bispectra constructed from pre- and post-reconstruction density fields.

We derived general one-loop ULPT expressions for the bispectrum and presented a detailed analysis of their infrared behavior. We showed that ULPT guarantees exact, nonperturbative IR cancellation when all density fields share the same displacement, thereby ensuring IR safety. By implementing the wiggle--no-wiggle decomposition at the level of the ULPT displacement-mapping sector, we further constructed IR-resummed bispectrum models within ULPT and clarified the analytic origin of BAO damping.

A key theoretical advance of this paper is the nonperturbative treatment of the wiggle--wiggle (${\cal O}(P_{\rm w}^2)$) contributions to the bispectrum. Unlike the power spectrum, the bispectrum already contains ${\cal O}(P_{\rm w}^2)$ terms at the lowest nontrivial order, namely at tree level, because it is constructed from products of two linear power spectra. A precise understanding of the BAO damping associated with these terms is therefore essential for building an accurate and systematically controlled bispectrum model. By deriving the IR-resummed bispectrum directly within the ULPT framework, we obtained a description of BAO damping that treats both ${\cal O}(P_{\rm w})$ and ${\cal O}(P_{\rm w}^2)$ contributions on equal footing and clarifies their distinct scale dependences in a controlled manner. Moreover, as in the case of the power spectrum, we find that for $\Lambda$CDM-like models the wiggle contribution inside the bispectrum mode-coupling terms can be safely neglected as a practical numerical approximation in the IR-resummed formulation, providing a useful simplification for future applications.

We further extended the IR-resummed bispectrum modeling systematically to all combinations of pre- and post-reconstruction density fields. For mixed configurations involving both pre- and post-reconstruction fields, we showed that the entire shape of the cross bispectrum is universally multiplied by an exponential damping factor whose characteristic scale is determined solely by the wavenumber of the single field whose displacement differs from the other two. These mixed bispectra also exhibit characteristic BAO-phase modulations that reflect reconstruction effects at the three-point level.

Our results establish ULPT as a consistent and predictive framework for bispectrum modeling that unifies nonlinear gravitational evolution, BAO damping, and reconstruction within a single analytic structure. Although the present study has focused on real-space dark matter, the ULPT formulation preserves this unified structure in the presence of galaxy bias and redshift-space distortions, allowing for straightforward and fully consistent extensions to biased tracers and redshift space. The analytic results presented here thus provide a solid theoretical foundation for future numerical implementations of the ULPT bispectrum. The remaining challenge is computational, namely the development of efficient and accurate algorithms for evaluating the ULPT bispectrum, which will be essential for fully exploiting its predictive power in upcoming precision large-scale structure surveys.

\begin{acknowledgments}
N.S. acknowledges financial support from JSPS KAKENHI Grant No. 25K07343, administratively hosted by the National Astronomical Observatory of Japan. The author acknowledges the use of \textit{ChatGPT} (OpenAI) for assistance in language refinement and literature exploration during the preparation of this manuscript.
\end{acknowledgments}

\appendix

\section{Intuitive derivation of the universal exponential factor in cross bispectra}
\label{sec:appendix_universal}

In this appendix, we provide an intuitive and concise derivation of the universal exponential damping factor that appears in cross bispectra involving density fields with different displacement fields, such as those constructed from pre- and post-reconstruction density fields.

We begin by recalling that, as discussed in detail in Sec.~\ref{sec:intuitive}, in the IR limit the nonlinear effects of the density field can be absorbed into a uniform coordinate transformation induced by a long-wavelength displacement field, so that the fluctuation is effectively described by evaluating the density field at a uniformly shifted position. Applying the same reasoning to reconstructed fields, the three-point correlation function in the IR limit given in Eq.~\eqref{eq:three_IR} takes the following forms.

For the pre--pre--post configuration, we have
\begin{align}
    & \langle \delta(\boldsymbol{x}_1)\,\delta(\boldsymbol{x}_2)\,\delta_{\rm rec}(\boldsymbol{x}_3) \rangle
    \nonumber \\
    & \xrightarrow[\text{IR}]{}
    \Big\langle
    \delta^{(1)}(\boldsymbol{x}_1-\overline{\boldsymbol{\Psi}}^{(1)})
    \delta^{(1)}(\boldsymbol{x}_2-\overline{\boldsymbol{\Psi}}^{(1)})
    \delta_{\rm rec}^{(2)}(\boldsymbol{x}_3-\overline{\boldsymbol{\Psi}}_{\rm rec}^{(1)})
    \Big\rangle \nonumber \\
    & \qquad + \text{2 perms.}
    \label{eq:three_IR_preprepost}
\end{align}
For the pre--post--post configuration, we similarly obtain
\begin{align}
    & \langle \delta(\boldsymbol{x}_1)\,\delta_{\rm rec}(\boldsymbol{x}_2)\,\delta_{\rm rec}(\boldsymbol{x}_3) \rangle
    \nonumber \\
    & \xrightarrow[\text{IR}]{}
    \Big\langle
    \delta^{(1)}(\boldsymbol{x}_1-\overline{\boldsymbol{\Psi}}^{(1)})
    \delta_{\rm rec}^{(1)}(\boldsymbol{x}_2-\overline{\boldsymbol{\Psi}}_{\rm rec}^{(1)})
    \delta_{\rm rec}^{(2)}(\boldsymbol{x}_3-\overline{\boldsymbol{\Psi}}_{\rm rec}^{(1)})
    \Big\rangle \nonumber \\
    & \qquad + \text{2 perms.}
    \label{eq:three_IR_prepostpost}
\end{align}
Here $\overline{\boldsymbol{\Psi}}_{\rm rec}^{(1)} = \overline{\boldsymbol{\Psi}}^{(1)} + \bar{\boldsymbol{t}}^{(1)}$ is the reconstructed long-wavelength displacement field in the IR limit, and $\bar{\boldsymbol{t}}^{(1)}$ denotes the reconstruction shift vector evaluated at the origin.

By statistical translational invariance, in the pre--pre--post case we may shift all coordinates by $\overline{\boldsymbol{\Psi}}^{(1)}$, while in the pre--post--post case we shift them by $\overline{\boldsymbol{\Psi}}_{\rm rec}^{(1)}$. This yields
\begin{align}
    & \langle \delta(\boldsymbol{x}_1)\,\delta(\boldsymbol{x}_2)\,\delta_{\rm rec}(\boldsymbol{x}_3) \rangle
    \nonumber \\
    & \xrightarrow[\text{IR}]{}
    \Big\langle
    \delta^{(1)}(\boldsymbol{x}_1)
    \delta^{(1)}(\boldsymbol{x}_2)
    \delta_{\rm rec}^{(2)}(\boldsymbol{x}_3-\overline{\boldsymbol{t}}^{(1)})
    \Big\rangle
    + \text{2 perms.}
    \label{eq:three_IR_preprepost_2}
\end{align}
\begin{align}
    & \langle \delta(\boldsymbol{x}_1)\,\delta_{\rm rec}(\boldsymbol{x}_2)\,\delta_{\rm rec}(\boldsymbol{x}_3) \rangle
    \nonumber \\
    & \xrightarrow[\text{IR}]{}
    \Big\langle
    \delta^{(1)}(\boldsymbol{x}_1+\bar{\boldsymbol{t}}^{(1)})
    \delta_{\rm rec}^{(1)}(\boldsymbol{x}_2)
    \delta_{\rm rec}^{(2)}(\boldsymbol{x}_3)
    \Big\rangle
    + \text{2 perms.}
    \label{eq:three_IR_prepostpost_2}
\end{align}
These expressions show that, in the pre--pre--post case, the contribution of the long-wavelength mode $\bar{\boldsymbol{t}}^{(1)}$ remains only in the post-reconstruction density field, while in the pre--post--post case it remains only in the pre-reconstruction field. Therefore, for cross correlations between density fields with different displacement fields, exact IR cancellation no longer holds.

In Fourier space, modifying Eq.~\eqref{eq:fourier_IR}, the pre--pre--post case becomes
\begin{align}
    & \langle \tilde{\delta}(\boldsymbol{k}_1)\tilde{\delta}(\boldsymbol{k}_2)\tilde{\delta}_{\rm rec}(\boldsymbol{k}_3) \rangle
    \nonumber \\
    & \xrightarrow[\text{IR}]{}
    \Big\langle
    e^{-i(\boldsymbol{k}_1+\boldsymbol{k}_2+\boldsymbol{k}_3)\cdot\overline{\boldsymbol{\Psi}}^{(1)}}
    e^{-i\boldsymbol{k}_3\cdot\bar{\boldsymbol{t}}^{(1)}}
    \Big\rangle
    \langle
    \tilde{\delta}^{(1)}(\boldsymbol{k}_1)\tilde{\delta}^{(1)}(\boldsymbol{k}_2)\tilde{\delta}_{\rm rec}^{(2)}(\boldsymbol{k}_3)
    \rangle
    \nonumber \\
    & \qquad + \text{2 perms.}
    \nonumber \\
    & =
    \big\langle e^{-i\boldsymbol{k}_3\cdot\bar{\boldsymbol{t}}^{(1)}} \big\rangle
    \langle
    \tilde{\delta}^{(1)}(\boldsymbol{k}_1)\tilde{\delta}^{(1)}(\boldsymbol{k}_2)\tilde{\delta}_{\rm rec}^{(2)}(\boldsymbol{k}_3)
    \rangle
    + \text{2 perms.}
    \label{eq:fourier_IR_preprepost}
\end{align}
while for the pre--post--post case we obtain
\begin{align}
    & \langle \tilde{\delta}(\boldsymbol{k}_1)\tilde{\delta}_{\rm rec}(\boldsymbol{k}_2)\tilde{\delta}_{\rm rec}(\boldsymbol{k}_3) \rangle
    \nonumber \\
    & \xrightarrow[\text{IR}]{}
    \Big\langle
    e^{-i(\boldsymbol{k}_1+\boldsymbol{k}_2+\boldsymbol{k}_3)\cdot\overline{\boldsymbol{\Psi}}^{(1)}}
    e^{-i(\boldsymbol{k}_2+\boldsymbol{k}_3)\cdot\bar{\boldsymbol{t}}^{(1)}}
    \Big\rangle
    \nonumber \\
    & \quad \times
    \langle
    \tilde{\delta}^{(1)}(\boldsymbol{k}_1)\tilde{\delta}_{\rm rec}^{(1)}(\boldsymbol{k}_2)\tilde{\delta}_{\rm rec}^{(2)}(\boldsymbol{k}_3)
    \rangle
    + \text{2 perms.}
    \nonumber \\
    & =
    \big\langle e^{i\boldsymbol{k}_1\cdot\bar{\boldsymbol{t}}^{(1)}} \big\rangle
    \langle
    \tilde{\delta}^{(1)}(\boldsymbol{k}_1)\tilde{\delta}_{\rm rec}^{(1)}(\boldsymbol{k}_2)\tilde{\delta}_{\rm rec}^{(2)}(\boldsymbol{k}_3)
    \rangle
    + \text{2 perms.}
    \label{eq:fourier_IR_prepostpost}
\end{align}
The remaining expectation values evaluate to
\begin{align}
    \big\langle e^{-i\boldsymbol{k}_3\cdot\bar{\boldsymbol{t}}^{(1)}} \big\rangle
    & = \exp\!\left[-\frac{1}{2}k_3^2\bar{\sigma}^{2}_{tt}\right], \nonumber \\
    \big\langle e^{i\boldsymbol{k}_1\cdot\bar{\boldsymbol{t}}^{(1)}} \big\rangle
    & = \exp\!\left[-\frac{1}{2}k_1^2\bar{\sigma}^{2}_{tt}\right].
\end{align}
Therefore, the pre--pre--post and pre--post--post cross bispectra acquire the universal exponential damping factors
\[
\exp\!\left[-\frac{1}{2}k_3^2\bar{\sigma}^{2}_{tt}\right],
\qquad
\exp\!\left[-\frac{1}{2}k_1^2\bar{\sigma}^{2}_{tt}\right],
\]
respectively. This result clearly demonstrates that the characteristic damping scale is determined solely by the wavenumber of the single density field whose large-scale displacement differs from the other two, thereby providing a direct physical explanation for the universal exponential factor appearing in the mixed pre/post bispectra.

%
% The \nocite command causes all entries in a bibliography to be printed out
% whether or not they are actually referenced in the text. This is appropriate
% for the sample file to show the different styles of references, but authors
% most likely will not want to use it.
%\nocite{*}
\bibliography{ms}% Produces the bibliography via BibTeX.

\end{document}